\documentclass[11pt,a4paper]{article}

\usepackage{float}
\usepackage{jheppub} 
                     
\usepackage{braket,slashed,bm}
\usepackage{array,multirow}
\usepackage[normalem]{ulem}
\usepackage{xcolor,cancel,youngtab}

\usepackage[T1]{fontenc} 

\usepackage{mathrsfs}

\usepackage{booktabs}
\usepackage{adjustbox}
\usepackage{mathtools}

\usepackage{soul}

\usepackage{tikz}
\usetikzlibrary{arrows,decorations.pathmorphing,backgrounds,positioning,fit,petri,automata,shadows,calendar,mindmap,
decorations.markings,calc,arrows.meta,bending}

\definecolor{labelkey}{rgb}{0,0.5,0.0}

\usepackage{xparse}
\usepackage{etoolbox}

\usepackage{feynmp}
\usepackage{hypcap}

\def\nn{\nonumber\\ }
\def\rd{{\rm d}}

\def\sc{{\overline s}}

\def\q{\mathsf{q}}

\renewcommand{\O}{\mathcal{O}}
\newcommand{\op}[3]{\O^{#2,#3}_{#1}}

\newcommand{\wc}[3]{L^{#2,#3}_{#1}}

\newcommand{\hc}{\mathrm{h.c.}}
\newcommand{\p}{\partial}

\newcommand{\nnn}{\nonumber\\[-0.4cm] }
\newcommand{\nns}{\nonumber\\*}

\NewDocumentCommand{\Op}{ m m O{} o }{
	\O^{\ifblank{#3}{}{#3,}#2 }_{\IfNoValueTF{#4}{#1}{\substack{#1\\#4}}}
}
\NewDocumentCommand{\lwc}{ m m O{} o }{
	L^{\ifblank{#3}{}{#3,}#2 }_{\IfNoValueTF{#4}{#1}{\substack{#1\\#4}}}
}
\NewDocumentCommand{\dlwc}{ m m O{} o }{
	{\dot L}^{\ifblank{#3}{}{#3,}#2 }_{\IfNoValueTF{#4}{#1}{\substack{#1\\#4}}}
}

{\count255=\time\divide\count255 by 60 \xdef\hourmin{\number\count255}
  \multiply\count255 by-60\advance\count255 by\time
  \xdef\hourmin{\hourmin:\ifnum\count255<10 0\fi\the\count255}}

\allowdisplaybreaks[1]

\title{Low-Energy Effective Field Theory below the Electroweak Scale: Anomalous Dimensions}

\author[a]{Elizabeth E.~Jenkins,}

\author[a]{Aneesh V.~Manohar,}

\author[a,1]{Peter Stoffer}\note{Corresponding author.}

\affiliation[a]{Department of Physics, University of California at San Diego, 9500 Gilman Drive,\\ La Jolla, CA 92093-0319, USA}

\abstract{We compute the  one-loop anomalous dimensions of the low-energy effective Lagrangian below the electroweak scale, up to terms of dimension six. The theory has 70 dimension-five and 3631 dimension-six Hermitian operators that preserve baryon and lepton number, as well as additional operators that violate baryon number and lepton number. The renormalization group equations for the quark and lepton masses and the QCD and QED gauge couplings are modified by dimension-five and dimension-six operator contributions. We compute the renormalization group equations from one insertion of dimension-five and dimension-six operators, as well as two insertions of dimension-five operators, to all terms of dimension less than or equal to six. The use of the equations of motion to eliminate operators can be ambiguous, and we show how to resolve this ambiguity by a careful use of field redefinitions.
}

\begin{document}
\maketitle



\section{Introduction}
\label{sec:Intro}

Current experimental data at the LHC is well-described by the Standard Model (SM), with electroweak gauge symmetry spontaneously broken by
a fundamental scalar doublet, and a Higgs boson of mass $\sim 125$~GeV.  In the absence of new particles up to energies of order $\sim 1$~TeV,
it is useful to test for new physics beyond the SM in a model-independent manner by constructing effective field theories containing only SM particles.
Standard Model Effective Field Theory (SMEFT) is the effective field theory based on the SM with all additional higher-dimensional operators built from
SM gauge multiplets, which include a single fundamental scalar doublet $H$~\cite{Buchmuller:1985jz,Grzadkowski:2010es,Jenkins:2013zja,Jenkins:2013wua,Alonso:2013hga}.
This EFT is currently under active study in order to quantify how closely high-energy LHC data
agrees with the predictions of the SM, and to search for new high-energy physics through the higher-dimensional operators which encode the effects of new physics at a scale $\Lambda$, above the electroweak symmetry-breaking scale, at which new particles are produced.  
A comprehensive review on SMEFT can be found in Ref.~\cite{Brivio:2017vri} and a \textsc{FeynRules} implementation of SMEFT has recently been presented in~\cite{Brivio:2017btx}.

Higgs Effective Field Theory (HEFT) is a more general EFT based on the SM with higher-dimensional operators
constructed using SM gauge multiplets, except that the requirement of SMEFT that the Higgs boson $h$ and the three Goldstone bosons of electroweak gauge-symmetry breaking transform as a complex scalar doublet is relaxed~\cite{Feruglio:1992wf,Grinstein:2007iv,Alonso:2012px,Buchalla:2013rka,Gavela:2014uta,Brivio:2016fzo}.
HEFT also is under active study~\cite{Gavela:2014uta,Alonso:2015fsp,Guo:2015isa,Buchalla:2017jlu,Alonso:2017tdy}, and fits to LHC data are being
performed to constrain the coefficients of its higher-dimensional operators~\cite{Brivio:2016fzo}.

An alternative and complementary way to search for new physics beyond the SM is to consider the low-energy effective field theory below the electroweak scale.
The low-energy effective field theory (LEFT) is constructed by integrating out the particles that acquire masses of the order of the electroweak scale in the spontaneously broken
high-energy effective field theory.  The particles with electroweak-scale masses are the $W^\pm$, $Z$, the top quark $t$ and the Higgs boson $h$.  
The LEFT has been extensively applied to $B$-meson and kaon physics, providing very accurate tests of the SM and constraints on physics beyond the SM (for reviews, see \cite{PDG,Buchalla:1995vs}).
In a recent paper~\cite{Jenkins:2017jig}, we constructed a complete set of independent operators for the LEFT below the electroweak scale up to dimension-six operators.  In this paper, we compute the complete one-loop anomalous dimensions of this LEFT operator basis.  The renormalization group equations (RGEs)
of the LEFT basis allow one to run the LEFT operator coefficients down to energies much smaller than the electroweak scale, where they can be compared to low-energy processes.
Together with the complete one-loop anomalous dimensions of the SMEFT operator basis up to dimension-six operators calculated 
in Ref.~\cite{Grojean:2013kd,Jenkins:2013zja,Jenkins:2013wua,Alonso:2013hga,Alonso:2014zka},\footnote{Some results for parts of the SMEFT anomalous-dimension matrix, with flavor neglected, also can be found in 
Refs.~\cite{Elias-Miro:2013gya,Elias-Miro:2013mua}.} and the tree-level matching of SMEFT operators onto the LEFT operators given
in Ref.~\cite{Jenkins:2017jig}, all low-energy operators in LEFT can be obtained using SMEFT as the high-energy theory to leading-log accuracy. 
Parts of the RGEs relevant for particular processes have been well-studied in the literature and are known to higher order~\cite{Buchalla:1995vs,Misiak:2004ew,Czakon:2006ss,Cirigliano:2012ab,Dekens:2013zca,Heeck:2013rpa,Pruna:2014asa,Bhattacharya:2015rsa,Aebischer:2015fzz,Davidson:2016edt,Crivellin:2017rmk,Bordone:2017anc,Misiak:2017woa,Cirigliano:2017azj,Celis:2017hod,Aebischer:2017gaw,Gonzalez-Alonso:2017iyc,Falkowski:2017pss}, but 
a systematic study of the entire RGE is new. In particular, we do not restrict the analysis to a scenario that derives from SMEFT as the high-energy theory, but consider all LEFT effects up to dimension six. These effects include terms quadratic in the dimension-five dipole-operator coefficients. Therefore, the one-loop anomalous dimensions of the LEFT operators derived in this work also can be used to compute low-energy processes using the more general HEFT as the high-energy effective field theory.
  
 The results given here allow one to systematically combine constraints on new physics at the LHC with constraints from low-energy measurements such as in $B$ decays, $\mu \to e \gamma$ transitions, electric and magnetic dipole moment measurements, etc., by relating the SMEFT Lagrangian at the scale $\mu$ of the order of a TeV to the low-energy Lagrangian at the scale $\mu$ of a few GeV or less.
  
The organization of this paper is as follows.  Section~\ref{sec:LEFT} gives a brief review of the LEFT operator basis up to dimension-six operators constructed in Ref.~\cite{Jenkins:2017jig}.
Section~\ref{sec:RGEcalculation} discusses interesting features of the one-loop anomalous dimensions in LEFT.  Some aspects of the LEFT power counting relevant for the RGEs are discussed. In addition, details on the use of the equations of motion and field redefinitions are explained.  Interesting cancellations occur due to approximate holomorphy of LEFT. 
Section~\ref{sec:Conclusions} presents our conclusions. 
The LEFT operator basis is reproduced in Tables~\ref{tab:oplist1} and~\ref{tab:oplist2} of Appendix~\ref{sec:LEFTBasis}.  
Appendix~\ref{sec:diagrams} displays the one-loop Feynman diagrams that are computed to obtain the LEFT renormalization group equations.
The explicit one-loop renormalization group equations of the LEFT operators are presented in Appendix~\ref{sec:RGE}.


\section{LEFT}
\label{sec:LEFT}

The gauge symmetry of LEFT is $SU(3) \times U(1)_Q$, and the matter content of the theory consists of the usual SM fermions, except
that there is no top quark in the theory, so the number of $u$-type quarks is $n_u=2$.  No scalar fields of the high-energy theory remain in LEFT.
In this paper, we follow the conventions of Ref.~\cite{Jenkins:2017jig}.  We use $\psi$ to denote a generic fermion field and $X$ to denote a generic field-strength tensor.  Operators 
are classified by their field content.  For example, the higher-dimension $d \ge 5$ operators of LEFT consist of dimension-five dipole operators $\psi^2 X$ and dimension-six operators
$X^3$ and $\psi^4$.

The QCD and QED Lagrangian is 
\begin{align}
	\label{eq:qcdqed}
	\mathcal{L}_{\rm QCD + QED} &= - \frac14 G_{\mu \nu}^A G^{A \mu \nu} -\frac14 F_{\mu \nu} F^{\mu\nu} + \theta_{\rm QCD} \frac{g^2}{32 \pi^2} G_{\mu \nu}^A \widetilde G^{A \mu \nu} +  \theta_{\rm QED} \frac{e^2}{32 \pi^2} F_{\mu \nu} \widetilde F^{\mu \nu} \nn
		&+ \sum_{\psi=u,d,e,\nu_L}\overline \psi i \slashed{D} \psi   - \left[ \sum_{\psi=u,d,e}  \overline \psi_{Rr} [M_\psi]_{rs} \psi_{Ls} + \text{h.c.} \right],
\end{align}
which contains QCD gauge interactions of $n_u=2$ $u$-type quarks and $n_d=3$ $d$-type quarks and QED gauge interactions of the $u$ and $d$ quarks and the $n_e=3$ charged leptons at dimension four, and Dirac-fermion mass terms for the $u$ quarks, $d$ quarks, and charged leptons $e$ at dimension three.  The $n_\nu =3$ left-handed neutrinos are gauge singlets with no mass term.  The gauge-covariant derivative is $D_\mu = \partial_\mu + i g T^A G^A_\mu + ie Q A_\mu$, where $g$ and $e$ are the QCD and QED gauge coupling constants, respectively.  The QCD and QED field strengths are $G^A_{\mu \nu}$ and $F_{\mu \nu}$, respectively.    At dimension four, $\theta$ terms for QCD and QED are included so that we can comment on holomorphy of the LEFT RGEs.  An approximate holomorphy was found for the SMEFT RGEs~\cite{Alonso:2014rga}.

Additional operators in LEFT beyond those in Eq.~(\ref{eq:qcdqed}) arise at dimension three
and at dimensions $d \ge 5$.  The complete LEFT operator basis up to dimension six is considered
in this work.
For $n_e=3$, $n_\nu=3$, $n_u=2$, and $n_d=3$, there are 5963 Hermitian operators in the LEFT operator basis up to dimension six. The complete set of operators is given in Appendix~\ref{sec:LEFTBasis}.

The dimension-three LEFT Lagrangian consists of the $\Delta L = \pm 2$ Majorana mass terms of the left-handed neutrinos,   
\begin{align}
	\mathcal{L}^{(3)}_{\slashed{L}} &= - \frac12 [M_\nu]_{rs} \mathcal{O}_{\substack{\nu \\ rs}} + \hc = -\frac12 [M_\nu]_{rs} (\nu_{Lr}^T C \nu_{Ls}) + \hc 
\end{align}
The Majorana neutrino mass matrix $M_\nu$ is symmetric in its generation indices, so $M_\nu^T = M_\nu$.  For $n_\nu =3$, there are six $\Delta L =2$ operators and six $\Delta L =-2$ conjugate operators.
The symmetry of the Majorana neutrino mass matrix is used to rewrite $M_\nu^* = M_\nu^\dagger$ in the RGEs.  However, it is important to emphasize
that $M_\nu$ as a Majorana mass matrix violates lepton number, whereas all the other fermion mass matrices appearing in Eq.~\eqref{eq:qcdqed} are Dirac mass matrices,
which do not violate baryon number or lepton number.

The dimension-five operators consist of $\Delta B = \Delta L = 0$ dipole operators for fermions $\psi=u$, $d$, $e$, which do not violate baryon number or lepton number,
and the $\Delta L = \pm 2$ dipole operators for the left-handed neutrinos.  The $\Delta B = \Delta L =0$ dimension-five LEFT Lagrangian
\begin{align}
	\mathcal{L}^{(5)} &= \sum_{\psi = e,u,d} \left( \lwc{\psi\gamma}{}[][rs] \ \mathcal{O}_{\substack{\psi \gamma \\ rs}} + \hc \right)
		+ \sum_{\psi=u,d} \left( \lwc{\psi G}{}[][rs] \ \mathcal{O}_{\substack{\psi G \\ rs}} + \hc \right)
\end{align}
contains electromagnetic dipole operators for the charged leptons $e$ and the charged $u$ and $d$ quarks and chromomagnetic dipole operators for the 
colored $u$ and $d$ quarks.  These $(\overline L R) X + \hc$ operators are defined in Table~\ref{tab:oplist1} of Appendix A.  For $n_e=3$, $n_u=2$ and $n_d=3$, there are 70 Hermitian operators.
The $\Delta L = \pm 2$ dimension-five LEFT Lagrangian
\begin{align}
	\label{eq:NeutrinoDipoles}
	\mathcal{L}^{(5)}_{\slashed{L}} &=  \lwc{\nu\gamma}{}[][rs] \ \mathcal{O}_{\substack{\nu \gamma \\ rs}} + \hc 
		=  \lwc{\nu\gamma}{}[][rs] (\nu_{Lr}^T C \sigma^{\mu \nu} \nu_{Ls}) F_{\mu \nu} + \hc 
\end{align}
contains electromagnetic dipole operators for the left-handed neutrinos.  These $(\nu \nu ) X + \hc$ operators also appear in Table~\ref{tab:oplist1}.  The neutrino dipole operators are antisymmetric in the flavor indices. For $n_\nu=3$, there are three $\Delta L=2$ dipole operators and three $\Delta L = -2$ conjugate operators.

The dimension-six operators also split into $\Delta B = \Delta L =0$ operators, which do not violate baryon number and lepton number,
and operator sectors that violate baryon number and/or lepton number.
The $\Delta B = \Delta L = 0$ operators consist of the two $X^3$ operators, $\mathcal{O}_G$ and $\mathcal{O}_{\widetilde G}$, and 78 four-fermion operator structures $\psi^4$,
which are further divided by their chiral structure into $(\overline L L) (\overline L L)$, $(\overline R R)(\overline R R)$, $(\overline L L) (\overline R R)$,
$(\overline L R)(\overline L R) + \hc$, and $(\overline L R)(\overline R L) + \hc$, appearing in Table~\ref{tab:oplist1}.  For $n_e=3$, $n_\nu=3$, $n_u=2$, and $n_d=3$, there are 3631 Hermitian operators.
The dimension-six operator sectors that violate baryon and/or lepton number consist of operators with $\Delta L = \pm 4$, $\Delta L = \pm 2$, $\Delta B = \Delta L = \pm 1$, and $\Delta B = - \Delta L = \pm 1$ appearing in Table~\ref{tab:oplist2} of Appendix A.  
For $n_e=3$, $n_\nu=3$, $n_u=2$, and $n_d=3$, there are six $\Delta L =  4$ operators, 600 $\Delta L =  2$ operators, 288 $\Delta B = \Delta L =  1$ operators, and 228 $\Delta B = - \Delta L =  1$ operators plus an equal number of conjugate operators, i.e.\ six $\Delta L =  -4$ operators, 600 $\Delta L =  -2$ operators, 288 $\Delta B = \Delta L =  -1$ operators, and 228 $\Delta B = - \Delta L =  -1$ operators.


\section{RGE Calculation}
\label{sec:RGEcalculation}

In this section, we present several features of the RGE calculation for LEFT. We comment on the power counting rule for LEFT, explain in more detail the use of the equations of motion in the calculation, and point out some interesting cancellations.
The main results of this article, the explicit RGEs, are given in Appendix~\ref{sec:RGE}.

\subsection{Power Counting}
\label{sec:PowerCounting}

The power-counting rule of the low-energy effective field theory below the electroweak scale is given by the canonical dimensions of the operators in the theory. The LEFT uses an expansion in inverse powers of the electroweak scale, $1/v$. The expansion is controlled by the small dimensionless parameters $p/v$ and $m/v$, where $p$ and $m$ are momenta and masses of the light SM particles that are the dynamical degrees of freedom in LEFT.

We write the LEFT Lagrangian schematically as
\begin{align}
	\label{eq:LEFTLagrangian}
	\mathcal{L}_\mathrm{LEFT} = \mathcal{L}_\mathrm{QCD+QED} + \mathcal{L}^{(3)}_{\slashed L} + \sum_{d\ge5} \sum_i \lwc{i}{(d)} \O_i^{(d)} \, ,
\end{align}
where the operator $\O_i^{(d)}$ has mass dimension $d$ and its operator coefficient $\lwc{i}{(d)}$ contains a suppression factor $1/v^{d-4}$.  A graph with insertions of operators of dimension $d_i \ge 5$ has the LEFT dimension
\begin{align}
	d = 4 + \sum_i ( d_i - 4 ) \, ,
\end{align}
because no positive powers of $v$ can be generated by the graph: all dynamical particles have masses that are parametrically smaller than $v$ and loop integrals do not generate powers of $v$ in dimensional regularization.

Obviously, if we consider the renormalization of LEFT up to and including dimension-six terms, we have to consider graphs with one insertion of a dimension-five or a dimension-six operator as well as graphs with two insertions of dimension-five operators.

In the special case that the LEFT derives from SMEFT as the high-energy theory, the SMEFT expansion in powers of $1/\Lambda$, where $\Lambda$ is the scale of new physics, is inherited by the LEFT. In particular, the dimension-five dipole terms of LEFT induced by SMEFT at the electroweak scale are of the order
\begin{align}
	\frac{1}{v} \frac{v^2}{\Lambda^2} = \frac{v}{\Lambda^2} \, .
\end{align}
Therefore, double-dipole insertions have the same SMEFT suppression as dimension-eight SMEFT contributions. However, such a strong suppression of double-insertions of dipole operators does not hold in the more general HEFT high-energy theory.  When the constraint that the Higgs boson be part of an electroweak doublet is relaxed in the high-energy theory, dipole operators are only suppressed by one power of $1/\Lambda$~\cite{Buchalla:2013rka,Brivio:2016fzo}, and the operators have primary dimension $d_p=5$ in the HEFT power 
counting of Ref.~\cite{Gavela:2016bzc}. Therefore, the effect of dipole operators at low energies is very interesting, because it can distinguish between a SMEFT and a HEFT high-energy theory.
Consequently, it is important to consistently take into account all contributions as dictated by the LEFT power-counting rule alone.

\subsection{Equations of Motion and Field Redefinitions}
\label{sec:EOM}

As is well known, in any EFT, operators that are related by the classical equations of motion (EOM) are redundant, and the elimination of redundant operators through the application of the EOM simply corresponds to a field redefinition, up to corrections that affect higher orders in the EFT expansion~\cite{Politzer:1980me,Georgi:1991ch,Arzt:1993gz}. The LEFT basis constructed in Ref.~\cite{Jenkins:2017jig} therefore excludes operators that are redundant due to the classical EOM, given by
\begin{align}
	\label{eq:EOM}
	i \slashed{D} \psi_r &= [M_\psi]_{rs} \psi_{Ls} + [M_\psi^\dagger]_{rs} \psi_{Rs} \, , \quad \psi = u, d, e \, , \nn
	(D_\mu G^{\mu\nu})^A &= g j^{A \nu} \, , \quad \p_\mu F^{\mu\nu} = e j_\mathrm{em}^\nu \, ,
\end{align}
where the QCD and QED currents are
\begin{align}
	j^{A \mu} = \sum_{\psi=u,d} \bar \psi T^A \gamma^\mu \psi \, , \quad
	j_\mathrm{em}^\mu = \sum_{\psi=u,d,e} \q_\psi\, \bar \psi  \gamma^\mu \psi \, ,
\end{align}
respectively.
The dimension-three Majorana neutrino mass terms change the left-handed neutrino EOM to
\begin{align}
	i \slashed\p \nu_{Lr} = [M_\nu^\dagger]_{rs} C {\bar\nu_{Ls}}^T \, .
\end{align}
When calculating the RGEs, one has to project the divergences that are generated in the loop calculation onto the LEFT basis by applying the EOM (and Fierz relations in the case of four-fermion structures) when necessary.

Since we are considering LEFT up to dimension-six operators, effects that are quadratic in dimension-five operator coefficients have to be taken into account.  Including these quadratic effects also affects the application of the EOM: in the case of a divergence generated by a loop diagram with an insertion of a dimension-five operator, the EOM including dimension-five corrections have to be employed when projecting onto the LEFT basis. The relevant EOM for $\psi=u, d, e$ including dimension-five operator corrections are
\begin{align}
	\label{eq:EOMdim5}
	i \slashed{D} \psi_r = [M_\psi]_{rs} \psi_{Ls} &+ [M_\psi^\dagger]_{rs} \psi_{Rs} - \left( \lwc{\psi\gamma}{}[][rs] \, \sigma^{\mu\nu} F_{\mu\nu}  \psi_{Rs} 
	+ \lwc{\psi\gamma}{*}[][sr] \, \sigma^{\mu\nu} F_{\mu\nu}  \psi_{Ls} \right) \nn
		& - \left( \lwc{\psi G}{}[][rs] \, \sigma^{\mu\nu} T^A G^A_{\mu\nu} \psi_{Rs} +\lwc{\psi G}{*}[][sr] \, \sigma^{\mu\nu} T^A G^A_{\mu\nu}  \psi_{Ls} \right)\, ,
\end{align}
where the gluonic dipoles in the second line are only present for $\psi = u, d$.
The application of the modified EOM is equivalent to a field redefinition, when terms quadratic in dimension-five operator coefficients are retained.

In Appendix~\ref{sec:diagrams}, we list all the Feynman diagrams that we computed to obtain the RGEs and indicate explicitly the cases where the EOM or field redefinitions have to be applied.
In the following, we discuss as an explicit example the double-dipole insertions into $\psi^2 X$ Green's functions ($X_{\mu\nu}$ is a field-strength tensor, either $G_{\mu\nu}$ or $F_{\mu \nu}$), in order to highlight some subtleties in the application of the EOM. The Feynman diagrams for these contributions are shown in Appendix~\ref{sec:2xDipoleToDipole}.  The dipole operators can be either left-chiral $\overline \psi \sigma^{\mu \nu}X_{\mu \nu} P_L \psi$ or right-chiral $\overline \psi \sigma^{\mu \nu}X_{\mu \nu} P_R \psi$.

In the case of the insertion of two left-chiral ($L$) or two right-chiral ($R$) operators, the calculation of the graphs in Appendix~\ref{sec:2xDipoleToDipole} is straightforward and results only in a contribution to the dipole operators. For example, the purely electromagnetic double-dipole insertions contribute to the RGEs of the electromagnetic dipoles as:
\begin{align}
	\dlwc{e\gamma}{}[][rs] &= - 12 e \q_e \lwc{e\gamma}{}[][rw] [M_e]_{wv} \lwc{e\gamma}{}[][vs] \, , \nn
	\dlwc{u\gamma}{}[][rs] &= - 12 e \q_u \lwc{u\gamma}{}[][rw] [M_u]_{wv} \lwc{u\gamma}{}[][vs] \, , \nn
	\dlwc{d\gamma}{}[][rs] &= - 12 e \q_d \lwc{d\gamma}{}[][rw] [M_d]_{wv} \lwc{d\gamma}{}[][vs] \, .
\end{align}
In the case of the insertion of both one $L$ and one $R$ dipole operator, the situation is more complicated: 
the divergence of the loop calculation has an explicit momentum dependence. We map this divergence onto a set of gauge-invariant counterterm operators containing covariant derivatives. As the covariant derivatives can either contribute one (from the $i g A_\mu $) or no gauge boson (from the $\partial_\mu$), the same counterterms have to cancel divergences of $\psi^2$ and $\psi^2 X$ one-loop Green's functions with double-dipole insertions. Therefore, one has to consider the $\psi^2 X$ diagrams in Appendix~\ref{sec:2xDipoleToDipole} together with the $\psi^2$ diagrams in Appendix~\ref{sec:2xDipoleToPsi2}.  When combined, these diagrams give a gauge-invariant set of counterterms.  After mapping the divergences onto the 
gauge-invariant counterterm operators, these intermediate counterterm operators can be converted to LEFT basis operators by using the EOM. 

For example, for the case of insertion of one $L$ and one $R$ electromagnetic dipole, we use the following basis of gauge-invariant Hermitian operators:
\begin{align}\label{3.9}
	&\lwc{\psi\gamma}{*}[][wr] \lwc{\psi\gamma}{}[][vs] \, \frac{1}{2} \left(  i\bar \psi_{Rr} \overleftarrow{\slashed D} [M^\dagger M]_{wv} \psi_{Rs} - i \bar \psi_{Rr} [M^\dagger M]_{wv} \slashed D \psi_{Rs} \right), \nn
	&\lwc{\psi\gamma}{*}[][wr] \lwc{\psi\gamma}{}[][vs] \, \frac{1}{2} \left(  i \bar \psi_{Rr} \overleftarrow{\slashed D}^3 \psi_{Rs} - i \bar \psi_{Rr} \slashed D^3 \psi_{Rs} \right) \delta_{wv} , \nn
	&\lwc{\psi\gamma}{*}[][wr] \lwc{\psi\gamma}{}[][vs] \, \frac{1}{2} \left( i \bar \psi_{Rr} \overleftarrow{\slashed D} \sigma^{\mu\nu} F_{\mu\nu} \psi_{Rs} - i \bar \psi_{Rr} \sigma^{\mu\nu} F_{\mu\nu} \slashed D \psi_{Rs} \right) \delta_{wv} , \nn
		&\lwc{\psi\gamma}{*}[][wr] \lwc{\psi\gamma}{}[][vs] \, \bar \psi_{Rr} (D_\mu F^{\mu\nu} ) \gamma_\nu \psi_{Rs} \; \delta_{wv}  ,
\end{align}
and a similar basis with opposite chiralities for the case of insertion of one $R$ and one $L$ electromagnetic dipole.  By applying the EOM, these operators are transformed into mass terms, dipole operators, and four-fermion operators. The one- and three-derivative operator coefficients are first fixed by the $\psi^2$ graphs in Appendix~\ref{sec:2xDipoleToPsi2}. The graphs in Appendix~\ref{sec:2xDipoleToDipole} then determine the remaining coefficients. The three-derivative operator can be written in several ways.  For example,
\begin{align}
	\label{3.23}
	O_1 &=\bar\psi (i \slashed D)^3 \psi \, , \qquad
	O_2 =\bar\psi (i \slashed D) D^2 \psi \, ,
\end{align}
are both $\psi^2 D^3$ operators.
The $\psi^2$ graphs give the same coefficient for either $O_1$ or $O_2$, since they pick out the $\partial$ part of $D$. However, the two operators have different $\psi^2 X$ matrix elements, since
\begin{align}
	\label{3.24}
	\slashed{D}\slashed{D} &= D^2 + \frac12 g \sigma^{\mu \nu} X_{\mu \nu} \, ,
\end{align}
so that the $\psi^2 D X$ coefficients in Eq.~(\ref{3.9}) depend on the choice of $O_1$ or $O_2$. Of course, the full Lagrangian is the same in either case, since Eq.~(\ref{3.24}) can be used to convert between the two forms. We have chosen to use $(i \slashed D)^3$ since it simplifies the use of the EOM.

Consider the three-derivative operator in Eq.~(\ref{3.9}), which can be written as
\begin{align}
	\label{3.10}
	\bar\psi_R  L^\dagger L (i \slashed D)^3 \psi_R \, ,
\end{align}
where the dipole coefficients $\lwc{\psi\gamma}{*}[][wr]$ and $\lwc{\psi\gamma}{}[][vs]$ are
represented by the matrices $L^\dagger$ and $L$. This operator illustrates a subtlety in eliminating derivatives using the EOM. There is an apparent ambiguity: one can transform the operator by integration by parts (since it is in the Lagrangian) so that one of the three derivatives acts on the other fermion,
\begin{align}
	\label{3.11}
	\bar\psi_R (-i\overleftarrow{\slashed D}) L^\dagger L (i\slashed D)^2 \psi_R \, .
\end{align}
The subsequent application of the EOM Eq.~(\ref{eq:EOM}) to Eqs.~(\ref{3.10}) and~(\ref{3.11}) leads to two different results,
differing by the flavor index structure:
\begin{align}
	\label{eq:D3EOM}
	\bar\psi_R  L^\dagger L (i\slashed D)^3 \psi_R &\stackrel{\mathrm{EOM}}{\mapsto}  \bar\psi_R L^\dagger L M M^\dagger M \psi_L  \, , \nn
	\bar\psi_R (-i\overleftarrow{\slashed D}) L^\dagger L (i\slashed D)^2 \psi_R 
		&\stackrel{\mathrm{EOM}}{\mapsto}  \bar\psi_L M^\dagger L^\dagger L M M^\dagger \psi_R  \, ,
\end{align}
where we also use matrix notation in flavor space for the mass matrices. We have two different forms for the final result, neither of which is Hermitian, and neither of which is correct.

The ambiguity is resolved by noting that in a field theory, what is allowed is the use of field redefinitions. Making a field redefinition is often referred to as ``using the EOM,'' because the two are sometimes equivalent. To understand this better, consider the simpler example of a one-derivative operator
\begin{align}
	\label{3.12}
	\delta L &=\epsilon \bar\psi_R H (i\slashed D) \psi_R 
\end{align}
added to the lowest order Lagrangian
\begin{align}
	\label{3.14}
	L &= \bar\psi (i\slashed D) \psi - \overline \psi_R M \psi_L - \overline \psi_L M^\dagger \psi_R \, .
\end{align}
In Eq.~(\ref{3.12}), $H$ is Hermitian, so that the operator $\delta L$ is Hermitian. Acting to the right with $i\slashed{D}$ and using the equations of motion gives
\begin{align}
	\label{3.15}
	\delta L &=\epsilon \bar\psi_R H M \psi_L \, ,
\end{align}
whereas acting to the left and using the equations of motion gives
\begin{align}
	\label{3.16}
	\delta L &= \epsilon \bar\psi_L M^\dagger H  \psi_R \, ,
\end{align}
neither of which is Hermitian. Instead of using the EOM, make a field redefinition
\begin{align}
	\label{3.17}
	\psi_R \to \psi_R + \epsilon A \psi_R  
\end{align}
where $A$ is an arbitrary matrix. Working to first order in $\epsilon$, $\delta L$ after the field redefinition on the lowest order Lagrangian is
\begin{align}
	\label{3.18}
	\delta L &=\epsilon \bar\psi_R H (i\slashed D) \psi_R 
  +\epsilon \overline \psi_R (A+A^\dagger) (i\slashed{D}) \psi_R - \epsilon \overline \psi_R A^\dagger M \psi_L
  -\epsilon \overline \psi_L M^\dagger A \psi_R \, .
\end{align}
The field redefinition Eq.~(\ref{3.17}) can be chosen to eliminate the $i\slashed{D}$ term,
\begin{align}
	\label{3.19}
	H + A + A^\dagger &=0 \, .
\end{align}
The general solution of Eq.~(\ref{3.19}) is
\begin{align}
	\label{3.20}
	A &= -\frac 12 H + i B \, ,
\end{align}
for Hermitian $B$. The resulting $\delta L$ is
\begin{align}
	\label{3.21}
	\delta L &=  \epsilon \overline \psi_R \left(\frac12 H + i B\right) M \psi_L
  + \epsilon \overline \psi_L M^\dagger  \left(\frac12 H - i B \right) \psi_R \, ,
\end{align}
which is Hermitian, in contrast to either Eq.~(\ref{3.15}) or Eq.~(\ref{3.16}). The simplest choice is to take $B=0$ in Eq.~(\ref{3.21}). There is freedom to choose an arbitrary value of $B$, however, which corresponds to the freedom to make a chiral rotation,
\begin{align}
	\label{3.22}
	\psi_R \to U \psi_R
\end{align}
with $U = \exp i \epsilon B$. The proper way to eliminate the derivative operator Eq.~(\ref{3.12}) is to make a field redefinition instead of naively using the EOM, which leads to an incorrect result. A correct result is also obtained if a manifestly Hermitian operator is used that differs from~\eqref{3.12} by a total derivative. However, the ambiguity persists even in this case.

The ambiguity in Eq.~(\ref{eq:D3EOM}) is similar to that for the one-derivative term.
The general chiral field redefinition that removes the $\slashed D^3$ terms leads to the following change of the Lagrangian:
\begin{align}
		\delta \mathcal{L} &= i \bar \psi_L (A_L + A_L^\dagger) \slashed D^3 \psi_L + i \bar \psi_R (A_R + A_R^\dagger) \slashed D^3 \psi_R  \nn
			&\quad + \bar \psi_L ( M^\dagger M M^\dagger A_R + A_L M^\dagger M M^\dagger ) \psi_R + \bar \psi_R ( A_R^\dagger M M^\dagger M + M M^\dagger M A_L^\dagger ) \psi_L  \nn
			&\quad + \bar \psi_L ( B^\dagger M M^\dagger - M^\dagger M B^\dagger ) \psi_R + \bar \psi_R ( M M^\dagger B  - B M^\dagger M ) \psi_L \nn
			&\quad + \bar \psi_L ( M^\dagger C_R - C_L M^\dagger ) \psi_R + \bar \psi_R ( M C_L - C_R M ) \psi_L \, ,
\end{align}
where $A_{L,R}$, $B$, $C_{L,R}$ are matrices in flavor space of order $\epsilon$, and terms of $\O(\epsilon^2)$ are neglected.  The matrices $C_{L,R} = - C_{L,R}^\dagger$ are 
anti-Hermitian.  Matrices $B$, $C_{L,R}$, and the anti-Hermitian parts of $A_{L,R}$ are chiral rotations as in Eq.~(\ref{3.22}). We eliminate the three-derivative terms using the choice $B=C_L=C_R=0$, $A_R=A_R^\dagger= - \frac{1}{2} L^\dagger L$ to eliminate the operator in Eq.~(\ref{3.10}). A different choice of matrices for eliminating the three-derivative operator is equivalent to combining the RGE evolution with a $\mu$-dependent chiral rotation. Such a transformation is always allowed, even in QED or QCD.

Note that similarly to the simpler example above, the use of a manifestly Hermitian operator (at the level of the Lagrangian instead of the action) before the application of the EOM leads to a correct result but again exhibits the apparent ambiguity that is explained by the chiral rotation:
\begin{align}
	\label{eq:D3EOMHermitian}
	\bar\psi_R  L^\dagger L (i\slashed D)^3 \psi_R &+ \bar\psi_R (-i \overleftarrow{\slashed D})^3 L^\dagger L \psi_R \nn
		&\stackrel{\mathrm{EOM}}{\mapsto}  \bar\psi_R L^\dagger L M M^\dagger M \psi_L + \bar\psi_L M^\dagger M M^\dagger L^\dagger L \psi_R  \, , \nn
	\bar\psi_R (-i\overleftarrow{\slashed D}) L^\dagger L (i\slashed D)^2 \psi_R &+ \bar\psi_R (-i\overleftarrow{\slashed D})^2 L^\dagger L (i\slashed D) \psi_R \nn
		&\stackrel{\mathrm{EOM}}{\mapsto}  \bar\psi_L M^\dagger L^\dagger L M M^\dagger \psi_R +  \bar\psi_R M M^\dagger L^\dagger L M \psi_L \, .
\end{align}

This discussion shows that the ambiguity due to integration by parts is just related to a field redefinition and therefore has no effect on physical observables.

\subsection{Cancellations and Holomorphy}
\label{sec:Cancellations}

When calculating the RGEs in a diagrammatic approach, many relations and cancellations can be observed that are simple consequences of gauge invariance.  For example, as explained above, the divergences of some $\psi^2$ and $\psi^2 X$ loop diagrams must be related as they derive from the same gauge-invariant operator. Consider the gluonic double-dipole insertions in $\psi^2$ two-point functions, shown in the second diagram of~\eqref{diag:2xDipoleToPsi2}. The color structure of these contributions is given by
\begin{align}
	T^A T^A = C_F \, .	
\end{align}
The related double-dipole insertions in $\psi^2 X$ three-point functions are given by the purely gluonic triangle diagrams in~\eqref{diag:Psi2GDoubleDipoleTriangles} and the additional QCD topologies in~\eqref{diag:Psi2GDoubleDipoleAdditional}. The color structure of the triangle diagrams is given by
\begin{align}
	T^B T^A T^B = \left( C_F - \frac{1}{2} C_A \right) T^A \, ,
\end{align}
while the additional topologies contain a color factor
\begin{align}
	f^{ABC} T^B T^C = \frac{i}{2} C_A T^A \, .
\end{align}
When summing the contributions of the two groups of $\psi^2 X$ diagrams to obtain gauge-invariant structures that also contribute to the $\psi^2$ Green's function, the terms proportional to $C_A$ cancel and the resulting divergence is proportional to $C_F$, as in the case of the $\psi^2$ Green's function. Of course, this cancellation is expected and a simple consequence of gauge invariance. Therefore, relations and cancellations of this type provide a consistency check for the calculation. Splitting the covariant derivative into different contributions that have to be recombined to form a gauge-invariant expression is inherent to the traditional approach of calculating the RGEs with a Feynman-diagram expansion. The splitting and recombination could be avoided by using functional methods and a covariant-derivative expansion~\cite{Cheyette:1985ue,Gaillard:1985uh,Henning:2014wua,Drozd:2015rsp,Henning:2016lyp,Fuentes-Martin:2016uol}.

A different type of cancellation is observed in the case of gauge couplings, in analogy to the holomorphy of the SMEFT RGEs found in Ref.~\cite{Alonso:2014rga}.
We define the following linear combinations:
\begin{align}
	\tau_\text{QCD} &= i \frac{4\pi}{g^2} + \frac{\theta_{\text{QCD}}}{2\pi} , \qquad \tau_\text{QED} = i \frac{4\pi}{e^2} + \frac{\theta_{\text{QED}}}{2\pi} \, .
\end{align}
The RGEs of $\tau_\text{QCD}$ and $\tau_\text{QED}$ are obtained from the RGEs of the gauge couplings\footnote{Note that the RGEs of the $\theta$ terms, which are total derivatives and usually give a vanishing contribution in perturbation theory, can be calculated by allowing for momentum insertion into the operators.} given in Appendix~\ref{sec:GaugeCouplingRGEs}:
\begin{align}
	\dot \tau_\text{QCD} &= 8 \pi i b_{0,g} + \frac{64 \pi i}{g} (\lwc{uG}{}[][rs] [M_u]_{sr} +\lwc{dG}{}[][rs] [M_d]_{sr} ) \nn
		&\quad -\frac{32 \pi i}{g^2} 2  \Bigl( [M_u]_{ts}  \lwc{u G}{}[][sr] [M_u]_{rp} \lwc{u G}{}[][pt] +  [M_d]_{ts}  \lwc{d G}{}[][sr] [M_d]_{rp} \lwc{d G}{}[][pt] \Bigr) \, , \\
\nnn
	\dot \tau_\text{QED} &= 8 \pi i b_{0,e}+\frac{128  \pi i}{e} (\q_e \lwc{e\gamma}{}[][rs] [M_e]_{sr}+N_c \q_u \lwc{u\gamma}{}[][rs] [M_u]_{sr} +N_c \q_d \lwc{d\gamma}{}[][rs] [M_d]_{sr} ) \nn
		&\quad -\frac{32 \pi i}{e^2} 4  \Bigl( 2 [M_\nu]_{pr}  \lwc{\nu \gamma}{*}[][sr] [M_\nu]_{st} \lwc{\nu \gamma}{*}[][pt] +[M_e]_{ts}  \lwc{e \gamma}{}[][sr] [M_e]_{rp} \lwc{e \gamma}{}[][pt]  \nn
		&\qquad\qquad +  N_c [M_u]_{ts}  \lwc{u \gamma}{}[][sr] [M_u]_{rp} \lwc{u \gamma}{}[][pt] +  N_c [M_d]_{ts}  \lwc{d \gamma}{}[][sr] [M_d]_{rp} \lwc{d \gamma}{}[][pt] \Bigr) \, .
\end{align}
i.e.\ their $\beta$-function only depends on the self-dual couplings $\lwc{uG}{}[][]$, $\lwc{dG}{}[][]$ or $\lwc{\nu\gamma}{*}[][]$, $\lwc{e\gamma}{}[][]$, $\lwc{u\gamma}{}[][]$, $\lwc{d\gamma}{}[][]$, respectively, but not the anti-self-dual couplings ($\lwc{uG}{*}[][]$, $\lwc{dG}{*}[][]$ or $\lwc{\nu\gamma}{}[][]$, $\lwc{e\gamma}{*}[][]$, $\lwc{u\gamma}{*}[][]$, $\lwc{d\gamma}{*}[][]$, respectively).\footnote{$\lwc{\nu\gamma}{*}[][]$ is a $\overline L R$ operator coefficient, as are $\lwc{e\gamma}{}[][]$, etc.} Therefore, these RGEs respect the holomorphy found in Ref.~\cite{Alonso:2014rga}. We note, however, that in the case of the dipole operators, holomorphy is violated by the double-dipole insertions, i.e.\ the LEFT RGEs for the self-dual dipole couplings, given in Appendix~\ref{sec:DipoleRGEs}, depend on products of the self-dual and anti-self-dual dipole operator coefficients. Ref.~\cite{Cheung:2015aba} explained the holomorphy found in 
Ref.~\cite{Alonso:2014rga} using amplitude methods. Their argument does not hold in the presence of mass terms, and the violation of holomorphy we find is proportional to the fermion mass matrix.\footnote{We would like to thank  C.-H.~Shen for discussions on this point.}

In the LEFT RGEs, another cancellation happens that was already observed in the SMEFT RGEs~\cite{Alonso:2014rga}: the insertions of the $X^3$ operators in $\psi^2 X$ Green's function, shown in diagram~\eqref{diag:X3inDipole}, generate divergences proportional to $\psi^2 X$ operators, but also a divergence proportional to the operator 
$(D_\mu G^{\mu\nu}_A)j^A_\nu$, where $j^A_\nu$ is the QCD fermion current. The $X^3$ insertions into $X^2$ Green's functions, shown in the diagrams~\eqref{diag:X3inX2}, generate another divergence proportional to the operator $(D_\mu G^{\mu\nu}_A)(D^\lambda G_{\lambda\nu}^A)$, which exactly cancels the $(D_\mu G^{\mu\nu}_A)j^A_\nu$ contribution from the $\psi^2 X$ Green's functions using the EOM. This cancellation explains the absence of an $X^3$ contribution to the RGEs of the $\psi^4$ operators.

More explicitly, the $X^3$ insertion into $X^2$ Green's functions actually only generates a divergence $(\p_\mu G^{\mu\nu}_A)(\p^\lambda G_{\lambda\nu}^A)$, because only two external gluons are present. Additional divergences are generated by the $X^3$ insertions into $X^3$ and $X^4$ Green's functions, which combine with the divergence from the $X^2$ Green's function to the full gauge-invariant operator $(D_\mu G^{\mu\nu}_A)(D^\lambda G_{\lambda\nu}^A)$. Again, this is a simple consequence of gauge invariance.

In addition to the contribution to the $(D_\mu G^{\mu\nu}_A)(D^\lambda G_{\lambda\nu}^A)$ operator, the $X^3$ insertions into $X^3$ (and higher multi-gluon) Green's functions generate divergences proportional to the $X^3$ operator itself.\footnote{As the $X^3$ operator starts with three gluons, this divergence does not show up in the $X^2$ Green's function.} The coefficient of these divergences can be fixed by calculating only the $X^3$ Green's functions, shown in diagrams~\eqref{diag:X3X3Bulbs} and~\eqref{diag:X3X3Triangles}. Since there exists no other possible divergent structure, the $X^4$ and higher multi-gluon Green's functions provide no additional information: their divergences are already fixed by gauge invariance and their calculation can only provide a cross check.

\subsection{Flavor Indices}
\label{sec:FlavorIndices}

In the mixing of higher-dimensional operators into lower-dimensional operators, explicit mass matrices appear. For the RGEs, we are only interested in the divergent pieces of the loop diagrams. We work in the weak-eigenstate basis: all flavor indices in this work refer to weak-eigenstate indices and the mass matrices have not been diagonalized.
Since we use the lowest-order Lagrangian
\begin{align}
	\mathcal{L} = \bar\psi ( i \slashed D - M_\psi P_L - M_\psi^\dagger P_R ) \psi  \, ,
\end{align}
the UV-divergent parts can be obtained by expanding the fermion propagators as
\begin{align}
	\frac{i \slashed k}{k^2} + \frac{i}{k^2} (M_\psi^\dagger P_L + M_\psi P_R) + \frac{i \slashed k}{(k^2)^2}( M_\psi M_\psi^\dagger P_L + M_\psi^\dagger M_\psi P_R ) + \ldots
\end{align}
Subsequent diagonalization of the mass matrices introduces explicit CKM matrix elements in the down-type quark sector (see the discussion in Ref.~\cite{Jenkins:2017jig}).

Many operators in the LEFT Lagrangian exhibit symmetries in the flavor indices. When writing the LEFT Lagrangian as in Eq.~\eqref{eq:LEFTLagrangian}, it is understood that the sum runs not only over the operator types but also over the flavor indices of Wilson coefficients and operators, including the redundant indices. For example, in $\mathcal{L}^{(5)}_{\slashed L}$ of Eq.~\eqref{eq:NeutrinoDipoles}, the sum over the indices $r$ and $s$ runs over all flavors, although there are only three independent $\Delta L=2$ dipole operators. The sum over the flavor indices implies that only the part of the Wilson coefficients that has the same flavor symmetry as the operator actually contributes.   For instance, only the part of $\lwc{\nu\gamma}{}[][rs]$ that is antisymmetric in $r\leftrightarrow s$ actually contributes to the sum, whereas the symmetric part cancels. Although this is not mandatory, we prefer to explicitly symmetrize the RGEs for the Wilson coefficients given in Appendix~\ref{sec:RGE}. The fact that the Wilson coefficients are symmetrized in the same way as the operators can be used to simplify some of the expressions.

The case of the $\Delta L = 4$ four-neutrino operator is special: it transforms as the $\Yboxdim{5pt}\yng(2,2)$ representation of the flavor-symmetry group, and satisfies the symmetry relations
\begin{align}
	\Op{\nu\nu}{LL}[S][prst] &= \Op{\nu\nu}{LL}[S][stpr] = \Op{\nu\nu}{LL}[S][prts] = \Op{\nu\nu}{LL}[S][rpst] = - \Op{\nu\nu}{LL}[S][psrt] - \Op{\nu\nu}{LL}[S][ptsr] \, .
\end{align}
We assume that these same symmetries and antisymmetries have been imposed on the Wilson coefficients, i.e.\ we assume that
\begin{align}
	\lwc{\nu\nu}{LL}[S][prst] &= \lwc{\nu\nu}{LL}[S][stpr] = \lwc{\nu\nu}{LL}[S][prts] = \lwc{\nu\nu}{LL}[S][rpst] = - \lwc{\nu\nu}{LL}[S][psrt] - \lwc{\nu\nu}{LL}[S][ptsr] \, .
\end{align}
This assumption allows us to simplify the contribution of this operator to the running of the neutrino mass, given by tadpole diagrams shown in Appendix~\ref{sec:Psi4Tadpoles}. Explicitly, we have performed the following simplification:
\begin{align}
\label{3.36}
	[\dot M_\nu]_{rs} &= 8 \wc{\substack{\nu \nu \\ v w r s}}{S}{LL}  [ M^\dagger_\nu M_\nu M^\dagger_\nu]_{wv} + 8 \wc{\substack{\nu \nu \\ r s v w}}{S}{LL}  [ M^\dagger_\nu M_\nu M^\dagger_\nu]_{wv} 
			- 4  \wc{\substack{\nu \nu \\ v s r w}}{S}{LL}  [ M^\dagger_\nu M_\nu M^\dagger_\nu]_{wv} \nn
		&\quad - 4  \wc{\substack{\nu \nu \\ r  w v  s}}{S}{LL}  [ M^\dagger_\nu M_\nu M^\dagger_\nu]_{wv} - 4  \wc{\substack{\nu \nu \\ s v r w}}{S}{LL}  [ M^\dagger_\nu M_\nu M^\dagger_\nu]_{wv} 
			- 4  \wc{\substack{\nu \nu \\ w r v  s}}{S}{LL}  [ M^\dagger_\nu M_\nu M^\dagger_\nu]_{wv} \nn
		&= 24 \wc{\substack{\nu \nu \\ v w r s}}{S}{LL}  [ M^\dagger_\nu M_\nu M^\dagger_\nu]_{wv} \, ,
\end{align}
using both the symmetries of $\lwc{\nu\nu}{LL}[S][prst]$ and $[M_\nu]_{rs}$.  Eq.~(\ref{3.36}) gives the first term in the neutrino mass RGE, Eq.~\eqref{eq:NuMassRGE}.

\subsection{Loop Calculation and Results}
\label{sec:Results}

In Appendix~\ref{sec:RGE}, we give the complete one-loop RGE for the LEFT Lagrangian up to dimension-six operators. We performed the calculation with dimensional regularization in the 
$\overline{\text{MS}}$ scheme, and we used background-field gauge~\cite{Abbott:1980hw}. Use of background-field gauge ensures that only divergences that are manifestly gauge invariant, instead of only BRST invariant due to gauge fixing, are generated.

We use the notation
\begin{align}
	\dot L \equiv 16\pi^2 \mu \frac{\rd}{\rd\mu} L \, .
\end{align}

The one-loop running of the QCD and QED coupling constants in pure QCD and QED is defined by
\begin{align}
	\dot g = - b_{0,g} g^3 \, , \qquad 
	\dot e = - b_{0,e} e^3 \, ,
\end{align}
respectively, where the coefficients of the QCD and QED $\beta$-functions are given by
\begin{align}
	\label{eq:betaFunctions}
	b_{0,g} &=  \frac{11}{3}N_c - \frac23\left(n_u+n_d \right) \, , \nn
	b_{0,e}  &=-\frac43 \left( \q_e^2 n_e + N_c \q_d^2 n_d +N_c \q_u^2 n_u \right)  \, ,
\end{align}
with $n_u=2$, $n_d=3$, $n_e=3$, $\q_u=2/3$, $\q_d=-1/3$, $\q_e=-1$.

Since the LEFT has mass terms, the running of the Wilson coefficients of the operators of dimension $d$ gets contributions proportional to Wilson coefficients of operators of dimension $\ge d$ times powers of the mass terms. In particular, this means that the SM $\beta$-functions in Eq.~\eqref{eq:betaFunctions} are modified. (A similar result was found for the SMEFT in Ref.~\cite{Alonso:2013hga}, where there were terms with positive powers of the Higgs mass $M_H^2$.)

In the expressions for the RGEs, we use the following abbreviations:
\begin{align}
	\zeta_e &= \frac{8}{3} \left( 2 \lwc{\nu \gamma}{}[][wv] \lwc{\nu \gamma}{*}[][wv]  + \lwc{e \gamma}{}[][wv] \lwc{e \gamma}{*}[][wv] + N_c \lwc{u \gamma}{}[][wv] \lwc{u \gamma}{*}[][wv] + N_c \lwc{d \gamma}{}[][wv] \lwc{d \gamma}{*}[][wv] \right) \, , \nn
	\zeta_g &= \frac{4}{3}  \left( \lwc{u G}{}[][wv] \lwc{u G}{*}[][wv] + \lwc{d G}{}[][wv] \lwc{d G}{*}[][wv] \right) \, ,
\end{align}
and the $SU(3)$ color factors:
\begin{align}
	C_A &= N_c \, , \qquad C_F = \frac{N_c^2-1}{2N_c} \, , \qquad C_1 = \frac{N_c^2 - 1}{4N_c^2}  \, , \qquad C_d = \frac{N_c^2 - 4}{N_c} \, .
\end{align}
$C_1$ and $C_d$ arise in the color identities for box graphs~\cite{Manohar:2000hj}.

Although we performed the full LEFT RGE calculation from scratch, we were able to cross-check large parts of the LEFT RGEs with a single-operator insertion using the SMEFT 
RGEs~\cite{Jenkins:2013zja,Jenkins:2013wua,Alonso:2013hga}, by excluding diagrams with heavy particles and fixing the Higgs field to its vacuum expectation value. In these cross-checks, we found that the $X^3$ contribution to the gluonic dipoles given in~\cite{Alonso:2013hga} has the wrong sign.\footnote{This sign error has been fixed on the website~\cite{SMEFT-RGE}.} We agree with the sign of Ref.~\cite{Morozov:1985ef}.

The majority of the diagrams in Appendix~\ref{sec:diagrams} consists of double insertions of dipole operators. To our knowledge, these contributions have not been considered before in the literature. We point out that some double-dipole contributions to four-fermion operator coefficients come with large numerical prefactors of 96 or 192. The RGEs also satisfy the naive dimensional analysis (NDA)~\cite{Manohar:1983md} counting rules of Ref.~\cite{Jenkins:2013sda}, where the dimension-five operator $g \psi^2 X$ (including one factor of the gauge coupling) has NDA weight $w=0$, the dimension-six operators $\psi^4$ have NDA weight $w=1$, and the dimension-six operators $g^3 X^3$ have NDA weight $w=-1$. The NDA weights explain the order in coupling constants of different contributions to the LEFT RGEs, in the same way as occurred for the SMEFT RGEs discussed in Ref.~\cite{Jenkins:2013sda}.


\section{Conclusions}
\label{sec:Conclusions}

In this paper, we have presented the complete one-loop renormalization-group equations for the low-energy effective field theory below the electroweak scale up to and including dimension-six operators. Together with the one-loop RGEs for the SMEFT given in Refs.~\cite{Jenkins:2013zja,Jenkins:2013wua,Alonso:2013hga} and the tree-level matching equations presented in Ref.~\cite{Jenkins:2017jig}, these RGEs allow one to evolve all the effects of physics beyond the SM consisting of new particles at some heavy scale $\Lambda$ down to low energies, where precision measurements are performed, to leading-log accuracy.  The results allow one to consistently combine high-energy constraints from the LHC with constraints from low-energy precision measurements to leading-log accuracy.
This procedure is valid under the assumption that the Higgs particle belongs to a fundamental electroweak doublet, as realized in SMEFT.  
The LEFT RGEs of the present work also are valid when the high-energy EFT is HEFT, with a nonlinear realization of the electroweak symmetry breaking, rather than SMEFT. 

The LEFT RGEs in Appendix~\ref{sec:RGE} include all dimension-six effects in the LEFT power counting, in particular contributions quadratic in dimension-five dipole operators. Dipole operators are of particular interest in the context of HEFT, where they appear as dimension-five operators suppressed by a factor of $1/\Lambda$, in contrast to the case of SMEFT, where the dipole operators are suppressed by a factor of $v/\Lambda^2$.  In SMEFT, the dimension-five dipole operators are generated from the dimension-six SMEFT dipole operators $\psi^2 X H$ upon spontaneous symmetry breaking of electroweak gauge symmetry.  Thus, they are suppressed by an additional factor of $(v/\Lambda)$ relative to the HEFT case.
Precision measurements of dipole operators will be important for distinguishing between SMEFT and HEFT as the appropriate high-energy effective field theory.

\section*{Acknowledgements}
\addcontentsline{toc}{section}{Acknowledgements}

\begin{sloppypar}
This work was supported in part by  DOE Grant No.\ DE-SC0009919.
P.S.\ is supported by a grant of the Swiss National Science Foundation (Project No.\ P300P2\_167751).
\end{sloppypar}


\appendix


\clearpage

\section{LEFT Operator Basis}
\label{sec:LEFTBasis}

In this appendix, we reproduce for convenience the list of LEFT operators up to dimension six from Ref.~\cite{Jenkins:2017jig}.  Weak-eigenstate indices of the operators are not shown---e.g.\ $\op{ee}{V}{LL}$ with the weak-eigenstate indices 
included is $\op{\substack{ ee \\ prst } }{V}{LL}$.

\begin{table}[H]
\capstart
\begin{adjustbox}{width=0.85\textwidth,center}
\begin{minipage}[t]{3cm}
\renewcommand{\arraystretch}{1.51}
\small
\begin{align*}
\begin{array}[t]{c|c}
\multicolumn{2}{c}{\boldsymbol{\nu \nu+\hc}} \\
\hline
\O_{\nu} & (\nu_{Lp}^T C \nu_{Lr})  \\
\end{array}
\end{align*}
\end{minipage}
%
\begin{minipage}[t]{3cm}
\renewcommand{\arraystretch}{1.51}
\small
\begin{align*}
\begin{array}[t]{c|c}
\multicolumn{2}{c}{\boldsymbol{(\nu \nu) X+\hc}} \\
\hline
\O_{\nu \gamma} & (\nu_{Lp}^T C   \sigma^{\mu \nu}  \nu_{Lr})  F_{\mu \nu}  \\
\end{array}
\end{align*}
\end{minipage}
\begin{minipage}[t]{3cm}
\renewcommand{\arraystretch}{1.51}
\small
\begin{align*}
\begin{array}[t]{c|c}
\multicolumn{2}{c}{\boldsymbol{(\overline L R ) X+\hc}} \\
\hline
\O_{e \gamma} & \bar e_{Lp}   \sigma^{\mu \nu} e_{Rr}\, F_{\mu \nu}  \\
\O_{u \gamma} & \bar u_{Lp}   \sigma^{\mu \nu}  u_{Rr}\, F_{\mu \nu}   \\
\O_{d \gamma} & \bar d_{Lp}  \sigma^{\mu \nu} d_{Rr}\, F_{\mu \nu}  \\
\O_{u G} & \bar u_{Lp}   \sigma^{\mu \nu}  T^A u_{Rr}\,  G_{\mu \nu}^A  \\
\O_{d G} & \bar d_{Lp}   \sigma^{\mu \nu} T^A d_{Rr}\,  G_{\mu \nu}^A \\
\end{array}
\end{align*}
\end{minipage}
\begin{minipage}[t]{3cm}
\renewcommand{\arraystretch}{1.51}
\small
\begin{align*}
\begin{array}[t]{c|c}
\multicolumn{2}{c}{\boldsymbol{X^3}} \\
\hline
\O_G     & f^{ABC} G_\mu^{A\nu} G_\nu^{B\rho} G_\rho^{C\mu}  \\
\O_{\widetilde G} & f^{ABC} \widetilde G_\mu^{A\nu} G_\nu^{B\rho} G_\rho^{C\mu}   \\
\end{array}
\end{align*}
\end{minipage}
\end{adjustbox}
%

%
\mbox{}\\[-1.25cm]

\begin{adjustbox}{width=1.05\textwidth,center}
\begin{minipage}[t]{3cm}
\renewcommand{\arraystretch}{1.51}
\small
\begin{align*}
\begin{array}[t]{c|c}
\multicolumn{2}{c}{\boldsymbol{(\overline L L)(\overline L  L)}} \\
\hline
\op{\nu\nu}{V}{LL} & (\bar \nu_{Lp}  \gamma^\mu \nu_{Lr} )(\bar \nu_{Ls} \gamma_\mu \nu_{Lt})   \\
\op{ee}{V}{LL}       & (\bar e_{Lp}  \gamma^\mu e_{Lr})(\bar e_{Ls} \gamma_\mu e_{Lt})   \\
\op{\nu e}{V}{LL}       & (\bar \nu_{Lp} \gamma^\mu \nu_{Lr})(\bar e_{Ls}  \gamma_\mu e_{Lt})  \\
\op{\nu u}{V}{LL}       & (\bar \nu_{Lp} \gamma^\mu \nu_{Lr}) (\bar u_{Ls}  \gamma_\mu u_{Lt})  \\
\op{\nu d}{V}{LL}       & (\bar \nu_{Lp} \gamma^\mu \nu_{Lr})(\bar d_{Ls} \gamma_\mu d_{Lt})     \\
\op{eu}{V}{LL}      & (\bar e_{Lp}  \gamma^\mu e_{Lr})(\bar u_{Ls} \gamma_\mu u_{Lt})   \\
\op{ed}{V}{LL}       & (\bar e_{Lp}  \gamma^\mu e_{Lr})(\bar d_{Ls} \gamma_\mu d_{Lt})  \\
\op{\nu edu}{V}{LL}      & (\bar \nu_{Lp} \gamma^\mu e_{Lr}) (\bar d_{Ls} \gamma_\mu u_{Lt})  + \hc   \\
\op{uu}{V}{LL}        & (\bar u_{Lp} \gamma^\mu u_{Lr})(\bar u_{Ls} \gamma_\mu u_{Lt})    \\
\op{dd}{V}{LL}   & (\bar d_{Lp} \gamma^\mu d_{Lr})(\bar d_{Ls} \gamma_\mu d_{Lt})    \\
\op{ud}{V1}{LL}     & (\bar u_{Lp} \gamma^\mu u_{Lr}) (\bar d_{Ls} \gamma_\mu d_{Lt})  \\
\op{ud}{V8}{LL}     & (\bar u_{Lp} \gamma^\mu T^A u_{Lr}) (\bar d_{Ls} \gamma_\mu T^A d_{Lt})   \\[-0.5cm]
\end{array}
\end{align*}
\renewcommand{\arraystretch}{1.51}
\small
\begin{align*}
\begin{array}[t]{c|c}
\multicolumn{2}{c}{\boldsymbol{(\overline R  R)(\overline R R)}} \\
\hline
\op{ee}{V}{RR}     & (\bar e_{Rp} \gamma^\mu e_{Rr})(\bar e_{Rs} \gamma_\mu e_{Rt})  \\
\op{eu}{V}{RR}       & (\bar e_{Rp}  \gamma^\mu e_{Rr})(\bar u_{Rs} \gamma_\mu u_{Rt})   \\
\op{ed}{V}{RR}     & (\bar e_{Rp} \gamma^\mu e_{Rr})  (\bar d_{Rs} \gamma_\mu d_{Rt})   \\
\op{uu}{V}{RR}      & (\bar u_{Rp} \gamma^\mu u_{Rr})(\bar u_{Rs} \gamma_\mu u_{Rt})  \\
\op{dd}{V}{RR}      & (\bar d_{Rp} \gamma^\mu d_{Rr})(\bar d_{Rs} \gamma_\mu d_{Rt})    \\
\op{ud}{V1}{RR}       & (\bar u_{Rp} \gamma^\mu u_{Rr}) (\bar d_{Rs} \gamma_\mu d_{Rt})  \\
\op{ud}{V8}{RR}    & (\bar u_{Rp} \gamma^\mu T^A u_{Rr}) (\bar d_{Rs} \gamma_\mu T^A d_{Rt})  \\
\end{array}
\end{align*}
\end{minipage}
%
%
\begin{minipage}[t]{3cm}
\renewcommand{\arraystretch}{1.51}
\small
\begin{align*}
\begin{array}[t]{c|c}
\multicolumn{2}{c}{\boldsymbol{(\overline L  L)(\overline R  R)}} \\
\hline
\op{\nu e}{V}{LR}     & (\bar \nu_{Lp} \gamma^\mu \nu_{Lr})(\bar e_{Rs}  \gamma_\mu e_{Rt})  \\
\op{ee}{V}{LR}       & (\bar e_{Lp}  \gamma^\mu e_{Lr})(\bar e_{Rs} \gamma_\mu e_{Rt}) \\
\op{\nu u}{V}{LR}         & (\bar \nu_{Lp} \gamma^\mu \nu_{Lr})(\bar u_{Rs}  \gamma_\mu u_{Rt})    \\
\op{\nu d}{V}{LR}         & (\bar \nu_{Lp} \gamma^\mu \nu_{Lr})(\bar d_{Rs} \gamma_\mu d_{Rt})   \\
\op{eu}{V}{LR}        & (\bar e_{Lp}  \gamma^\mu e_{Lr})(\bar u_{Rs} \gamma_\mu u_{Rt})   \\
\op{ed}{V}{LR}        & (\bar e_{Lp}  \gamma^\mu e_{Lr})(\bar d_{Rs} \gamma_\mu d_{Rt})   \\
\op{ue}{V}{LR}        & (\bar u_{Lp} \gamma^\mu u_{Lr})(\bar e_{Rs}  \gamma_\mu e_{Rt})   \\
\op{de}{V}{LR}         & (\bar d_{Lp} \gamma^\mu d_{Lr}) (\bar e_{Rs} \gamma_\mu e_{Rt})   \\
\op{\nu edu}{V}{LR}        & (\bar \nu_{Lp} \gamma^\mu e_{Lr})(\bar d_{Rs} \gamma_\mu u_{Rt})  +\hc \\
\op{uu}{V1}{LR}        & (\bar u_{Lp} \gamma^\mu u_{Lr})(\bar u_{Rs} \gamma_\mu u_{Rt})   \\
\op{uu}{V8}{LR}       & (\bar u_{Lp} \gamma^\mu T^A u_{Lr})(\bar u_{Rs} \gamma_\mu T^A u_{Rt})    \\ 
\op{ud}{V1}{LR}       & (\bar u_{Lp} \gamma^\mu u_{Lr}) (\bar d_{Rs} \gamma_\mu d_{Rt})  \\
\op{ud}{V8}{LR}       & (\bar u_{Lp} \gamma^\mu T^A u_{Lr})  (\bar d_{Rs} \gamma_\mu T^A d_{Rt})  \\
\op{du}{V1}{LR}       & (\bar d_{Lp} \gamma^\mu d_{Lr})(\bar u_{Rs} \gamma_\mu u_{Rt})   \\
\op{du}{V8}{LR}       & (\bar d_{Lp} \gamma^\mu T^A d_{Lr})(\bar u_{Rs} \gamma_\mu T^A u_{Rt}) \\
\op{dd}{V1}{LR}      & (\bar d_{Lp} \gamma^\mu d_{Lr})(\bar d_{Rs} \gamma_\mu d_{Rt})  \\
\op{dd}{V8}{LR}   & (\bar d_{Lp} \gamma^\mu T^A d_{Lr})(\bar d_{Rs} \gamma_\mu T^A d_{Rt}) \\
\op{uddu}{V1}{LR}   & (\bar u_{Lp} \gamma^\mu d_{Lr})(\bar d_{Rs} \gamma_\mu u_{Rt})  + \hc  \\
\op{uddu}{V8}{LR}      & (\bar u_{Lp} \gamma^\mu T^A d_{Lr})(\bar d_{Rs} \gamma_\mu T^A  u_{Rt})  + \hc \\
\end{array}
\end{align*}
\end{minipage}

\begin{minipage}[t]{3cm}
\renewcommand{\arraystretch}{1.51}
\small
\begin{align*}
\begin{array}[t]{c|c}
\multicolumn{2}{c}{\boldsymbol{(\overline L R)(\overline L R)+\hc}} \\
\hline
\op{ee}{S}{RR} 		& (\bar e_{Lp}   e_{Rr}) (\bar e_{Ls} e_{Rt})   \\
\op{eu}{S}{RR}  & (\bar e_{Lp}   e_{Rr}) (\bar u_{Ls} u_{Rt})   \\
\op{eu}{T}{RR} & (\bar e_{Lp}   \sigma^{\mu \nu}   e_{Rr}) (\bar u_{Ls}  \sigma_{\mu \nu}  u_{Rt})  \\
\op{ed}{S}{RR}  & (\bar e_{Lp} e_{Rr})(\bar d_{Ls} d_{Rt})  \\
\op{ed}{T}{RR} & (\bar e_{Lp} \sigma^{\mu \nu} e_{Rr}) (\bar d_{Ls} \sigma_{\mu \nu} d_{Rt})   \\
\op{\nu edu}{S}{RR} & (\bar   \nu_{Lp} e_{Rr})  (\bar d_{Ls} u_{Rt} ) \\
\op{\nu edu}{T}{RR} &  (\bar  \nu_{Lp}  \sigma^{\mu \nu} e_{Rr} )  (\bar  d_{Ls}  \sigma_{\mu \nu} u_{Rt} )   \\
\op{uu}{S1}{RR}  & (\bar u_{Lp}   u_{Rr}) (\bar u_{Ls} u_{Rt})  \\
\op{uu}{S8}{RR}   & (\bar u_{Lp}   T^A u_{Rr}) (\bar u_{Ls} T^A u_{Rt})  \\
\op{ud}{S1}{RR}   & (\bar u_{Lp} u_{Rr})  (\bar d_{Ls} d_{Rt})   \\
\op{ud}{S8}{RR}  & (\bar u_{Lp} T^A u_{Rr})  (\bar d_{Ls} T^A d_{Rt})  \\
\op{dd}{S1}{RR}   & (\bar d_{Lp} d_{Rr}) (\bar d_{Ls} d_{Rt}) \\
\op{dd}{S8}{RR}  & (\bar d_{Lp} T^A d_{Rr}) (\bar d_{Ls} T^A d_{Rt})  \\
\op{uddu}{S1}{RR} &  (\bar u_{Lp} d_{Rr}) (\bar d_{Ls}  u_{Rt})   \\
\op{uddu}{S8}{RR}  &  (\bar u_{Lp} T^A d_{Rr}) (\bar d_{Ls}  T^A u_{Rt})  \\[-0.5cm]
\end{array}
\end{align*}
\renewcommand{\arraystretch}{1.51}
\small
\begin{align*}
\begin{array}[t]{c|c}
\multicolumn{2}{c}{\boldsymbol{(\overline L R)(\overline R L) +\hc}} \\
\hline
\op{eu}{S}{RL}  & (\bar e_{Lp} e_{Rr}) (\bar u_{Rs}  u_{Lt})  \\
\op{ed}{S}{RL} & (\bar e_{Lp} e_{Rr}) (\bar d_{Rs} d_{Lt}) \\
\op{\nu edu}{S}{RL}  & (\bar \nu_{Lp} e_{Rr}) (\bar d_{Rs}  u_{Lt})  \\
\end{array}
\end{align*}
\end{minipage}
\end{adjustbox}
\setlength{\belowcaptionskip}{-3cm}
\caption{LEFT operators of dimension three and five, as well as LEFT operators of dimension six that conserve baryon and lepton number, reproduced from Ref.~\cite{Jenkins:2017jig}.}
\label{tab:oplist1}
\end{table}

\begin{table}
\capstart
%
\centering
\begin{minipage}[t]{3cm}
\renewcommand{\arraystretch}{1.5}
\small
\begin{align*}
\begin{array}[t]{c|c}
\multicolumn{2}{c}{\boldsymbol{\Delta L = 4 + \hc}}  \\
\hline
\op{\nu\nu}{S}{LL} &  (\nu_{Lp}^T C \nu_{Lr}^{}) (\nu_{Ls}^T C \nu_{Lt}^{} )  \\
\end{array}
\end{align*}
\end{minipage}

\begin{adjustbox}{width=\textwidth,center}
\begin{minipage}[t]{3cm}
\renewcommand{\arraystretch}{1.5}
\small
\begin{align*}
\begin{array}[t]{c|c}
\multicolumn{2}{c}{\boldsymbol{\Delta L =2 + \hc}}  \\
\hline
\op{\nu e}{S}{LL}  &  (\nu_{Lp}^T C \nu_{Lr}) (\bar e_{Rs} e_{Lt})   \\
\op{\nu e}{T}{LL} &  (\nu_{Lp}^T C \sigma^{\mu \nu} \nu_{Lr}) (\bar e_{Rs}\sigma_{\mu \nu} e_{Lt} )  \\
\op{\nu e}{S}{LR} &  (\nu_{Lp}^T C \nu_{Lr}) (\bar e_{Ls} e_{Rt} )  \\
\op{\nu u}{S}{LL}  &  (\nu_{Lp}^T C \nu_{Lr}) (\bar u_{Rs} u_{Lt} )  \\
\op{\nu u}{T}{LL}  &  (\nu_{Lp}^T C \sigma^{\mu \nu} \nu_{Lr}) (\bar u_{Rs} \sigma_{\mu \nu} u_{Lt} ) \\
\op{\nu u}{S}{LR}  &  (\nu_{Lp}^T C \nu_{Lr}) (\bar u_{Ls} u_{Rt} )  \\
\op{\nu d}{S}{LL}   &  (\nu_{Lp}^T C \nu_{Lr}) (\bar d_{Rs} d_{Lt} ) \\
\op{\nu d}{T}{LL}   &  (\nu_{Lp}^T C \sigma^{\mu \nu}  \nu_{Lr}) (\bar d_{Rs} \sigma_{\mu \nu} d_{Lt} ) \\
\op{\nu d}{S}{LR}  &  (\nu_{Lp}^T C \nu_{Lr}) (\bar d_{Ls} d_{Rt} ) \\
\op{\nu edu}{S}{LL} &  (\nu_{Lp}^T C e_{Lr}) (\bar d_{Rs} u_{Lt} )  \\
\op{\nu edu}{T}{LL}  & (\nu_{Lp}^T C  \sigma^{\mu \nu} e_{Lr}) (\bar d_{Rs}  \sigma_{\mu \nu} u_{Lt} ) \\
\op{\nu edu}{S}{LR}   & (\nu_{Lp}^T C e_{Lr}) (\bar d_{Ls} u_{Rt} ) \\
\op{\nu edu}{V}{RL}   & (\nu_{Lp}^T C \gamma^\mu e_{Rr}) (\bar d_{Ls} \gamma_\mu u_{Lt} )  \\
\op{\nu edu}{V}{RR}   & (\nu_{Lp}^T C \gamma^\mu e_{Rr}) (\bar d_{Rs} \gamma_\mu u_{Rt} )  \\
\end{array}
\end{align*}
\end{minipage}
%
\begin{minipage}[t]{3cm}
\renewcommand{\arraystretch}{1.5}
\small
\begin{align*}
\begin{array}[t]{c|c}
\multicolumn{2}{c}{\boldsymbol{\Delta B = \Delta L = 1 + \hc}} \\
\hline
\op{udd}{S}{LL} &  \epsilon_{\alpha\beta\gamma}  (u_{Lp}^{\alpha T} C d_{Lr}^{\beta}) (d_{Ls}^{\gamma T} C \nu_{Lt}^{})   \\
\op{duu}{S}{LL} & \epsilon_{\alpha\beta\gamma}  (d_{Lp}^{\alpha T} C u_{Lr}^{\beta}) (u_{Ls}^{\gamma T} C e_{Lt}^{})  \\
\op{uud}{S}{LR} & \epsilon_{\alpha\beta\gamma}  (u_{Lp}^{\alpha T} C u_{Lr}^{\beta}) (d_{Rs}^{\gamma T} C e_{Rt}^{})  \\
\op{duu}{S}{LR} & \epsilon_{\alpha\beta\gamma}  (d_{Lp}^{\alpha T} C u_{Lr}^{\beta}) (u_{Rs}^{\gamma T} C e_{Rt}^{})   \\
\op{uud}{S}{RL} & \epsilon_{\alpha\beta\gamma}  (u_{Rp}^{\alpha T} C u_{Rr}^{\beta}) (d_{Ls}^{\gamma T} C e_{Lt}^{})   \\
\op{duu}{S}{RL} & \epsilon_{\alpha\beta\gamma}  (d_{Rp}^{\alpha T} C u_{Rr}^{\beta}) (u_{Ls}^{\gamma T} C e_{Lt}^{})   \\
\op{dud}{S}{RL} & \epsilon_{\alpha\beta\gamma}  (d_{Rp}^{\alpha T} C u_{Rr}^{\beta}) (d_{Ls}^{\gamma T} C \nu_{Lt}^{})   \\
\op{ddu}{S}{RL} & \epsilon_{\alpha\beta\gamma}  (d_{Rp}^{\alpha T} C d_{Rr}^{\beta}) (u_{Ls}^{\gamma T} C \nu_{Lt}^{})   \\
\op{duu}{S}{RR}  & \epsilon_{\alpha\beta\gamma}  (d_{Rp}^{\alpha T} C u_{Rr}^{\beta}) (u_{Rs}^{\gamma T} C e_{Rt}^{})  \\
\end{array}
\end{align*}
\end{minipage}
%
\begin{minipage}[t]{3cm}
\renewcommand{\arraystretch}{1.5}
\small
\begin{align*}
\begin{array}[t]{c|c}
\multicolumn{2}{c}{\boldsymbol{\Delta B = - \Delta L = 1 + \hc}}  \\
\hline
\op{ddd}{S}{LL} & \epsilon_{\alpha\beta\gamma}  (d_{Lp}^{\alpha T} C d_{Lr}^{\beta}) (\bar e_{Rs}^{} d_{Lt}^\gamma )  \\
\op{udd}{S}{LR}  & \epsilon_{\alpha\beta\gamma}  (u_{Lp}^{\alpha T} C d_{Lr}^{\beta}) (\bar \nu_{Ls}^{} d_{Rt}^\gamma )  \\
\op{ddu}{S}{LR} & \epsilon_{\alpha\beta\gamma}  (d_{Lp}^{\alpha T} C d_{Lr}^{\beta})  (\bar \nu_{Ls}^{} u_{Rt}^\gamma )  \\
\op{ddd}{S}{LR} & \epsilon_{\alpha\beta\gamma}  (d_{Lp}^{\alpha T} C d_{Lr}^{\beta}) (\bar e_{Ls}^{} d_{Rt}^\gamma ) \\
\op{ddd}{S}{RL}  & \epsilon_{\alpha\beta\gamma}  (d_{Rp}^{\alpha T} C d_{Rr}^{\beta}) (\bar e_{Rs}^{} d_{Lt}^\gamma )  \\
\op{udd}{S}{RR}  & \epsilon_{\alpha\beta\gamma}  (u_{Rp}^{\alpha T} C d_{Rr}^{\beta}) (\bar \nu_{Ls}^{} d_{Rt}^\gamma )  \\
\op{ddd}{S}{RR}  & \epsilon_{\alpha\beta\gamma}  (d_{Rp}^{\alpha T} C d_{Rr}^{\beta}) (\bar e_{Ls}^{} d_{Rt}^\gamma )  \\
\end{array}
\end{align*}
\end{minipage}
\end{adjustbox}
%
\caption{LEFT operators of dimension six that violate baryon and/or lepton number, reproduced from Ref.~\cite{Jenkins:2017jig}.}
\label{tab:oplist2}
\end{table}


\clearpage

\section{Diagrams}
\label{sec:diagrams}

In this appendix, we list all one-particle-irreducible one-loop diagrams that we computed to obtain the complete RGEs of the LEFT parameters. Through the application of the EOM, contributions to the running of Wilson coefficients of operators with different fields than the external fields of the calculated diagram are generated. The diagrams are ordered into groups of the different four-point, three-point, and two-point functions. In each group, the diagrams are ordered according to the operator insertions. The classes of operators that are needed to renormalize the diagrams after the application of the EOM are listed explicitly.

Note that for the lepton- and baryon-number-violating operators and for certain neutrino operators, some fermion lines have to be reversed.
The inserted dipole and $X^3$ operators are denoted by a black square, while the four-fermion operators are denoted by two black dots (for the two fermion bilinears).

\begin{fmffile}{diags/rgediags}
		
\subsection{Fermion Four-Point Functions}

\subsubsection[$\psi^4$: Single Insertion of a Four-Fermion Operator]{\boldmath $\psi^4$: Single Insertion of a Four-Fermion Operator}

\paragraph{External-leg corrections}

\begin{align}
	\begin{gathered}
		\begin{fmfgraph}(60,50)
			\fmfleft{i1,i2} \fmfright{o1,o2}
			\fmf{fermion}{i1,v1}
			\fmf{fermion}{v1,o1}
			\fmf{fermion}{i2,v3}
			\fmf{fermion}{v3,o2}
			\fmf{phantom,tension=8}{v1,v3}
			\fmfdot{v1,v3}
		\end{fmfgraph}
	\end{gathered} \times \text{ external-leg correction}
\end{align}
External-leg corrections contribute to the self-renormalization, i.e.\ only $\psi^4$ counterterms are needed. 

\paragraph{Vertex corrections}

\begin{align}
	\begin{gathered}
		\begin{fmfgraph}(60,50)
			\fmfleft{i1,i2} \fmfright{o1,o2}
			\fmf{fermion}{i1,v1}
			\fmf{fermion}{v1,o1}
			\fmf{fermion,tension=2}{i2,v2,v3}
			\fmf{fermion,tension=2}{v3,v4,o2}
			\fmf{phantom,tension=8}{v1,v3}
			\fmffreeze
			\fmf{photon,left=0.5}{v2,v4}
			\fmfdot{v1,v3}
		\end{fmfgraph}
		\begin{fmfgraph}(60,50)
			\fmfleft{i1,i2} \fmfright{o1,o2}
			\fmf{fermion,tension=2}{i1,v1,v2}
			\fmf{fermion,tension=2}{v2,v3,o1}
			\fmf{fermion}{i2,v4}
			\fmf{fermion}{v4,o2}
			\fmf{phantom,tension=8}{v2,v4}
			\fmffreeze
			\fmf{photon,right=0.5}{v1,v3}
			\fmfdot{v2,v4}
		\end{fmfgraph}
		\begin{fmfgraph}(60,50)
			\fmfleft{i1,i2} \fmfright{o1,o2}
			\fmf{fermion,tension=2}{i1,v1,v2}
			\fmf{fermion}{v2,o1}
			\fmf{fermion,tension=2}{i2,v3,v4}
			\fmf{fermion}{v4,o2}
			\fmf{phantom,tension=8}{v2,v4}
			\fmffreeze
			\fmf{photon,left=0.5}{v1,v3}
			\fmfdot{v2,v4}
		\end{fmfgraph}
		\begin{fmfgraph}(60,50)
			\fmfleft{i1,i2} \fmfright{o1,o2}
			\fmf{fermion}{i1,v1}
			\fmf{fermion,tension=2}{v1,v2,o1}
			\fmf{fermion}{i2,v3}
			\fmf{fermion,tension=2}{v3,v4,o2}
			\fmf{phantom,tension=8}{v1,v3}
			\fmffreeze
			\fmf{photon,right=0.5}{v2,v4}
			\fmfdot{v1,v3}
		\end{fmfgraph}
		\begin{fmfgraph}(60,50)
			\fmfleft{i1,i2} \fmfright{o1,o2}
			\fmf{fermion,tension=2}{i1,v1,v2}
			\fmf{fermion}{v2,o1}
			\fmf{fermion}{i2,v3}
			\fmf{fermion,tension=2}{v3,v4,o2}
			\fmf{phantom,tension=8}{v2,v3}
			\fmffreeze
			\fmf{photon,right=0.75}{v1,v4}
			\fmfdot{v2,v3}
		\end{fmfgraph}
		\begin{fmfgraph}(60,50)
			\fmfleft{i1,i2} \fmfright{o1,o2}
			\fmf{fermion}{i1,v1}
			\fmf{fermion,tension=2}{v1,v2,o1}
			\fmf{fermion,tension=2}{i2,v3,v4}
			\fmf{fermion}{v4,o2}
			\fmf{phantom,tension=8}{v1,v4}
			\fmffreeze
			\fmf{photon,left=0.75}{v2,v3}
			\fmfdot{v1,v4}
		\end{fmfgraph}
	\end{gathered}
\end{align}
\begin{align}
	\begin{gathered}
		\begin{fmfgraph}(60,50)
			\fmfset{curly_len}{2mm}
			\fmfleft{i1,i2} \fmfright{o1,o2}
			\fmf{fermion}{i1,v1}
			\fmf{fermion}{v1,o1}
			\fmf{fermion,tension=2}{i2,v2,v3}
			\fmf{fermion,tension=2}{v3,v4,o2}
			\fmf{phantom,tension=8}{v1,v3}
			\fmffreeze
			\fmf{gluon,right=0.5}{v4,v2}
			\fmfdot{v1,v3}
		\end{fmfgraph}
		\begin{fmfgraph}(60,50)
			\fmfset{curly_len}{2mm}
			\fmfleft{i1,i2} \fmfright{o1,o2}
			\fmf{fermion,tension=2}{i1,v1,v2}
			\fmf{fermion,tension=2}{v2,v3,o1}
			\fmf{fermion}{i2,v4}
			\fmf{fermion}{v4,o2}
			\fmf{phantom,tension=8}{v2,v4}
			\fmffreeze
			\fmf{gluon,right=0.5}{v1,v3}
			\fmfdot{v2,v4}
		\end{fmfgraph}
		\begin{fmfgraph}(60,50)
			\fmfset{curly_len}{2mm}
			\fmfleft{i1,i2} \fmfright{o1,o2}
			\fmf{fermion,tension=2}{i1,v1,v2}
			\fmf{fermion}{v2,o1}
			\fmf{fermion,tension=2}{i2,v3,v4}
			\fmf{fermion}{v4,o2}
			\fmf{phantom,tension=8}{v2,v4}
			\fmffreeze
			\fmf{gluon,right=0.5}{v3,v1}
			\fmfdot{v2,v4}
		\end{fmfgraph}
		\begin{fmfgraph}(60,50)
			\fmfset{curly_len}{2mm}
			\fmfleft{i1,i2} \fmfright{o1,o2}
			\fmf{fermion}{i1,v1}
			\fmf{fermion,tension=2}{v1,v2,o1}
			\fmf{fermion}{i2,v3}
			\fmf{fermion,tension=2}{v3,v4,o2}
			\fmf{phantom,tension=8}{v1,v3}
			\fmffreeze
			\fmf{gluon,right=0.5}{v2,v4}
			\fmfdot{v1,v3}
		\end{fmfgraph}
		\begin{fmfgraph}(60,50)
			\fmfset{curly_len}{2mm}
			\fmfleft{i1,i2} \fmfright{o1,o2}
			\fmf{fermion,tension=2}{i1,v1,v2}
			\fmf{fermion}{v2,o1}
			\fmf{fermion}{i2,v3}
			\fmf{fermion,tension=2}{v3,v4,o2}
			\fmf{phantom,tension=8}{v2,v3}
			\fmffreeze
			\fmf{gluon,right=0.75}{v1,v4}
			\fmfdot{v2,v3}
		\end{fmfgraph}
		\begin{fmfgraph}(60,50)
			\fmfset{curly_len}{2mm}
			\fmfleft{i1,i2} \fmfright{o1,o2}
			\fmf{fermion}{i1,v1}
			\fmf{fermion,tension=2}{v1,v2,o1}
			\fmf{fermion,tension=2}{i2,v3,v4}
			\fmf{fermion}{v4,o2}
			\fmf{phantom,tension=8}{v1,v4}
			\fmffreeze
			\fmf{gluon,right=0.75}{v3,v2}
			\fmfdot{v1,v4}
		\end{fmfgraph}
	\end{gathered}
\end{align}
The QED and QCD vertex corrections contribute to the self-renormalization and mixing between $\psi^4$ operators.

\clearpage

\subsubsection[$2\times \psi^2 X$: Double Insertion of Dipole Operators]{\boldmath $2\times \psi^2 X$: Double Insertion of Dipole Operators}

The $\psi^4$ bulb, triangle, and box diagrams only generate $\psi^4$ divergences.

\paragraph{Bulb and triangle diagrams}
		
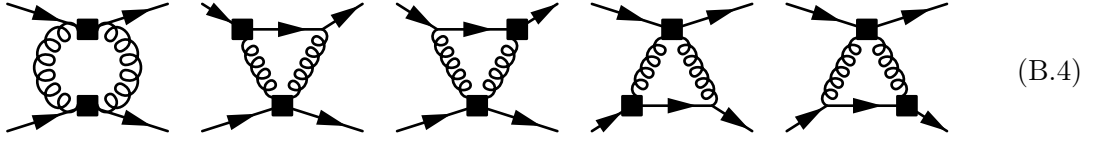
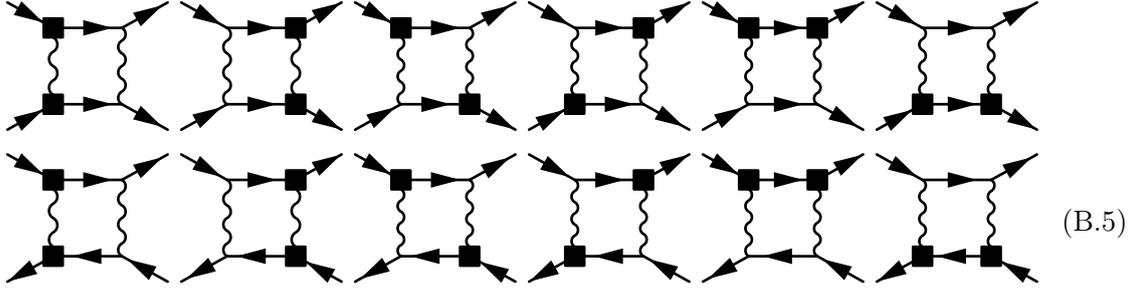
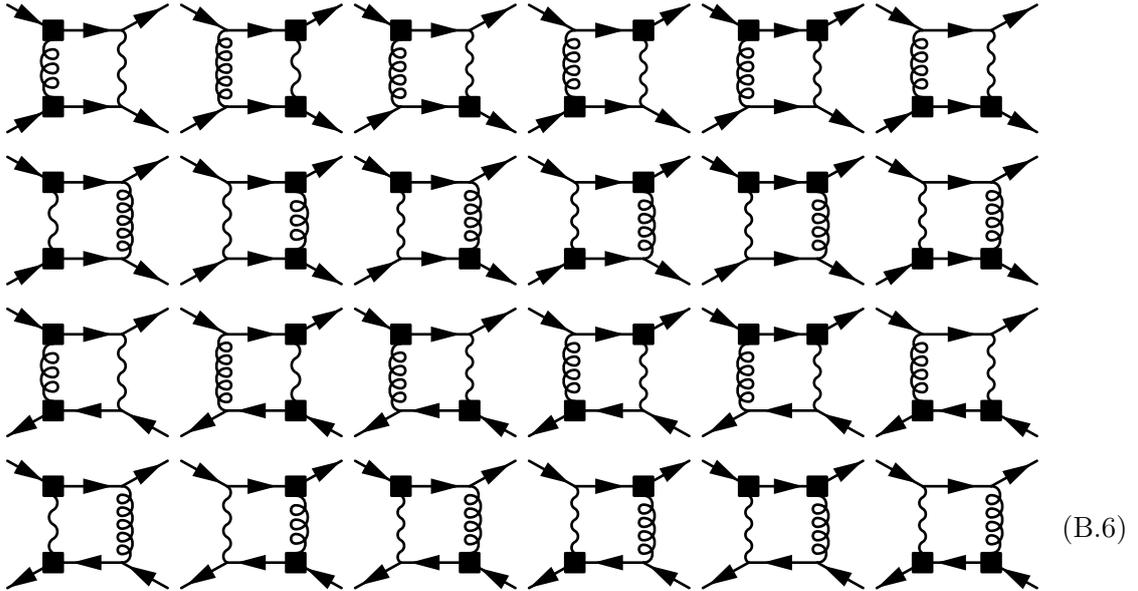
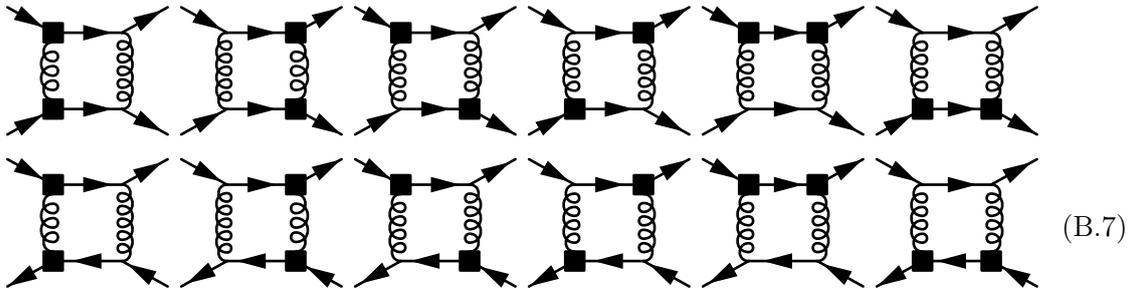
\begin{align}
	\begin{gathered}
		\begin{fmfgraph}(60,60)
			\fmfset{curly_len}{2mm}
			\fmftop{t1,t2} \fmfbottom{b1,b2}
			\fmf{fermion,tension=3}{t1,v1,t2}
			\fmf{fermion,tension=3}{b1,v2,b2}
			\fmf{gluon,right}{v1,v2}
			\fmf{gluon,right}{v2,v1}
			\fmfv{decoration.shape=square,decoration.size=2.5mm}{v1,v2}
		\end{fmfgraph}
	\end{gathered} \quad
	\begin{gathered}
		\begin{fmfgraph}(60,60)
			\fmfset{curly_len}{2mm}
			\fmftop{t1,t2} \fmfbottom{b1,b2}
			\fmf{fermion,tension=3}{t1,v1}
			\fmf{fermion}{v1,v3}
			\fmf{fermion,tension=3}{v3,t2}
			\fmf{fermion,tension=3}{b1,v2,b2}
			\fmf{gluon}{v1,v2}
			\fmf{gluon}{v2,v3}
			\fmfv{decoration.shape=square,decoration.size=2.5mm}{v1,v2}
		\end{fmfgraph}
	\end{gathered} \quad
	\begin{gathered}
		\begin{fmfgraph}(60,60)
			\fmfset{curly_len}{2mm}
			\fmftop{t1,t2} \fmfbottom{b1,b2}
			\fmf{fermion,tension=3}{t1,v1}
			\fmf{fermion}{v1,v3}
			\fmf{fermion,tension=3}{v3,t2}
			\fmf{fermion,tension=3}{b1,v2,b2}
			\fmf{gluon}{v1,v2}
			\fmf{gluon}{v2,v3}
			\fmfv{decoration.shape=square,decoration.size=2.5mm}{v3,v2}
		\end{fmfgraph}
	\end{gathered} \quad
	\begin{gathered}
		\begin{fmfgraph}(60,60)
			\fmfset{curly_len}{2mm}
			\fmftop{t1,t2} \fmfbottom{b1,b2}
			\fmf{fermion,tension=3}{b1,v1}
			\fmf{fermion}{v1,v3}
			\fmf{fermion,tension=3}{v3,b2}
			\fmf{fermion,tension=3}{t1,v2,t2}
			\fmf{gluon}{v2,v1}
			\fmf{gluon}{v3,v2}
			\fmfv{decoration.shape=square,decoration.size=2.5mm}{v1,v2}
		\end{fmfgraph}
	\end{gathered} \quad
	\begin{gathered}
		\begin{fmfgraph}(60,60)
			\fmfset{curly_len}{2mm}
			\fmftop{t1,t2} \fmfbottom{b1,b2}
			\fmf{fermion,tension=3}{b1,v1}
			\fmf{fermion}{v1,v3}
			\fmf{fermion,tension=3}{v3,b2}
			\fmf{fermion,tension=3}{t1,v2,t2}
			\fmf{gluon}{v2,v1}
			\fmf{gluon}{v3,v2}
			\fmfv{decoration.shape=square,decoration.size=2.5mm}{v3,v2}
		\end{fmfgraph}
	\end{gathered}
\end{align}

\paragraph{Box diagrams}
		
\begin{align}
	\begin{gathered}
		\begin{fmfgraph}(60,60)
			\fmfset{curly_len}{2mm}
			\fmftop{t1,t2} \fmfbottom{b1,b2}
			\fmf{fermion,tension=3}{t1,v1}
			\fmf{fermion,tension=2}{v1,v2}
			\fmf{fermion,tension=3}{v2,t2}
			\fmf{fermion,tension=3}{b1,v3}
			\fmf{fermion,tension=2}{v3,v4}
			\fmf{fermion,tension=3}{v4,b2}
			\fmf{photon}{v1,v3}
			\fmf{photon}{v4,v2}
			\fmfv{decoration.shape=square,decoration.size=2.5mm}{v1,v3}
		\end{fmfgraph}
	\end{gathered} \;
	\begin{gathered}
		\begin{fmfgraph}(60,60)
			\fmfset{curly_len}{2mm}
			\fmftop{t1,t2} \fmfbottom{b1,b2}
			\fmf{fermion,tension=3}{t1,v1}
			\fmf{fermion,tension=2}{v1,v2}
			\fmf{fermion,tension=3}{v2,t2}
			\fmf{fermion,tension=3}{b1,v3}
			\fmf{fermion,tension=2}{v3,v4}
			\fmf{fermion,tension=3}{v4,b2}
			\fmf{photon}{v1,v3}
			\fmf{photon}{v4,v2}
			\fmfv{decoration.shape=square,decoration.size=2.5mm}{v2,v4}
		\end{fmfgraph}
	\end{gathered} \;
	\begin{gathered}
		\begin{fmfgraph}(60,60)
			\fmfset{curly_len}{2mm}
			\fmftop{t1,t2} \fmfbottom{b1,b2}
			\fmf{fermion,tension=3}{t1,v1}
			\fmf{fermion,tension=2}{v1,v2}
			\fmf{fermion,tension=3}{v2,t2}
			\fmf{fermion,tension=3}{b1,v3}
			\fmf{fermion,tension=2}{v3,v4}
			\fmf{fermion,tension=3}{v4,b2}
			\fmf{photon}{v1,v3}
			\fmf{photon}{v4,v2}
			\fmfv{decoration.shape=square,decoration.size=2.5mm}{v1,v4}
		\end{fmfgraph}
	\end{gathered} \;
	\begin{gathered}
		\begin{fmfgraph}(60,60)
			\fmfset{curly_len}{2mm}
			\fmftop{t1,t2} \fmfbottom{b1,b2}
			\fmf{fermion,tension=3}{t1,v1}
			\fmf{fermion,tension=2}{v1,v2}
			\fmf{fermion,tension=3}{v2,t2}
			\fmf{fermion,tension=3}{b1,v3}
			\fmf{fermion,tension=2}{v3,v4}
			\fmf{fermion,tension=3}{v4,b2}
			\fmf{photon}{v1,v3}
			\fmf{photon}{v4,v2}
			\fmfv{decoration.shape=square,decoration.size=2.5mm}{v2,v3}
		\end{fmfgraph}
	\end{gathered} \;
	\begin{gathered}
		\begin{fmfgraph}(60,60)
			\fmfset{curly_len}{2mm}
			\fmftop{t1,t2} \fmfbottom{b1,b2}
			\fmf{fermion,tension=3}{t1,v1}
			\fmf{fermion,tension=2}{v1,v2}
			\fmf{fermion,tension=3}{v2,t2}
			\fmf{fermion,tension=3}{b1,v3}
			\fmf{fermion,tension=2}{v3,v4}
			\fmf{fermion,tension=3}{v4,b2}
			\fmf{photon}{v1,v3}
			\fmf{photon}{v4,v2}
			\fmfv{decoration.shape=square,decoration.size=2.5mm}{v1,v2}
		\end{fmfgraph}
	\end{gathered} \;
	\begin{gathered}
		\begin{fmfgraph}(60,60)
			\fmfset{curly_len}{2mm}
			\fmftop{t1,t2} \fmfbottom{b1,b2}
			\fmf{fermion,tension=3}{t1,v1}
			\fmf{fermion,tension=2}{v1,v2}
			\fmf{fermion,tension=3}{v2,t2}
			\fmf{fermion,tension=3}{b1,v3}
			\fmf{fermion,tension=2}{v3,v4}
			\fmf{fermion,tension=3}{v4,b2}
			\fmf{photon}{v1,v3}
			\fmf{photon}{v4,v2}
			\fmfv{decoration.shape=square,decoration.size=2.5mm}{v3,v4}
		\end{fmfgraph}
	\end{gathered}
	\nonumber \\[-0.4cm]
	\begin{gathered}
		\begin{fmfgraph}(60,60)
			\fmfset{curly_len}{2mm}
			\fmftop{t1,t2} \fmfbottom{b1,b2}
			\fmf{fermion,tension=3}{t1,v1}
			\fmf{fermion,tension=2}{v1,v2}
			\fmf{fermion,tension=3}{v2,t2}
			\fmf{fermion,tension=3}{b2,v4}
			\fmf{fermion,tension=2}{v4,v3}
			\fmf{fermion,tension=3}{v3,b1}
			\fmf{photon}{v1,v3}
			\fmf{photon}{v4,v2}
			\fmfv{decoration.shape=square,decoration.size=2.5mm}{v1,v3}
		\end{fmfgraph}
	\end{gathered} \;
	\begin{gathered}
		\begin{fmfgraph}(60,60)
			\fmfset{curly_len}{2mm}
			\fmftop{t1,t2} \fmfbottom{b1,b2}
			\fmf{fermion,tension=3}{t1,v1}
			\fmf{fermion,tension=2}{v1,v2}
			\fmf{fermion,tension=3}{v2,t2}
			\fmf{fermion,tension=3}{b2,v4}
			\fmf{fermion,tension=2}{v4,v3}
			\fmf{fermion,tension=3}{v3,b1}
			\fmf{photon}{v1,v3}
			\fmf{photon}{v4,v2}
			\fmfv{decoration.shape=square,decoration.size=2.5mm}{v2,v4}
		\end{fmfgraph}
	\end{gathered} \;
	\begin{gathered}
		\begin{fmfgraph}(60,60)
			\fmfset{curly_len}{2mm}
			\fmftop{t1,t2} \fmfbottom{b1,b2}
			\fmf{fermion,tension=3}{t1,v1}
			\fmf{fermion,tension=2}{v1,v2}
			\fmf{fermion,tension=3}{v2,t2}
			\fmf{fermion,tension=3}{b2,v4}
			\fmf{fermion,tension=2}{v4,v3}
			\fmf{fermion,tension=3}{v3,b1}
			\fmf{photon}{v1,v3}
			\fmf{photon}{v4,v2}
			\fmfv{decoration.shape=square,decoration.size=2.5mm}{v1,v4}
		\end{fmfgraph}
	\end{gathered} \;
	\begin{gathered}
		\begin{fmfgraph}(60,60)
			\fmfset{curly_len}{2mm}
			\fmftop{t1,t2} \fmfbottom{b1,b2}
			\fmf{fermion,tension=3}{t1,v1}
			\fmf{fermion,tension=2}{v1,v2}
			\fmf{fermion,tension=3}{v2,t2}
			\fmf{fermion,tension=3}{b2,v4}
			\fmf{fermion,tension=2}{v4,v3}
			\fmf{fermion,tension=3}{v3,b1}
			\fmf{photon}{v1,v3}
			\fmf{photon}{v4,v2}
			\fmfv{decoration.shape=square,decoration.size=2.5mm}{v2,v3}
		\end{fmfgraph}
	\end{gathered} \;
	\begin{gathered}
		\begin{fmfgraph}(60,60)
			\fmfset{curly_len}{2mm}
			\fmftop{t1,t2} \fmfbottom{b1,b2}
			\fmf{fermion,tension=3}{t1,v1}
			\fmf{fermion,tension=2}{v1,v2}
			\fmf{fermion,tension=3}{v2,t2}
			\fmf{fermion,tension=3}{b2,v4}
			\fmf{fermion,tension=2}{v4,v3}
			\fmf{fermion,tension=3}{v3,b1}
			\fmf{photon}{v1,v3}
			\fmf{photon}{v4,v2}
			\fmfv{decoration.shape=square,decoration.size=2.5mm}{v1,v2}
		\end{fmfgraph}
	\end{gathered} \;
	\begin{gathered}
		\begin{fmfgraph}(60,60)
			\fmfset{curly_len}{2mm}
			\fmftop{t1,t2} \fmfbottom{b1,b2}
			\fmf{fermion,tension=3}{t1,v1}
			\fmf{fermion,tension=2}{v1,v2}
			\fmf{fermion,tension=3}{v2,t2}
			\fmf{fermion,tension=3}{b2,v4}
			\fmf{fermion,tension=2}{v4,v3}
			\fmf{fermion,tension=3}{v3,b1}
			\fmf{photon}{v1,v3}
			\fmf{photon}{v4,v2}
			\fmfv{decoration.shape=square,decoration.size=2.5mm}{v3,v4}
		\end{fmfgraph}
	\end{gathered}
\\[0.15cm] 
	\begin{gathered}
		\begin{fmfgraph}(60,60)
			\fmfset{curly_len}{2mm}
			\fmftop{t1,t2} \fmfbottom{b1,b2}
			\fmf{fermion,tension=3}{t1,v1}
			\fmf{fermion,tension=2}{v1,v2}
			\fmf{fermion,tension=3}{v2,t2}
			\fmf{fermion,tension=3}{b1,v3}
			\fmf{fermion,tension=2}{v3,v4}
			\fmf{fermion,tension=3}{v4,b2}
			\fmf{gluon}{v1,v3}
			\fmf{photon}{v4,v2}
			\fmfv{decoration.shape=square,decoration.size=2.5mm}{v1,v3}
		\end{fmfgraph}
	\end{gathered} \;
	\begin{gathered}
		\begin{fmfgraph}(60,60)
			\fmfset{curly_len}{2mm}
			\fmftop{t1,t2} \fmfbottom{b1,b2}
			\fmf{fermion,tension=3}{t1,v1}
			\fmf{fermion,tension=2}{v1,v2}
			\fmf{fermion,tension=3}{v2,t2}
			\fmf{fermion,tension=3}{b1,v3}
			\fmf{fermion,tension=2}{v3,v4}
			\fmf{fermion,tension=3}{v4,b2}
			\fmf{gluon}{v1,v3}
			\fmf{photon}{v4,v2}
			\fmfv{decoration.shape=square,decoration.size=2.5mm}{v2,v4}
		\end{fmfgraph}
	\end{gathered} \;
	\begin{gathered}
		\begin{fmfgraph}(60,60)
			\fmfset{curly_len}{2mm}
			\fmftop{t1,t2} \fmfbottom{b1,b2}
			\fmf{fermion,tension=3}{t1,v1}
			\fmf{fermion,tension=2}{v1,v2}
			\fmf{fermion,tension=3}{v2,t2}
			\fmf{fermion,tension=3}{b1,v3}
			\fmf{fermion,tension=2}{v3,v4}
			\fmf{fermion,tension=3}{v4,b2}
			\fmf{gluon}{v1,v3}
			\fmf{photon}{v4,v2}
			\fmfv{decoration.shape=square,decoration.size=2.5mm}{v1,v4}
		\end{fmfgraph}
	\end{gathered} \;
	\begin{gathered}
		\begin{fmfgraph}(60,60)
			\fmfset{curly_len}{2mm}
			\fmftop{t1,t2} \fmfbottom{b1,b2}
			\fmf{fermion,tension=3}{t1,v1}
			\fmf{fermion,tension=2}{v1,v2}
			\fmf{fermion,tension=3}{v2,t2}
			\fmf{fermion,tension=3}{b1,v3}
			\fmf{fermion,tension=2}{v3,v4}
			\fmf{fermion,tension=3}{v4,b2}
			\fmf{gluon}{v1,v3}
			\fmf{photon}{v4,v2}
			\fmfv{decoration.shape=square,decoration.size=2.5mm}{v2,v3}
		\end{fmfgraph}
	\end{gathered} \;
	\begin{gathered}
		\begin{fmfgraph}(60,60)
			\fmfset{curly_len}{2mm}
			\fmftop{t1,t2} \fmfbottom{b1,b2}
			\fmf{fermion,tension=3}{t1,v1}
			\fmf{fermion,tension=2}{v1,v2}
			\fmf{fermion,tension=3}{v2,t2}
			\fmf{fermion,tension=3}{b1,v3}
			\fmf{fermion,tension=2}{v3,v4}
			\fmf{fermion,tension=3}{v4,b2}
			\fmf{gluon}{v1,v3}
			\fmf{photon}{v4,v2}
			\fmfv{decoration.shape=square,decoration.size=2.5mm}{v1,v2}
		\end{fmfgraph}
	\end{gathered} \;
	\begin{gathered}
		\begin{fmfgraph}(60,60)
			\fmfset{curly_len}{2mm}
			\fmftop{t1,t2} \fmfbottom{b1,b2}
			\fmf{fermion,tension=3}{t1,v1}
			\fmf{fermion,tension=2}{v1,v2}
			\fmf{fermion,tension=3}{v2,t2}
			\fmf{fermion,tension=3}{b1,v3}
			\fmf{fermion,tension=2}{v3,v4}
			\fmf{fermion,tension=3}{v4,b2}
			\fmf{gluon}{v1,v3}
			\fmf{photon}{v4,v2}
			\fmfv{decoration.shape=square,decoration.size=2.5mm}{v3,v4}
		\end{fmfgraph}
	\end{gathered}
	\nonumber \\[-0.4cm]
	\begin{gathered}
		\begin{fmfgraph}(60,60)
			\fmfset{curly_len}{2mm}
			\fmftop{t1,t2} \fmfbottom{b1,b2}
			\fmf{fermion,tension=3}{t1,v1}
			\fmf{fermion,tension=2}{v1,v2}
			\fmf{fermion,tension=3}{v2,t2}
			\fmf{fermion,tension=3}{b1,v3}
			\fmf{fermion,tension=2}{v3,v4}
			\fmf{fermion,tension=3}{v4,b2}
			\fmf{photon}{v1,v3}
			\fmf{gluon}{v4,v2}
			\fmfv{decoration.shape=square,decoration.size=2.5mm}{v1,v3}
		\end{fmfgraph}
	\end{gathered} \;
	\begin{gathered}
		\begin{fmfgraph}(60,60)
			\fmfset{curly_len}{2mm}
			\fmftop{t1,t2} \fmfbottom{b1,b2}
			\fmf{fermion,tension=3}{t1,v1}
			\fmf{fermion,tension=2}{v1,v2}
			\fmf{fermion,tension=3}{v2,t2}
			\fmf{fermion,tension=3}{b1,v3}
			\fmf{fermion,tension=2}{v3,v4}
			\fmf{fermion,tension=3}{v4,b2}
			\fmf{photon}{v1,v3}
			\fmf{gluon}{v4,v2}
			\fmfv{decoration.shape=square,decoration.size=2.5mm}{v2,v4}
		\end{fmfgraph}
	\end{gathered} \;
	\begin{gathered}
		\begin{fmfgraph}(60,60)
			\fmfset{curly_len}{2mm}
			\fmftop{t1,t2} \fmfbottom{b1,b2}
			\fmf{fermion,tension=3}{t1,v1}
			\fmf{fermion,tension=2}{v1,v2}
			\fmf{fermion,tension=3}{v2,t2}
			\fmf{fermion,tension=3}{b1,v3}
			\fmf{fermion,tension=2}{v3,v4}
			\fmf{fermion,tension=3}{v4,b2}
			\fmf{photon}{v1,v3}
			\fmf{gluon}{v4,v2}
			\fmfv{decoration.shape=square,decoration.size=2.5mm}{v1,v4}
		\end{fmfgraph}
	\end{gathered} \;
	\begin{gathered}
		\begin{fmfgraph}(60,60)
			\fmfset{curly_len}{2mm}
			\fmftop{t1,t2} \fmfbottom{b1,b2}
			\fmf{fermion,tension=3}{t1,v1}
			\fmf{fermion,tension=2}{v1,v2}
			\fmf{fermion,tension=3}{v2,t2}
			\fmf{fermion,tension=3}{b1,v3}
			\fmf{fermion,tension=2}{v3,v4}
			\fmf{fermion,tension=3}{v4,b2}
			\fmf{photon}{v1,v3}
			\fmf{gluon}{v4,v2}
			\fmfv{decoration.shape=square,decoration.size=2.5mm}{v2,v3}
		\end{fmfgraph}
	\end{gathered} \;
	\begin{gathered}
		\begin{fmfgraph}(60,60)
			\fmfset{curly_len}{2mm}
			\fmftop{t1,t2} \fmfbottom{b1,b2}
			\fmf{fermion,tension=3}{t1,v1}
			\fmf{fermion,tension=2}{v1,v2}
			\fmf{fermion,tension=3}{v2,t2}
			\fmf{fermion,tension=3}{b1,v3}
			\fmf{fermion,tension=2}{v3,v4}
			\fmf{fermion,tension=3}{v4,b2}
			\fmf{photon}{v1,v3}
			\fmf{gluon}{v4,v2}
			\fmfv{decoration.shape=square,decoration.size=2.5mm}{v1,v2}
		\end{fmfgraph}
	\end{gathered} \;
	\begin{gathered}
		\begin{fmfgraph}(60,60)
			\fmfset{curly_len}{2mm}
			\fmftop{t1,t2} \fmfbottom{b1,b2}
			\fmf{fermion,tension=3}{t1,v1}
			\fmf{fermion,tension=2}{v1,v2}
			\fmf{fermion,tension=3}{v2,t2}
			\fmf{fermion,tension=3}{b1,v3}
			\fmf{fermion,tension=2}{v3,v4}
			\fmf{fermion,tension=3}{v4,b2}
			\fmf{photon}{v1,v3}
			\fmf{gluon}{v4,v2}
			\fmfv{decoration.shape=square,decoration.size=2.5mm}{v3,v4}
		\end{fmfgraph}
	\end{gathered}
	\nonumber \\[-0.4cm]
	\begin{gathered}
		\begin{fmfgraph}(60,60)
			\fmfset{curly_len}{2mm}
			\fmftop{t1,t2} \fmfbottom{b1,b2}
			\fmf{fermion,tension=3}{t1,v1}
			\fmf{fermion,tension=2}{v1,v2}
			\fmf{fermion,tension=3}{v2,t2}
			\fmf{fermion,tension=3}{b2,v4}
			\fmf{fermion,tension=2}{v4,v3}
			\fmf{fermion,tension=3}{v3,b1}
			\fmf{gluon}{v1,v3}
			\fmf{photon}{v4,v2}
			\fmfv{decoration.shape=square,decoration.size=2.5mm}{v1,v3}
		\end{fmfgraph}
	\end{gathered} \;
	\begin{gathered}
		\begin{fmfgraph}(60,60)
			\fmfset{curly_len}{2mm}
			\fmftop{t1,t2} \fmfbottom{b1,b2}
			\fmf{fermion,tension=3}{t1,v1}
			\fmf{fermion,tension=2}{v1,v2}
			\fmf{fermion,tension=3}{v2,t2}
			\fmf{fermion,tension=3}{b2,v4}
			\fmf{fermion,tension=2}{v4,v3}
			\fmf{fermion,tension=3}{v3,b1}
			\fmf{gluon}{v1,v3}
			\fmf{photon}{v4,v2}
			\fmfv{decoration.shape=square,decoration.size=2.5mm}{v2,v4}
		\end{fmfgraph}
	\end{gathered} \;
	\begin{gathered}
		\begin{fmfgraph}(60,60)
			\fmfset{curly_len}{2mm}
			\fmftop{t1,t2} \fmfbottom{b1,b2}
			\fmf{fermion,tension=3}{t1,v1}
			\fmf{fermion,tension=2}{v1,v2}
			\fmf{fermion,tension=3}{v2,t2}
			\fmf{fermion,tension=3}{b2,v4}
			\fmf{fermion,tension=2}{v4,v3}
			\fmf{fermion,tension=3}{v3,b1}
			\fmf{gluon}{v1,v3}
			\fmf{photon}{v4,v2}
			\fmfv{decoration.shape=square,decoration.size=2.5mm}{v1,v4}
		\end{fmfgraph}
	\end{gathered} \;
	\begin{gathered}
		\begin{fmfgraph}(60,60)
			\fmfset{curly_len}{2mm}
			\fmftop{t1,t2} \fmfbottom{b1,b2}
			\fmf{fermion,tension=3}{t1,v1}
			\fmf{fermion,tension=2}{v1,v2}
			\fmf{fermion,tension=3}{v2,t2}
			\fmf{fermion,tension=3}{b2,v4}
			\fmf{fermion,tension=2}{v4,v3}
			\fmf{fermion,tension=3}{v3,b1}
			\fmf{gluon}{v1,v3}
			\fmf{photon}{v4,v2}
			\fmfv{decoration.shape=square,decoration.size=2.5mm}{v2,v3}
		\end{fmfgraph}
	\end{gathered} \;
	\begin{gathered}
		\begin{fmfgraph}(60,60)
			\fmfset{curly_len}{2mm}
			\fmftop{t1,t2} \fmfbottom{b1,b2}
			\fmf{fermion,tension=3}{t1,v1}
			\fmf{fermion,tension=2}{v1,v2}
			\fmf{fermion,tension=3}{v2,t2}
			\fmf{fermion,tension=3}{b2,v4}
			\fmf{fermion,tension=2}{v4,v3}
			\fmf{fermion,tension=3}{v3,b1}
			\fmf{gluon}{v1,v3}
			\fmf{photon}{v4,v2}
			\fmfv{decoration.shape=square,decoration.size=2.5mm}{v1,v2}
		\end{fmfgraph}
	\end{gathered} \;
	\begin{gathered}
		\begin{fmfgraph}(60,60)
			\fmfset{curly_len}{2mm}
			\fmftop{t1,t2} \fmfbottom{b1,b2}
			\fmf{fermion,tension=3}{t1,v1}
			\fmf{fermion,tension=2}{v1,v2}
			\fmf{fermion,tension=3}{v2,t2}
			\fmf{fermion,tension=3}{b2,v4}
			\fmf{fermion,tension=2}{v4,v3}
			\fmf{fermion,tension=3}{v3,b1}
			\fmf{gluon}{v1,v3}
			\fmf{photon}{v4,v2}
			\fmfv{decoration.shape=square,decoration.size=2.5mm}{v3,v4}
		\end{fmfgraph}
	\end{gathered}
	\nonumber \\[-0.4cm]
	\begin{gathered}
		\begin{fmfgraph}(60,60)
			\fmfset{curly_len}{2mm}
			\fmftop{t1,t2} \fmfbottom{b1,b2}
			\fmf{fermion,tension=3}{t1,v1}
			\fmf{fermion,tension=2}{v1,v2}
			\fmf{fermion,tension=3}{v2,t2}
			\fmf{fermion,tension=3}{b2,v4}
			\fmf{fermion,tension=2}{v4,v3}
			\fmf{fermion,tension=3}{v3,b1}
			\fmf{photon}{v1,v3}
			\fmf{gluon}{v4,v2}
			\fmfv{decoration.shape=square,decoration.size=2.5mm}{v1,v3}
		\end{fmfgraph}
	\end{gathered} \;
	\begin{gathered}
		\begin{fmfgraph}(60,60)
			\fmfset{curly_len}{2mm}
			\fmftop{t1,t2} \fmfbottom{b1,b2}
			\fmf{fermion,tension=3}{t1,v1}
			\fmf{fermion,tension=2}{v1,v2}
			\fmf{fermion,tension=3}{v2,t2}
			\fmf{fermion,tension=3}{b2,v4}
			\fmf{fermion,tension=2}{v4,v3}
			\fmf{fermion,tension=3}{v3,b1}
			\fmf{photon}{v1,v3}
			\fmf{gluon}{v4,v2}
			\fmfv{decoration.shape=square,decoration.size=2.5mm}{v2,v4}
		\end{fmfgraph}
	\end{gathered} \;
	\begin{gathered}
		\begin{fmfgraph}(60,60)
			\fmfset{curly_len}{2mm}
			\fmftop{t1,t2} \fmfbottom{b1,b2}
			\fmf{fermion,tension=3}{t1,v1}
			\fmf{fermion,tension=2}{v1,v2}
			\fmf{fermion,tension=3}{v2,t2}
			\fmf{fermion,tension=3}{b2,v4}
			\fmf{fermion,tension=2}{v4,v3}
			\fmf{fermion,tension=3}{v3,b1}
			\fmf{photon}{v1,v3}
			\fmf{gluon}{v4,v2}
			\fmfv{decoration.shape=square,decoration.size=2.5mm}{v1,v4}
		\end{fmfgraph}
	\end{gathered} \;
	\begin{gathered}
		\begin{fmfgraph}(60,60)
			\fmfset{curly_len}{2mm}
			\fmftop{t1,t2} \fmfbottom{b1,b2}
			\fmf{fermion,tension=3}{t1,v1}
			\fmf{fermion,tension=2}{v1,v2}
			\fmf{fermion,tension=3}{v2,t2}
			\fmf{fermion,tension=3}{b2,v4}
			\fmf{fermion,tension=2}{v4,v3}
			\fmf{fermion,tension=3}{v3,b1}
			\fmf{photon}{v1,v3}
			\fmf{gluon}{v4,v2}
			\fmfv{decoration.shape=square,decoration.size=2.5mm}{v2,v3}
		\end{fmfgraph}
	\end{gathered} \;
	\begin{gathered}
		\begin{fmfgraph}(60,60)
			\fmfset{curly_len}{2mm}
			\fmftop{t1,t2} \fmfbottom{b1,b2}
			\fmf{fermion,tension=3}{t1,v1}
			\fmf{fermion,tension=2}{v1,v2}
			\fmf{fermion,tension=3}{v2,t2}
			\fmf{fermion,tension=3}{b2,v4}
			\fmf{fermion,tension=2}{v4,v3}
			\fmf{fermion,tension=3}{v3,b1}
			\fmf{photon}{v1,v3}
			\fmf{gluon}{v4,v2}
			\fmfv{decoration.shape=square,decoration.size=2.5mm}{v1,v2}
		\end{fmfgraph}
	\end{gathered} \;
	\begin{gathered}
		\begin{fmfgraph}(60,60)
			\fmfset{curly_len}{2mm}
			\fmftop{t1,t2} \fmfbottom{b1,b2}
			\fmf{fermion,tension=3}{t1,v1}
			\fmf{fermion,tension=2}{v1,v2}
			\fmf{fermion,tension=3}{v2,t2}
			\fmf{fermion,tension=3}{b2,v4}
			\fmf{fermion,tension=2}{v4,v3}
			\fmf{fermion,tension=3}{v3,b1}
			\fmf{photon}{v1,v3}
			\fmf{gluon}{v4,v2}
			\fmfv{decoration.shape=square,decoration.size=2.5mm}{v3,v4}
		\end{fmfgraph}
	\end{gathered}
\\[0.15cm] 
	\begin{gathered}
		\begin{fmfgraph}(60,60)
			\fmfset{curly_len}{2mm}
			\fmftop{t1,t2} \fmfbottom{b1,b2}
			\fmf{fermion,tension=3}{t1,v1}
			\fmf{fermion,tension=2}{v1,v2}
			\fmf{fermion,tension=3}{v2,t2}
			\fmf{fermion,tension=3}{b1,v3}
			\fmf{fermion,tension=2}{v3,v4}
			\fmf{fermion,tension=3}{v4,b2}
			\fmf{gluon}{v1,v3}
			\fmf{gluon}{v4,v2}
			\fmfv{decoration.shape=square,decoration.size=2.5mm}{v1,v3}
		\end{fmfgraph}
	\end{gathered} \;
	\begin{gathered}
		\begin{fmfgraph}(60,60)
			\fmfset{curly_len}{2mm}
			\fmftop{t1,t2} \fmfbottom{b1,b2}
			\fmf{fermion,tension=3}{t1,v1}
			\fmf{fermion,tension=2}{v1,v2}
			\fmf{fermion,tension=3}{v2,t2}
			\fmf{fermion,tension=3}{b1,v3}
			\fmf{fermion,tension=2}{v3,v4}
			\fmf{fermion,tension=3}{v4,b2}
			\fmf{gluon}{v1,v3}
			\fmf{gluon}{v4,v2}
			\fmfv{decoration.shape=square,decoration.size=2.5mm}{v2,v4}
		\end{fmfgraph}
	\end{gathered} \;
	\begin{gathered}
		\begin{fmfgraph}(60,60)
			\fmfset{curly_len}{2mm}
			\fmftop{t1,t2} \fmfbottom{b1,b2}
			\fmf{fermion,tension=3}{t1,v1}
			\fmf{fermion,tension=2}{v1,v2}
			\fmf{fermion,tension=3}{v2,t2}
			\fmf{fermion,tension=3}{b1,v3}
			\fmf{fermion,tension=2}{v3,v4}
			\fmf{fermion,tension=3}{v4,b2}
			\fmf{gluon}{v1,v3}
			\fmf{gluon}{v4,v2}
			\fmfv{decoration.shape=square,decoration.size=2.5mm}{v1,v4}
		\end{fmfgraph}
	\end{gathered} \;
	\begin{gathered}
		\begin{fmfgraph}(60,60)
			\fmfset{curly_len}{2mm}
			\fmftop{t1,t2} \fmfbottom{b1,b2}
			\fmf{fermion,tension=3}{t1,v1}
			\fmf{fermion,tension=2}{v1,v2}
			\fmf{fermion,tension=3}{v2,t2}
			\fmf{fermion,tension=3}{b1,v3}
			\fmf{fermion,tension=2}{v3,v4}
			\fmf{fermion,tension=3}{v4,b2}
			\fmf{gluon}{v1,v3}
			\fmf{gluon}{v4,v2}
			\fmfv{decoration.shape=square,decoration.size=2.5mm}{v2,v3}
		\end{fmfgraph}
	\end{gathered} \;
	\begin{gathered}
		\begin{fmfgraph}(60,60)
			\fmfset{curly_len}{2mm}
			\fmftop{t1,t2} \fmfbottom{b1,b2}
			\fmf{fermion,tension=3}{t1,v1}
			\fmf{fermion,tension=2}{v1,v2}
			\fmf{fermion,tension=3}{v2,t2}
			\fmf{fermion,tension=3}{b1,v3}
			\fmf{fermion,tension=2}{v3,v4}
			\fmf{fermion,tension=3}{v4,b2}
			\fmf{gluon}{v1,v3}
			\fmf{gluon}{v4,v2}
			\fmfv{decoration.shape=square,decoration.size=2.5mm}{v1,v2}
		\end{fmfgraph}
	\end{gathered} \;
	\begin{gathered}
		\begin{fmfgraph}(60,60)
			\fmfset{curly_len}{2mm}
			\fmftop{t1,t2} \fmfbottom{b1,b2}
			\fmf{fermion,tension=3}{t1,v1}
			\fmf{fermion,tension=2}{v1,v2}
			\fmf{fermion,tension=3}{v2,t2}
			\fmf{fermion,tension=3}{b1,v3}
			\fmf{fermion,tension=2}{v3,v4}
			\fmf{fermion,tension=3}{v4,b2}
			\fmf{gluon}{v1,v3}
			\fmf{gluon}{v4,v2}
			\fmfv{decoration.shape=square,decoration.size=2.5mm}{v3,v4}
		\end{fmfgraph}
	\end{gathered}
	\nonumber \\[-0.4cm]
	\begin{gathered}
		\begin{fmfgraph}(60,60)
			\fmfset{curly_len}{2mm}
			\fmftop{t1,t2} \fmfbottom{b1,b2}
			\fmf{fermion,tension=3}{t1,v1}
			\fmf{fermion,tension=2}{v1,v2}
			\fmf{fermion,tension=3}{v2,t2}
			\fmf{fermion,tension=3}{b2,v4}
			\fmf{fermion,tension=2}{v4,v3}
			\fmf{fermion,tension=3}{v3,b1}
			\fmf{gluon}{v1,v3}
			\fmf{gluon}{v4,v2}
			\fmfv{decoration.shape=square,decoration.size=2.5mm}{v1,v3}
		\end{fmfgraph}
	\end{gathered} \;
	\begin{gathered}
		\begin{fmfgraph}(60,60)
			\fmfset{curly_len}{2mm}
			\fmftop{t1,t2} \fmfbottom{b1,b2}
			\fmf{fermion,tension=3}{t1,v1}
			\fmf{fermion,tension=2}{v1,v2}
			\fmf{fermion,tension=3}{v2,t2}
			\fmf{fermion,tension=3}{b2,v4}
			\fmf{fermion,tension=2}{v4,v3}
			\fmf{fermion,tension=3}{v3,b1}
			\fmf{gluon}{v1,v3}
			\fmf{gluon}{v4,v2}
			\fmfv{decoration.shape=square,decoration.size=2.5mm}{v2,v4}
		\end{fmfgraph}
	\end{gathered} \;
	\begin{gathered}
		\begin{fmfgraph}(60,60)
			\fmfset{curly_len}{2mm}
			\fmftop{t1,t2} \fmfbottom{b1,b2}
			\fmf{fermion,tension=3}{t1,v1}
			\fmf{fermion,tension=2}{v1,v2}
			\fmf{fermion,tension=3}{v2,t2}
			\fmf{fermion,tension=3}{b2,v4}
			\fmf{fermion,tension=2}{v4,v3}
			\fmf{fermion,tension=3}{v3,b1}
			\fmf{gluon}{v1,v3}
			\fmf{gluon}{v4,v2}
			\fmfv{decoration.shape=square,decoration.size=2.5mm}{v1,v4}
		\end{fmfgraph}
	\end{gathered} \;
	\begin{gathered}
		\begin{fmfgraph}(60,60)
			\fmfset{curly_len}{2mm}
			\fmftop{t1,t2} \fmfbottom{b1,b2}
			\fmf{fermion,tension=3}{t1,v1}
			\fmf{fermion,tension=2}{v1,v2}
			\fmf{fermion,tension=3}{v2,t2}
			\fmf{fermion,tension=3}{b2,v4}
			\fmf{fermion,tension=2}{v4,v3}
			\fmf{fermion,tension=3}{v3,b1}
			\fmf{gluon}{v1,v3}
			\fmf{gluon}{v4,v2}
			\fmfv{decoration.shape=square,decoration.size=2.5mm}{v2,v3}
		\end{fmfgraph}
	\end{gathered} \;
	\begin{gathered}
		\begin{fmfgraph}(60,60)
			\fmfset{curly_len}{2mm}
			\fmftop{t1,t2} \fmfbottom{b1,b2}
			\fmf{fermion,tension=3}{t1,v1}
			\fmf{fermion,tension=2}{v1,v2}
			\fmf{fermion,tension=3}{v2,t2}
			\fmf{fermion,tension=3}{b2,v4}
			\fmf{fermion,tension=2}{v4,v3}
			\fmf{fermion,tension=3}{v3,b1}
			\fmf{gluon}{v1,v3}
			\fmf{gluon}{v4,v2}
			\fmfv{decoration.shape=square,decoration.size=2.5mm}{v1,v2}
		\end{fmfgraph}
	\end{gathered} \;
	\begin{gathered}
		\begin{fmfgraph}(60,60)
			\fmfset{curly_len}{2mm}
			\fmftop{t1,t2} \fmfbottom{b1,b2}
			\fmf{fermion,tension=3}{t1,v1}
			\fmf{fermion,tension=2}{v1,v2}
			\fmf{fermion,tension=3}{v2,t2}
			\fmf{fermion,tension=3}{b2,v4}
			\fmf{fermion,tension=2}{v4,v3}
			\fmf{fermion,tension=3}{v3,b1}
			\fmf{gluon}{v1,v3}
			\fmf{gluon}{v4,v2}
			\fmfv{decoration.shape=square,decoration.size=2.5mm}{v3,v4}
		\end{fmfgraph}
	\end{gathered}
\end{align}

\subsection{Gauge-Boson Three-Point Functions}

\subsubsection[$X^3$ Insertion]{\boldmath $X^3$ Insertion}

\paragraph{External-leg corrections}

\begin{align}
	\begin{gathered}
		\begin{fmfgraph}(58,50)
			\fmfset{curly_len}{2mm}
			\fmftop{t1} \fmfbottom{b1,b2}
			\fmf{gluon,tension=3}{t1,v1}
			\fmf{gluon,tension=3}{b1,v1}
			\fmf{gluon,tension=3}{v1,b2}
			\fmfv{decoration.shape=square,decoration.size=2.5mm}{v1}
		\end{fmfgraph}
	\end{gathered} \times \text{ external-leg correction}
\end{align}
External-leg corrections contribute to the self-renormalization of the $X^3$ operators.

\paragraph{Tadpole}

\begin{align}
	\begin{gathered}
		\begin{fmfgraph}(58,50)
			\fmfset{curly_len}{2mm}
			\fmftop{t1,t2,t3} \fmfbottom{b1}
			\fmf{gluon}{t1,v1}
			\fmf{gluon}{t2,v1}
			\fmf{gluon}{t3,v1}
			\fmf{gluon,right}{v1,v2,v1}
			\fmf{phantom,tension=6}{b1,v2}
			\fmfv{decoration.shape=square,decoration.size=2.5mm}{v1}
		\end{fmfgraph}
	\end{gathered}
\end{align}
The $X^3$ tadpole diagram is a scaleless integral that vanishes in dimensional regularization.

\paragraph{Bulb and triangle diagrams}

\begin{gather}
	\label{diag:X3X3Bulbs}
	\begin{gathered}
		\begin{fmfgraph}(58,50)
			\fmfset{curly_len}{2mm}
			\fmftop{t1} \fmfbottom{b1,b2}
			\fmf{gluon,tension=3}{t1,v1}
			\fmf{gluon,right}{v1,v2,v1}
			\fmf{gluon,tension=3}{b1,v2,b2}
			\fmfv{decoration.shape=square,decoration.size=2.5mm}{v1}
		\end{fmfgraph}
	\end{gathered} \;
	\begin{gathered}
		\begin{fmfgraph}(58,50)
			\fmfset{curly_len}{2mm}
			\fmftop{t1} \fmfbottom{b1,b2}
			\fmf{gluon,tension=3}{b1,v1}
			\fmf{gluon,right}{v1,v2,v1}
			\fmf{gluon,tension=3}{b2,v2,t1}
			\fmfv{decoration.shape=square,decoration.size=2.5mm}{v1}
		\end{fmfgraph}
	\end{gathered} \;\;
	\begin{gathered}
		\begin{fmfgraph}(58,50)
			\fmfset{curly_len}{2mm}
			\fmftop{t1} \fmfbottom{b1,b2}
			\fmf{gluon,tension=3}{b2,v1}
			\fmf{gluon,right}{v1,v2,v1}
			\fmf{gluon,tension=3}{t1,v2,b1}
			\fmfv{decoration.shape=square,decoration.size=2.5mm}{v1}
		\end{fmfgraph}
	\end{gathered} \;\;
	\begin{gathered}
		\begin{fmfgraph}(58,50)
			\fmfset{curly_len}{2mm}
			\fmftop{t1} \fmfbottom{b1,b2}
			\fmf{gluon,tension=3}{t1,v1}
			\fmf{gluon,right}{v1,v2,v1}
			\fmf{gluon,tension=3}{b1,v2,b2}
			\fmfv{decoration.shape=square,decoration.size=2.5mm}{v2}
		\end{fmfgraph}
	\end{gathered} \;
	\begin{gathered}
		\begin{fmfgraph}(58,50)
			\fmfset{curly_len}{2mm}
			\fmftop{t1} \fmfbottom{b1,b2}
			\fmf{gluon,tension=3}{b1,v1}
			\fmf{gluon,right}{v1,v2,v1}
			\fmf{gluon,tension=3}{b2,v2,t1}
			\fmfv{decoration.shape=square,decoration.size=2.5mm}{v2}
		\end{fmfgraph}
	\end{gathered} \;\;
	\begin{gathered}
		\begin{fmfgraph}(58,50)
			\fmfset{curly_len}{2mm}
			\fmftop{t1} \fmfbottom{b1,b2}
			\fmf{gluon,tension=3}{b2,v1}
			\fmf{gluon,right}{v1,v2,v1}
			\fmf{gluon,tension=3}{t1,v2,b1}
			\fmfv{decoration.shape=square,decoration.size=2.5mm}{v2}
		\end{fmfgraph}
	\end{gathered}
\\ 
	\label{diag:X3X3Triangles}
	\begin{gathered}
		\begin{fmfgraph}(60,52)
			\fmfset{curly_len}{2mm}
			\fmftop{t1} \fmfbottom{b1,b2}
			\fmf{gluon,tension=3}{t1,v1}
			\fmf{gluon,right=0.25}{v1,v2,v3,v1}
			\fmf{gluon,tension=3}{b1,v2}
			\fmf{gluon,tension=3}{b2,v3}
			\fmfv{decoration.shape=square,decoration.size=2.5mm}{v1}
		\end{fmfgraph}
	\end{gathered} \quad
	\begin{gathered}
		\begin{fmfgraph}(60,52)
			\fmfset{curly_len}{2mm}
			\fmftop{t1} \fmfbottom{b1,b2}
			\fmf{gluon,tension=3}{t1,v1}
			\fmf{gluon,right=0.25}{v1,v2,v3,v1}
			\fmf{gluon,tension=3}{b1,v2}
			\fmf{gluon,tension=3}{b2,v3}
			\fmfv{decoration.shape=square,decoration.size=2.5mm}{v2}
		\end{fmfgraph}
	\end{gathered} \quad
	\begin{gathered}
		\begin{fmfgraph}(60,52)
			\fmfset{curly_len}{2mm}
			\fmftop{t1} \fmfbottom{b1,b2}
			\fmf{gluon,tension=3}{t1,v1}
			\fmf{gluon,right=0.25}{v1,v2,v3,v1}
			\fmf{gluon,tension=3}{b1,v2}
			\fmf{gluon,tension=3}{b2,v3}
			\fmfv{decoration.shape=square,decoration.size=2.5mm}{v3}
		\end{fmfgraph}
	\end{gathered}
\end{gather}
The $X^3$ bulb and triangle diagrams generate divergences proportional to the $X^3$ operators, as well as an additional piece proportional to $(D_\mu G^{\mu\nu}_A)(D^\lambda G_{\lambda\nu}^A)$. After applying the EOM, this piece cancels with a contribution from the $X^3$ insertions in the $\psi^2 X$ Green's functions.

\subsubsection[$2\times \psi^2 X$: Double Insertion of Dipole Operators]{\boldmath $2\times \psi^2 X$: Double Insertion of Dipole Operators}

\paragraph{Bulb and triangle diagrams}

\begin{gather}
	\begin{gathered}
		\begin{fmfgraph}(60,52)
			\fmfset{curly_len}{2mm}
			\fmftop{t1} \fmfbottom{b1,b2}
			\fmf{gluon,tension=3}{t1,v1}
			\fmf{fermion,right}{v1,v2,v1}
			\fmf{gluon,tension=3}{b1,v2,b2}
			\fmfv{decoration.shape=square,decoration.size=2.5mm}{v1,v2}
		\end{fmfgraph}
	\end{gathered} \quad
	\begin{gathered}
		\begin{fmfgraph}(60,52)
			\fmfset{curly_len}{2mm}
			\fmftop{t1} \fmfbottom{b1,b2}
			\fmf{gluon,tension=3}{b1,v1}
			\fmf{fermion,right}{v1,v2,v1}
			\fmf{gluon,tension=3}{t1,v2,b2}
			\fmfv{decoration.shape=square,decoration.size=2.5mm}{v1,v2}
		\end{fmfgraph}
	\end{gathered} \quad
	\begin{gathered}
		\begin{fmfgraph}(60,52)
			\fmfset{curly_len}{2mm}
			\fmftop{t1} \fmfbottom{b1,b2}
			\fmf{gluon,tension=3}{v1,b2}
			\fmf{fermion,right}{v1,v2,v1}
			\fmf{gluon,tension=3}{t1,v2,b1}
			\fmfv{decoration.shape=square,decoration.size=2.5mm}{v1,v2}
		\end{fmfgraph}
	\end{gathered}
\\ 
	\begin{gathered}
		\begin{fmfgraph}(58,50)
			\fmfset{curly_len}{2mm}
			\fmftop{t1} \fmfbottom{b1,b2}
			\fmf{gluon,tension=2.5}{t1,v1}
			\fmf{fermion,right=0.55}{v1,v2,v3,v1}
			\fmf{gluon,tension=2.5}{b1,v2}
			\fmf{gluon,tension=2.5}{b2,v3}
			\fmfv{decoration.shape=square,decoration.size=2.5mm}{v2,v3}
		\end{fmfgraph}
	\end{gathered} \;
	\begin{gathered}
		\begin{fmfgraph}(58,50)
			\fmfset{curly_len}{2mm}
			\fmftop{t1} \fmfbottom{b1,b2}
			\fmf{gluon,tension=2.5}{t1,v1}
			\fmf{fermion,right=0.55}{v1,v2,v3,v1}
			\fmf{gluon,tension=2.5}{b1,v2}
			\fmf{gluon,tension=2.5}{b2,v3}
			\fmfv{decoration.shape=square,decoration.size=2.5mm}{v1,v3}
		\end{fmfgraph}
	\end{gathered} \;
	\begin{gathered}
		\begin{fmfgraph}(58,50)
			\fmfset{curly_len}{2mm}
			\fmftop{t1} \fmfbottom{b1,b2}
			\fmf{gluon,tension=2.5}{t1,v1}
			\fmf{fermion,right=0.55}{v1,v2,v3,v1}
			\fmf{gluon,tension=2.5}{b1,v2}
			\fmf{gluon,tension=2.5}{b2,v3}
			\fmfv{decoration.shape=square,decoration.size=2.5mm}{v1,v2}
		\end{fmfgraph}
	\end{gathered} \;\;
	\begin{gathered}
		\begin{fmfgraph}(58,50)
			\fmfset{curly_len}{2mm}
			\fmftop{t1} \fmfbottom{b1,b2}
			\fmf{gluon,tension=2.5}{t1,v1}
			\fmf{fermion,left=0.55}{v1,v3,v2,v1}
			\fmf{gluon,tension=2.5}{b1,v2}
			\fmf{gluon,tension=2.5}{b2,v3}
			\fmfv{decoration.shape=square,decoration.size=2.5mm}{v2,v3}
		\end{fmfgraph}
	\end{gathered} \;
	\begin{gathered}
		\begin{fmfgraph}(58,50)
			\fmfset{curly_len}{2mm}
			\fmftop{t1} \fmfbottom{b1,b2}
			\fmf{gluon,tension=2.5}{t1,v1}
			\fmf{fermion,left=0.55}{v1,v3,v2,v1}
			\fmf{gluon,tension=2.5}{b1,v2}
			\fmf{gluon,tension=2.5}{b2,v3}
			\fmfv{decoration.shape=square,decoration.size=2.5mm}{v1,v3}
		\end{fmfgraph}
	\end{gathered} \;
	\begin{gathered}
		\begin{fmfgraph}(58,50)
			\fmfset{curly_len}{2mm}
			\fmftop{t1} \fmfbottom{b1,b2}
			\fmf{gluon,tension=2.5}{t1,v1}
			\fmf{fermion,left=0.55}{v1,v3,v2,v1}
			\fmf{gluon,tension=2.5}{b1,v2}
			\fmf{gluon,tension=2.5}{b2,v3}
			\fmfv{decoration.shape=square,decoration.size=2.5mm}{v1,v2}
		\end{fmfgraph}
	\end{gathered}
\end{gather}
The divergences generated by the double-dipole insertions in $X^3$ Green's functions are identical to the ones generated by the insertions into $X^2$ Green's functions in Sect.~\ref{sec:2xDipoleToX2} and required by gauge invariance. No divergence proportional to an $X^3$ operator is generated.

\subsection[$\psi^2 X$ Three-Point Functions]{\boldmath $\psi^2 X$ Three-Point Functions}

\subsubsection[$\psi^2 X$: Single Insertion of a Dipole Operator]{\boldmath $\psi^2 X$: Single Insertion of a Dipole Operator}

\paragraph{External-leg corrections}

\begin{align}
	\left(
	\begin{gathered}
		\begin{fmfgraph}(58,50)
			\fmfbottom{b1,b2} \fmftop{t1}
			\fmf{fermion}{b1,v1,b2}
			\fmf{photon}{v1,t1}
			\fmfv{decoration.shape=square,decoration.size=2.5mm}{v1}
		\end{fmfgraph}
	\end{gathered} \quad
	\begin{gathered}
		\begin{fmfgraph}(58,50)
			\fmfbottom{b1,b2} \fmftop{t1}
			\fmf{fermion}{b1,v1,b2}
			\fmf{gluon}{v1,t1}
			\fmfv{decoration.shape=square,decoration.size=2.5mm}{v1}
		\end{fmfgraph}
	\end{gathered} 
	\right) \times \text{ external-leg correction}
\end{align}
External-leg corrections contribute to the self-renormalization of the $\psi^2 X$ operators.

\paragraph{Triangle vertex corrections}
\begin{gather}
	\begin{gathered}
		\begin{fmfgraph}(52,55)
			\fmfset{curly_len}{2mm}
			\fmftop{t1} \fmfbottom{b1,b2}
			\fmf{photon,tension=2}{t1,v1}
			\fmf{fermion,tension=2}{b1,v2}
			\fmf{fermion}{v2,v1,v3}
			\fmf{fermion,tension=2}{v3,b2}
			\fmffreeze
			\fmf{photon}{v2,v3}
			\fmfv{decoration.shape=square,decoration.size=2.5mm}{v1}
		\end{fmfgraph}
	\end{gathered} \quad
	\begin{gathered}
		\begin{fmfgraph}(52,55)
			\fmfset{curly_len}{2mm}
			\fmftop{t1} \fmfbottom{b1,b2}
			\fmf{photon,tension=2}{t1,v1}
			\fmf{fermion,tension=2}{b1,v2}
			\fmf{fermion}{v2,v1,v3}
			\fmf{fermion,tension=2}{v3,b2}
			\fmffreeze
			\fmf{photon}{v2,v3}
			\fmfv{decoration.shape=square,decoration.size=2.5mm}{v2}
		\end{fmfgraph}
	\end{gathered} \quad
	\begin{gathered}
		\begin{fmfgraph}(52,55)
			\fmfset{curly_len}{2mm}
			\fmftop{t1} \fmfbottom{b1,b2}
			\fmf{photon,tension=2}{t1,v1}
			\fmf{fermion,tension=2}{b1,v2}
			\fmf{fermion}{v2,v1,v3}
			\fmf{fermion,tension=2}{v3,b2}
			\fmffreeze
			\fmf{photon}{v2,v3}
			\fmfv{decoration.shape=square,decoration.size=2.5mm}{v3}
		\end{fmfgraph}
	\end{gathered} \quad
	\begin{gathered}
		\begin{fmfgraph}(52,55)
			\fmfset{curly_len}{2mm}
			\fmftop{t1} \fmfbottom{b1,b2}
			\fmf{photon,tension=2}{t1,v1}
			\fmf{fermion,tension=2}{b1,v2}
			\fmf{fermion}{v2,v1,v3}
			\fmf{fermion,tension=2}{v3,b2}
			\fmffreeze
			\fmf{gluon}{v2,v3}
			\fmfv{decoration.shape=square,decoration.size=2.5mm}{v1}
		\end{fmfgraph}
	\end{gathered} \quad
	\begin{gathered}
		\begin{fmfgraph}(52,55)
			\fmfset{curly_len}{2mm}
			\fmftop{t1} \fmfbottom{b1,b2}
			\fmf{photon,tension=2}{t1,v1}
			\fmf{fermion,tension=2}{b1,v2}
			\fmf{fermion}{v2,v1,v3}
			\fmf{fermion,tension=2}{v3,b2}
			\fmffreeze
			\fmf{gluon}{v2,v3}
			\fmfv{decoration.shape=square,decoration.size=2.5mm}{v2}
		\end{fmfgraph}
	\end{gathered} \quad
	\begin{gathered}
		\begin{fmfgraph}(52,55)
			\fmfset{curly_len}{2mm}
			\fmftop{t1} \fmfbottom{b1,b2}
			\fmf{photon,tension=2}{t1,v1}
			\fmf{fermion,tension=2}{b1,v2}
			\fmf{fermion}{v2,v1,v3}
			\fmf{fermion,tension=2}{v3,b2}
			\fmffreeze
			\fmf{gluon}{v2,v3}
			\fmfv{decoration.shape=square,decoration.size=2.5mm}{v3}
		\end{fmfgraph}
	\end{gathered}
\\ 
	\begin{gathered}
		\begin{fmfgraph}(52,55)
			\fmfset{curly_len}{2mm}
			\fmftop{t1} \fmfbottom{b1,b2}
			\fmf{gluon,tension=2}{t1,v1}
			\fmf{fermion,tension=2}{b1,v2}
			\fmf{fermion}{v2,v1,v3}
			\fmf{fermion,tension=2}{v3,b2}
			\fmffreeze
			\fmf{photon}{v2,v3}
			\fmfv{decoration.shape=square,decoration.size=2.5mm}{v1}
		\end{fmfgraph}
	\end{gathered} \quad
	\begin{gathered}
		\begin{fmfgraph}(52,55)
			\fmfset{curly_len}{2mm}
			\fmftop{t1} \fmfbottom{b1,b2}
			\fmf{gluon,tension=2}{t1,v1}
			\fmf{fermion,tension=2}{b1,v2}
			\fmf{fermion}{v2,v1,v3}
			\fmf{fermion,tension=2}{v3,b2}
			\fmffreeze
			\fmf{photon}{v2,v3}
			\fmfv{decoration.shape=square,decoration.size=2.5mm}{v2}
		\end{fmfgraph}
	\end{gathered} \quad
	\begin{gathered}
		\begin{fmfgraph}(52,55)
			\fmfset{curly_len}{2mm}
			\fmftop{t1} \fmfbottom{b1,b2}
			\fmf{gluon,tension=2}{t1,v1}
			\fmf{fermion,tension=2}{b1,v2}
			\fmf{fermion}{v2,v1,v3}
			\fmf{fermion,tension=2}{v3,b2}
			\fmffreeze
			\fmf{photon}{v2,v3}
			\fmfv{decoration.shape=square,decoration.size=2.5mm}{v3}
		\end{fmfgraph}
	\end{gathered} \quad
	\begin{gathered}
		\begin{fmfgraph}(52,55)
			\fmfset{curly_len}{2mm}
			\fmftop{t1} \fmfbottom{b1,b2}
			\fmf{gluon,tension=2}{t1,v1}
			\fmf{fermion,tension=2}{b1,v2}
			\fmf{fermion}{v2,v1,v3}
			\fmf{fermion,tension=2}{v3,b2}
			\fmffreeze
			\fmf{gluon}{v2,v3}
			\fmfv{decoration.shape=square,decoration.size=2.5mm}{v1}
		\end{fmfgraph}
	\end{gathered} \quad
	\begin{gathered}
		\begin{fmfgraph}(52,55)
			\fmfset{curly_len}{2mm}
			\fmftop{t1} \fmfbottom{b1,b2}
			\fmf{gluon,tension=2}{t1,v1}
			\fmf{fermion,tension=2}{b1,v2}
			\fmf{fermion}{v2,v1,v3}
			\fmf{fermion,tension=2}{v3,b2}
			\fmffreeze
			\fmf{gluon}{v2,v3}
			\fmfv{decoration.shape=square,decoration.size=2.5mm}{v2}
		\end{fmfgraph}
	\end{gathered} \quad
	\begin{gathered}
		\begin{fmfgraph}(52,55)
			\fmfset{curly_len}{2mm}
			\fmftop{t1} \fmfbottom{b1,b2}
			\fmf{gluon,tension=2}{t1,v1}
			\fmf{fermion,tension=2}{b1,v2}
			\fmf{fermion}{v2,v1,v3}
			\fmf{fermion,tension=2}{v3,b2}
			\fmffreeze
			\fmf{gluon}{v2,v3}
			\fmfv{decoration.shape=square,decoration.size=2.5mm}{v3}
		\end{fmfgraph}
	\end{gathered}
\end{gather}
The triangle diagrams generate not only divergences proportional to $\psi^2 X$ operators, but also $\psi^2$ divergences through the EOM. These contributions have to be considered in combination with the insertions into $\psi^2$ Green's functions. The EOM including dimension-five corrections have to be used, i.e.\ the EOM generate not only mass terms, but also $\psi^2 X$ terms with coefficients quadratic in the dipole Wilson coefficients.

\paragraph{Additional QCD topologies}

\begin{align}
	\begin{gathered}
		\begin{fmfgraph}(60,60)
			\fmfset{curly_len}{2mm}
			\fmftop{t1} \fmfbottom{b1,b2}
			\fmf{gluon,tension=2}{t1,v1}
			\fmf{fermion,tension=2}{b1,v2}
			\fmf{gluon}{v3,v1,v2}
			\fmf{fermion,tension=2}{v3,b2}
			\fmffreeze
			\fmf{fermion}{v2,v3}
			\fmfv{decoration.shape=square,decoration.size=2.5mm}{v2}
		\end{fmfgraph}
	\end{gathered} \quad
	\begin{gathered}
		\begin{fmfgraph}(60,60)
			\fmfset{curly_len}{2mm}
			\fmftop{t1} \fmfbottom{b1,b2}
			\fmf{gluon,tension=2}{t1,v1}
			\fmf{fermion,tension=2}{b1,v2}
			\fmf{gluon}{v3,v1,v2}
			\fmf{fermion,tension=2}{v3,b2}
			\fmffreeze
			\fmf{fermion}{v2,v3}
			\fmfv{decoration.shape=square,decoration.size=2.5mm}{v3}
		\end{fmfgraph}
	\end{gathered} \quad
	\begin{gathered}
		\begin{fmfgraph}(60,60)
			\fmfset{curly_len}{2mm}
			\fmftop{t1,t2} \fmfbottom{b1,b2}
			\fmf{gluon}{t1,v1}
			\fmf{phantom}{t2,v2}
			\fmf{fermion,tension=2}{b1,v1}
			\fmf{fermion}{v1,v2}
			\fmf{fermion,tension=2}{v2,b2}
			\fmf{gluon,right}{v2,v1}
			\fmfv{decoration.shape=square,decoration.size=2.5mm}{v1}
		\end{fmfgraph}
	\end{gathered} \quad
	\begin{gathered}
		\begin{fmfgraph}(60,60)
			\fmfset{curly_len}{2mm}
			\fmftop{t1,t2} \fmfbottom{b1,b2}
			\fmf{phantom}{t1,v1}
			\fmf{gluon}{v2,t2}
			\fmf{fermion,tension=2}{b1,v1}
			\fmf{fermion}{v1,v2}
			\fmf{fermion,tension=2}{v2,b2}
			\fmf{gluon,right}{v2,v1}
			\fmfv{decoration.shape=square,decoration.size=2.5mm}{v2}
		\end{fmfgraph}
	\end{gathered} \quad
	\begin{gathered}
		\begin{fmfgraph}(60,60)
			\fmfset{curly_len}{2mm}
			\fmftop{t1} \fmfbottom{b1,b2}
			\fmf{gluon,tension=2}{t1,v1}
			\fmf{gluon,right}{v1,v2,v1}
			\fmf{fermion,tension=2}{b1,v2,b2}
			\fmfv{decoration.shape=square,decoration.size=2.5mm}{v2}
		\end{fmfgraph}
	\end{gathered}
\end{align}
These diagrams contribute through the EOM to the $\psi^2$ and $\psi^2 X$ terms.

\subsubsection[$2\times\psi^2 X$: Double Insertion of Dipole Operators]{\boldmath $2\times\psi^2 X$: Double Insertion of Dipole Operators}
\label{sec:2xDipoleToDipole}

\paragraph{Triangle vertex corrections}

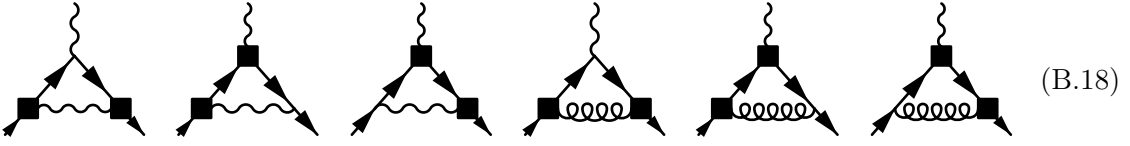
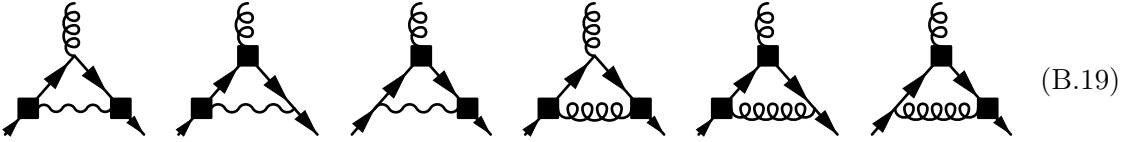
\begin{gather}
	\begin{gathered}
		\begin{fmfgraph}(52,55)
			\fmfset{curly_len}{2mm}
			\fmftop{t1} \fmfbottom{b1,b2}
			\fmf{photon,tension=2}{t1,v1}
			\fmf{fermion,tension=2}{b1,v2}
			\fmf{fermion}{v2,v1,v3}
			\fmf{fermion,tension=2}{v3,b2}
			\fmffreeze
			\fmf{photon}{v2,v3}
			\fmfv{decoration.shape=square,decoration.size=2.5mm}{v2,v3}
		\end{fmfgraph}
	\end{gathered} \quad
	\begin{gathered}
		\begin{fmfgraph}(52,55)
			\fmfset{curly_len}{2mm}
			\fmftop{t1} \fmfbottom{b1,b2}
			\fmf{photon,tension=2}{t1,v1}
			\fmf{fermion,tension=2}{b1,v2}
			\fmf{fermion}{v2,v1,v3}
			\fmf{fermion,tension=2}{v3,b2}
			\fmffreeze
			\fmf{photon}{v2,v3}
			\fmfv{decoration.shape=square,decoration.size=2.5mm}{v1,v2}
		\end{fmfgraph}
	\end{gathered} \quad
	\begin{gathered}
		\begin{fmfgraph}(52,55)
			\fmfset{curly_len}{2mm}
			\fmftop{t1} \fmfbottom{b1,b2}
			\fmf{photon,tension=2}{t1,v1}
			\fmf{fermion,tension=2}{b1,v2}
			\fmf{fermion}{v2,v1,v3}
			\fmf{fermion,tension=2}{v3,b2}
			\fmffreeze
			\fmf{photon}{v2,v3}
			\fmfv{decoration.shape=square,decoration.size=2.5mm}{v1,v3}
		\end{fmfgraph}
	\end{gathered} \quad
	\begin{gathered}
		\begin{fmfgraph}(52,55)
			\fmfset{curly_len}{2mm}
			\fmftop{t1} \fmfbottom{b1,b2}
			\fmf{photon,tension=2}{t1,v1}
			\fmf{fermion,tension=2}{b1,v2}
			\fmf{fermion}{v2,v1,v3}
			\fmf{fermion,tension=2}{v3,b2}
			\fmffreeze
			\fmf{gluon}{v2,v3}
			\fmfv{decoration.shape=square,decoration.size=2.5mm}{v2,v3}
		\end{fmfgraph}
	\end{gathered} \quad
	\begin{gathered}
		\begin{fmfgraph}(52,55)
			\fmfset{curly_len}{2mm}
			\fmftop{t1} \fmfbottom{b1,b2}
			\fmf{photon,tension=2}{t1,v1}
			\fmf{fermion,tension=2}{b1,v2}
			\fmf{fermion}{v2,v1,v3}
			\fmf{fermion,tension=2}{v3,b2}
			\fmffreeze
			\fmf{gluon}{v2,v3}
			\fmfv{decoration.shape=square,decoration.size=2.5mm}{v1,v2}
		\end{fmfgraph}
	\end{gathered} \quad
	\begin{gathered}
		\begin{fmfgraph}(52,55)
			\fmfset{curly_len}{2mm}
			\fmftop{t1} \fmfbottom{b1,b2}
			\fmf{photon,tension=2}{t1,v1}
			\fmf{fermion,tension=2}{b1,v2}
			\fmf{fermion}{v2,v1,v3}
			\fmf{fermion,tension=2}{v3,b2}
			\fmffreeze
			\fmf{gluon}{v2,v3}
			\fmfv{decoration.shape=square,decoration.size=2.5mm}{v1,v3}
		\end{fmfgraph}
	\end{gathered}
\\ 
	\label{diag:Psi2GDoubleDipoleTriangles}
	\begin{gathered}
		\begin{fmfgraph}(52,55)
			\fmfset{curly_len}{2mm}
			\fmftop{t1} \fmfbottom{b1,b2}
			\fmf{gluon,tension=2}{t1,v1}
			\fmf{fermion,tension=2}{b1,v2}
			\fmf{fermion}{v2,v1,v3}
			\fmf{fermion,tension=2}{v3,b2}
			\fmffreeze
			\fmf{photon}{v2,v3}
			\fmfv{decoration.shape=square,decoration.size=2.5mm}{v2,v3}
		\end{fmfgraph}
	\end{gathered} \quad
	\begin{gathered}
		\begin{fmfgraph}(52,55)
			\fmfset{curly_len}{2mm}
			\fmftop{t1} \fmfbottom{b1,b2}
			\fmf{gluon,tension=2}{t1,v1}
			\fmf{fermion,tension=2}{b1,v2}
			\fmf{fermion}{v2,v1,v3}
			\fmf{fermion,tension=2}{v3,b2}
			\fmffreeze
			\fmf{photon}{v2,v3}
			\fmfv{decoration.shape=square,decoration.size=2.5mm}{v1,v2}
		\end{fmfgraph}
	\end{gathered} \quad
	\begin{gathered}
		\begin{fmfgraph}(52,55)
			\fmfset{curly_len}{2mm}
			\fmftop{t1} \fmfbottom{b1,b2}
			\fmf{gluon,tension=2}{t1,v1}
			\fmf{fermion,tension=2}{b1,v2}
			\fmf{fermion}{v2,v1,v3}
			\fmf{fermion,tension=2}{v3,b2}
			\fmffreeze
			\fmf{photon}{v2,v3}
			\fmfv{decoration.shape=square,decoration.size=2.5mm}{v1,v3}
		\end{fmfgraph}
	\end{gathered} \quad
	\begin{gathered}
		\begin{fmfgraph}(52,55)
			\fmfset{curly_len}{2mm}
			\fmftop{t1} \fmfbottom{b1,b2}
			\fmf{gluon,tension=2}{t1,v1}
			\fmf{fermion,tension=2}{b1,v2}
			\fmf{fermion}{v2,v1,v3}
			\fmf{fermion,tension=2}{v3,b2}
			\fmffreeze
			\fmf{gluon}{v2,v3}
			\fmfv{decoration.shape=square,decoration.size=2.5mm}{v2,v3}
		\end{fmfgraph}
	\end{gathered} \quad
	\begin{gathered}
		\begin{fmfgraph}(52,55)
			\fmfset{curly_len}{2mm}
			\fmftop{t1} \fmfbottom{b1,b2}
			\fmf{gluon,tension=2}{t1,v1}
			\fmf{fermion,tension=2}{b1,v2}
			\fmf{fermion}{v2,v1,v3}
			\fmf{fermion,tension=2}{v3,b2}
			\fmffreeze
			\fmf{gluon}{v2,v3}
			\fmfv{decoration.shape=square,decoration.size=2.5mm}{v1,v2}
		\end{fmfgraph}
	\end{gathered} \quad
	\begin{gathered}
		\begin{fmfgraph}(52,55)
			\fmfset{curly_len}{2mm}
			\fmftop{t1} \fmfbottom{b1,b2}
			\fmf{gluon,tension=2}{t1,v1}
			\fmf{fermion,tension=2}{b1,v2}
			\fmf{fermion}{v2,v1,v3}
			\fmf{fermion,tension=2}{v3,b2}
			\fmffreeze
			\fmf{gluon}{v2,v3}
			\fmfv{decoration.shape=square,decoration.size=2.5mm}{v1,v3}
		\end{fmfgraph}
	\end{gathered}
\end{gather}
These diagrams generate $\psi^2 X$ divergences as well as $\psi^2$ and $\psi^4$ divergences through the EOM. Again, these contributions have to be considered in combination with the insertions into $\psi^2$ Green's functions.

\paragraph{Additional QCD topologies}

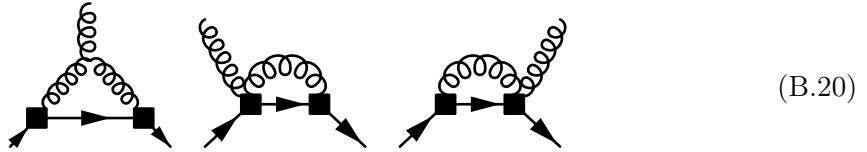
\begin{align}
	\label{diag:Psi2GDoubleDipoleAdditional}
	\begin{gathered}
		\begin{fmfgraph}(60,60)
			\fmfset{curly_len}{2mm}
			\fmftop{t1} \fmfbottom{b1,b2}
			\fmf{gluon,tension=2}{t1,v1}
			\fmf{fermion,tension=2}{b1,v2}
			\fmf{gluon}{v3,v1,v2}
			\fmf{fermion,tension=2}{v3,b2}
			\fmffreeze
			\fmf{fermion}{v2,v3}
			\fmfv{decoration.shape=square,decoration.size=2.5mm}{v2,v3}
		\end{fmfgraph}
	\end{gathered} \quad
	\begin{gathered}
		\begin{fmfgraph}(60,60)
			\fmfset{curly_len}{2mm}
			\fmftop{t1,t2} \fmfbottom{b1,b2}
			\fmf{gluon}{t1,v1}
			\fmf{phantom}{t2,v2}
			\fmf{fermion,tension=2}{b1,v1}
			\fmf{fermion}{v1,v2}
			\fmf{fermion,tension=2}{v2,b2}
			\fmf{gluon,right}{v2,v1}
			\fmfv{decoration.shape=square,decoration.size=2.5mm}{v1,v2}
		\end{fmfgraph}
	\end{gathered} \quad
	\begin{gathered}
		\begin{fmfgraph}(60,60)
			\fmfset{curly_len}{2mm}
			\fmftop{t1,t2} \fmfbottom{b1,b2}
			\fmf{phantom}{t1,v1}
			\fmf{gluon}{v2,t2}
			\fmf{fermion,tension=2}{b1,v1}
			\fmf{fermion}{v1,v2}
			\fmf{fermion,tension=2}{v2,b2}
			\fmf{gluon,right}{v2,v1}
			\fmfv{decoration.shape=square,decoration.size=2.5mm}{v1,v2}
		\end{fmfgraph}
	\end{gathered}
\end{align}
Depending on the chirality of the inserted operators, these diagrams generate $\psi^2 X$ divergences ($L-L$ and $R-R$ insertions) or EOM $\psi^2$ and $\psi^4$ divergences ($L-R$ and $R-L$ insertions).

\subsubsection[$X^3$ Insertion]{\boldmath $X^3$ Insertion}

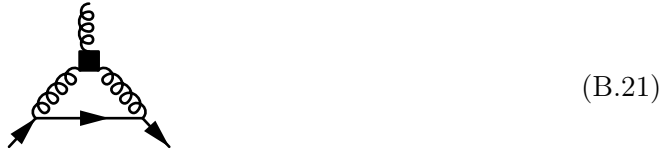
\begin{align}
	\label{diag:X3inDipole}
	\begin{gathered}
		\begin{fmfgraph}(60,60)
			\fmfset{curly_len}{2mm}
			\fmftop{t1} \fmfbottom{b1,b2}
			\fmf{gluon,tension=2}{t1,v1}
			\fmf{fermion,tension=2}{b1,v2}
			\fmf{gluon}{v3,v1,v2}
			\fmf{fermion,tension=2}{v3,b2}
			\fmffreeze
			\fmf{fermion}{v2,v3}
			\fmfv{decoration.shape=square,decoration.size=2.5mm}{v1}
		\end{fmfgraph}
	\end{gathered}
\end{align}
This diagram generates a $\psi^2 X$ divergence as well as a $(D_\mu G^{\mu\nu}_A) j^A_\nu$ divergence that cancels with a contribution from the $X^3$ insertions in the $X^3$ and $X^2$ Green's functions after the application of the EOM, see Sect.~\ref{sec:Cancellations}.

\subsubsection[$\psi^4$: Penguin Diagrams]{\boldmath $\psi^4$: Penguin Diagrams}

\begin{align}
	\begin{gathered}
		\begin{fmfgraph}(60,50)
			\fmfleft{i1,i2} \fmfright{o1,o2} \fmftop{t1}
			\fmf{fermion,tension=2}{i1,v1,o1}
			\fmf{fermion,right,tension=0.75}{v3,v2,v3}
			\fmf{phantom,tension=6.5}{v1,v3}
			\fmf{phantom}{i2,v2,o2}
			\fmf{photon}{v2,t1}
			\fmfdot{v1,v3}
		\end{fmfgraph}
	\end{gathered} \quad
	\begin{gathered}
		\begin{fmfgraph}(60,50)
			\fmfleft{i1,i2} \fmfright{o1,o2} \fmftop{t1}
			\fmf{fermion}{i1,v1}
			\fmf{fermion}{v3,o1}
			\fmf{fermion,left,tension=0.75}{v1,v2,v3}
			\fmf{phantom,tension=3.5}{v1,v3}
			\fmf{phantom}{i2,v2,o2}
			\fmf{photon,tension=0.1}{v2,t1}
			\fmfdot{v1,v3}
		\end{fmfgraph}
	\end{gathered} \quad
	\begin{gathered}
		\begin{fmfgraph}(60,50)
			\fmfset{curly_len}{2mm}
			\fmfleft{i1,i2} \fmfright{o1,o2} \fmftop{t1}
			\fmf{fermion,tension=2}{i1,v1,o1}
			\fmf{fermion,right,tension=0.75}{v3,v2,v3}
			\fmf{phantom,tension=6.5}{v1,v3}
			\fmf{phantom}{i2,v2,o2}
			\fmf{gluon}{v2,t1}
			\fmfdot{v1,v3}
		\end{fmfgraph}
	\end{gathered} \quad
	\begin{gathered}
		\begin{fmfgraph}(60,50)
			\fmfset{curly_len}{2mm}
			\fmfleft{i1,i2} \fmfright{o1,o2} \fmftop{t1}
			\fmf{fermion}{i1,v1}
			\fmf{fermion}{v3,o1}
			\fmf{fermion,left,tension=0.75}{v1,v2,v3}
			\fmf{phantom,tension=3.5}{v1,v3}
			\fmf{phantom}{i2,v2,o2}
			\fmf{gluon,tension=0.1}{v2,t1}
			\fmfdot{v1,v3}
		\end{fmfgraph}
	\end{gathered}
\end{align}
The penguin diagrams either generate $\psi^2 X$ divergences (in the case of $S$ and $T$ insertions) or $\psi^4$ divergences through the EOM (in the case of $V$ insertions).

\subsection{Fermion Two-Point Functions}

\subsubsection[$\psi^2 X$: Single Insertion of a Dipole Operator]{\boldmath $\psi^2 X$: Single Insertion of a Dipole Operator}

\begin{align}
	\begin{gathered}
		\begin{fmfgraph}(60,40)
			\fmfset{curly_len}{2mm}
			\fmftop{t1,t2} \fmfbottom{b1,b2}
			\fmf{phantom,tension=0.5}{t1,v1}
			\fmf{phantom,tension=0.5}{t2,v2}
			\fmf{fermion,tension=2}{b1,v1}
			\fmf{fermion}{v1,v2}
			\fmf{fermion,tension=2}{v2,b2}
			\fmf{photon,right=2}{v2,v1}
			\fmfv{decoration.shape=square,decoration.size=2.5mm}{v1}
		\end{fmfgraph}
	\end{gathered} \quad
	\begin{gathered}
		\begin{fmfgraph}(60,40)
			\fmfset{curly_len}{2mm}
			\fmftop{t1,t2} \fmfbottom{b1,b2}
			\fmf{phantom,tension=0.5}{t1,v1}
			\fmf{phantom,tension=0.5}{t2,v2}
			\fmf{fermion,tension=2}{b1,v1}
			\fmf{fermion}{v1,v2}
			\fmf{fermion,tension=2}{v2,b2}
			\fmf{photon,right=2}{v2,v1}
			\fmfv{decoration.shape=square,decoration.size=2.5mm}{v2}
		\end{fmfgraph}
	\end{gathered} \quad
	\begin{gathered}
		\begin{fmfgraph}(60,40)
			\fmfset{curly_len}{2mm}
			\fmftop{t1,t2} \fmfbottom{b1,b2}
			\fmf{phantom,tension=0.5}{t1,v1}
			\fmf{phantom,tension=0.5}{t2,v2}
			\fmf{fermion,tension=2}{b1,v1}
			\fmf{fermion}{v1,v2}
			\fmf{fermion,tension=2}{v2,b2}
			\fmf{gluon,right=2}{v2,v1}
			\fmfv{decoration.shape=square,decoration.size=2.5mm}{v1}
		\end{fmfgraph}
	\end{gathered} \quad
	\begin{gathered}
		\begin{fmfgraph}(60,40)
			\fmfset{curly_len}{2mm}
			\fmftop{t1,t2} \fmfbottom{b1,b2}
			\fmf{phantom,tension=0.5}{t1,v1}
			\fmf{phantom,tension=0.5}{t2,v2}
			\fmf{fermion,tension=2}{b1,v1}
			\fmf{fermion}{v1,v2}
			\fmf{fermion,tension=2}{v2,b2}
			\fmf{gluon,right=2}{v2,v1}
			\fmfv{decoration.shape=square,decoration.size=2.5mm}{v2}
		\end{fmfgraph}
	\end{gathered} \quad
	\begin{gathered}
		\begin{fmfgraph}(60,40)
			\fmfset{curly_len}{2mm}
			\fmftop{t1,t2} \fmfbottom{b1,b2}
			\fmf{phantom,tension=0.5}{t1,v1}
			\fmf{phantom,tension=0.5}{t2,v2}
			\fmf{fermion,tension=2}{b1,v1}
			\fmf{fermion,tension=2}{v2,b2}
			\fmf{gluon,right=15}{v2,v1}
			\fmf{phantom,tension=20}{v1,v2}
			\fmfv{decoration.shape=square,decoration.size=2.5mm}{v2}
		\end{fmfgraph}
	\end{gathered}
\end{align}
These diagrams complement the insertions in $\psi^2 X$ Green's functions and generate $\psi^2$ divergences as well as $\psi^2 X$ divergences (due to the dimension-five correction to the EOM).

\subsubsection[$2\times\psi^2 X$: Double Insertion of Dipole Operators]{\boldmath $2\times\psi^2 X$: Double Insertion of Dipole Operators}
\label{sec:2xDipoleToPsi2}

\begin{align}
	\label{diag:2xDipoleToPsi2}
	\begin{gathered}
		\begin{fmfgraph}(60,40)
			\fmfset{curly_len}{2mm}
			\fmftop{t1,t2} \fmfbottom{b1,b2}
			\fmf{phantom,tension=0.5}{t1,v1}
			\fmf{phantom,tension=0.5}{t2,v2}
			\fmf{fermion,tension=2}{b1,v1}
			\fmf{fermion}{v1,v2}
			\fmf{fermion,tension=2}{v2,b2}
			\fmf{photon,right=2}{v2,v1}
			\fmfv{decoration.shape=square,decoration.size=2.5mm}{v1,v2}
		\end{fmfgraph}
	\end{gathered} \quad
	\begin{gathered}
		\begin{fmfgraph}(60,40)
			\fmfset{curly_len}{2mm}
			\fmftop{t1,t2} \fmfbottom{b1,b2}
			\fmf{phantom,tension=0.5}{t1,v1}
			\fmf{phantom,tension=0.5}{t2,v2}
			\fmf{fermion,tension=2}{b1,v1}
			\fmf{fermion}{v1,v2}
			\fmf{fermion,tension=2}{v2,b2}
			\fmf{gluon,right=2}{v2,v1}
			\fmfv{decoration.shape=square,decoration.size=2.5mm}{v1,v2}
		\end{fmfgraph}
	\end{gathered}
\end{align}
These diagrams complement the insertions in $\psi^2 X$ Green's functions and generate $\psi^2$ divergences (direct and EOM).

\subsubsection[$\psi^4$: Tadpole Diagrams]{\boldmath $\psi^4$: Tadpole Diagrams}
\label{sec:Psi4Tadpoles}

\begin{align}
	\begin{gathered}
		\begin{fmfgraph}(60,50)
			\fmfleft{i1,i2} \fmfright{o1,o2} \fmftop{t1}
			\fmf{fermion,tension=5}{i1,v1,o1}
			\fmf{phantom}{i2,v2,o2}
			\fmf{phantom,tension=13}{v1,v2}
			\fmffreeze
			\fmf{phantom}{i1,v3}
			\fmf{phantom,tension=20}{v3,v2}
			\fmf{phantom}{o1,v4}
			\fmf{phantom,tension=20}{v4,v2}
			\fmf{fermion,left=30}{v3,v4}
			\fmfdot{v1,v2}
		\end{fmfgraph}
	\end{gathered} \quad
	\begin{gathered}
		\begin{fmfgraph}(60,50)
			\fmfleft{i1,i2} \fmfright{o1,o2} \fmftop{t1}
			\fmf{fermion}{i1,v1}
			\fmf{fermion}{v3,o1}
			\fmf{fermion,left=12}{v1,v3}
			\fmf{phantom,tension=2.6}{v1,v3}
			\fmf{phantom,tension=0.2}{v1,t1,v3}
			\fmfdot{v1,v3}
		\end{fmfgraph}
	\end{gathered}
\end{align}
The $\psi^4$ tadpole graphs generate $\psi^2$ divergences.

\subsection{Gauge-Boson Two-Point Functions}
\label{sec:X2Diagrams}

\subsubsection[$\psi^2 X$: Single Insertion of a Dipole Operator]{\boldmath $\psi^2 X$: Single Insertion of a Dipole Operator}

\paragraph{Tadpole diagram}

\begin{align}
	\begin{gathered}
		\begin{fmfgraph}(60,40)
			\fmfset{curly_len}{2mm}
			\fmftop{t1,t2} \fmfbottom{b1,b2}
			\fmf{phantom,tension=0.5}{t1,v1}
			\fmf{phantom,tension=0.5}{t2,v2}
			\fmf{gluon,tension=2}{b1,v1}
			\fmf{gluon,tension=2}{v2,b2}
			\fmf{fermion,left=15}{v1,v2}
			\fmf{phantom,tension=20}{v1,v2}
			\fmfv{decoration.shape=square,decoration.size=2.5mm}{v2}
		\end{fmfgraph}
	\end{gathered}
\end{align}
This tadpole diagram vanishes (it is proportional to the trace of a color generator).

\paragraph{Bulb diagrams}
\begin{align}
	\begin{gathered}
		\begin{fmfgraph}(60,40)
			\fmfset{curly_len}{2mm}
			\fmfleft{l1} \fmfright{r1}
			\fmf{photon,tension=3}{l1,v1}
			\fmf{fermion,right}{v1,v2,v1}
			\fmf{photon,tension=3}{v2,r1}
			\fmfv{decoration.shape=square,decoration.size=2.5mm}{v1}
		\end{fmfgraph}
	\end{gathered} \quad
	\begin{gathered}
		\begin{fmfgraph}(60,40)
			\fmfset{curly_len}{2mm}
			\fmfleft{l1} \fmfright{r1}
			\fmf{photon,tension=3}{l1,v1}
			\fmf{fermion,right}{v1,v2,v1}
			\fmf{photon,tension=3}{v2,r1}
			\fmfv{decoration.shape=square,decoration.size=2.5mm}{v2}
		\end{fmfgraph}
	\end{gathered} \quad
	\begin{gathered}
		\begin{fmfgraph}(60,40)
			\fmfset{curly_len}{2mm}
			\fmfleft{l1} \fmfright{r1}
			\fmf{gluon,tension=3}{l1,v1}
			\fmf{fermion,right}{v1,v2,v1}
			\fmf{gluon,tension=3}{v2,r1}
			\fmfv{decoration.shape=square,decoration.size=2.5mm}{v1}
		\end{fmfgraph}
	\end{gathered} \quad
	\begin{gathered}
		\begin{fmfgraph}(60,40)
			\fmfset{curly_len}{2mm}
			\fmfleft{l1} \fmfright{r1}
			\fmf{gluon,tension=3}{l1,v1}
			\fmf{fermion,right}{v1,v2,v1}
			\fmf{gluon,tension=3}{v2,r1}
			\fmfv{decoration.shape=square,decoration.size=2.5mm}{v2}
		\end{fmfgraph}
	\end{gathered}
\end{align}
These diagrams generate $X^2$ divergences that have to be reabsorbed by field redefinitions (or the EOM). This generates contributions to the RGEs of the gauge couplings and the $\psi^2 X$ Wilson coefficients.

\subsubsection[$2\times\psi^2 X$: Double Insertion of Dipole Operators]{\boldmath $2\times\psi^2 X$: Double Insertion of Dipole Operators}
\label{sec:2xDipoleToX2}

\begin{align}
	\begin{gathered}
		\begin{fmfgraph}(60,40)
			\fmfset{curly_len}{2mm}
			\fmfleft{l1} \fmfright{r1}
			\fmf{photon,tension=3}{l1,v1}
			\fmf{fermion,right}{v1,v2,v1}
			\fmf{photon,tension=3}{v2,r1}
			\fmfv{decoration.shape=square,decoration.size=2.5mm}{v1,v2}
		\end{fmfgraph}
	\end{gathered} \quad
	\begin{gathered}
		\begin{fmfgraph}(60,40)
			\fmfset{curly_len}{2mm}
			\fmfleft{l1} \fmfright{r1}
			\fmf{gluon,tension=3}{l1,v1}
			\fmf{fermion,right}{v1,v2,v1}
			\fmf{gluon,tension=3}{v2,r1}
			\fmfv{decoration.shape=square,decoration.size=2.5mm}{v1,v2}
		\end{fmfgraph}
	\end{gathered}
\end{align}
The $R-R$ and $L-L$ insertions generate $X^2$ divergences that have to be reabsorbed by field redefinitions (or the EOM) and contribute to the RGEs of the gauge couplings. The $R-L$ and $L-R$ insertions generate $\psi^4$ divergences via EOM.

\subsubsection[$X^3$ Insertion]{\boldmath $X^3$ Insertion}

\begin{align}
	\label{diag:X3inX2}
	\begin{gathered}
		\begin{fmfgraph}(60,40)
			\fmfset{curly_len}{2mm}
			\fmftop{t1,t2} \fmfbottom{b1,b2}
			\fmf{phantom,tension=0.5}{t1,v1}
			\fmf{phantom,tension=0.5}{t2,v2}
			\fmf{gluon,tension=2}{b1,v1}
			\fmf{gluon,tension=2}{v2,b2}
			\fmf{gluon,right=15}{v2,v1}
			\fmf{phantom,tension=20}{v1,v2}
			\fmfv{decoration.shape=square,decoration.size=2.5mm}{v2}
		\end{fmfgraph}
	\end{gathered} \quad
	\begin{gathered}
		\begin{fmfgraph}(60,40)
			\fmfset{curly_len}{2mm}
			\fmfleft{l1} \fmfright{r1}
			\fmf{gluon,tension=3}{l1,v1}
			\fmf{gluon,right}{v1,v2,v1}
			\fmf{gluon,tension=3}{v2,r1}
			\fmfv{decoration.shape=square,decoration.size=2.5mm}{v1}
		\end{fmfgraph}
	\end{gathered} \quad
	\begin{gathered}
		\begin{fmfgraph}(60,40)
			\fmfset{curly_len}{2mm}
			\fmfleft{l1} \fmfright{r1}
			\fmf{gluon,tension=3}{l1,v1}
			\fmf{gluon,right}{v1,v2,v1}
			\fmf{gluon,tension=3}{v2,r1}
			\fmfv{decoration.shape=square,decoration.size=2.5mm}{v2}
		\end{fmfgraph}
	\end{gathered}
\end{align}
These diagrams generate divergences proportional to $(D_\mu G^{\mu\nu}_A)(D^\lambda G_{\lambda\nu}^A)$, that also show up in the $X^3$ Green's functions. After applying the EOM, they cancel with a contribution from the $X^3$ insertions in the $\psi^2 X$ Green's functions.

\end{fmffile}%



\section{RGE}
\label{sec:RGE}

The results of the original publication contained a few typos and mistakes, which are corrected in the following. The complete one-loop LEFT RGEs up to dimension six have been cross-checked in Ref.~\cite{Naterop:2023dek}, making use of a much higher degree of automation for the loop calculation.

\subsection{Dimension 3: Masses}

The RGEs for the dimension-three mass terms are
\begin{align}
	\label{eq:NuMassRGE}
	[\dot M_\nu]_{rs} &=
		24 \lwc{\nu \nu}{LL}[S][v w r s]  [ M^\dagger_\nu M_\nu M^\dagger_\nu]_{wv} + 16 \lwc{\nu \nu}{LL}[V][wrvs] [M_\nu M_\nu^\dagger M_\nu]_{wv} \nn
		&\quad + 8 \lwc{\nu e}{LL}[S][rsvw] [M_e^\dagger M_e M_e^\dagger]_{wv} + 8 \lwc{\nu e}{LR}[S][rsvw] [M_e M_e^\dagger M_e]_{wv}  \nn
		&\quad + 8 N_c \lwc{\nu u}{LL}[S][rsvw] [M_u^\dagger M_u M_u^\dagger]_{wv} + 8 N_c \lwc{\nu u}{LR}[S][rsvw] [M_u M_u^\dagger M_u]_{wv}  \nn
		&\quad + 8 N_c \lwc{\nu d}{LL}[S][rsvw] [M_d^\dagger M_d M_d^\dagger]_{wv} + 8 N_c \lwc{\nu d}{LR}[S][rsvw] [M_d M_d^\dagger M_d]_{wv}  \nn
		&\quad -48 \lwc{\nu\gamma}{}[][rw] \lwc{\nu\gamma}{*}[][uv] [M_\nu^\dagger M_\nu]_{wv} [M_\nu]_{us} + 16 \lwc{\nu\gamma}{}[][rw] \lwc{\nu\gamma}{*}[][vw] [M_\nu M_\nu^\dagger M_\nu]_{vs} \nn
		&\quad -48 \lwc{\nu\gamma}{*}[][wu] \lwc{\nu\gamma}{}[][vs] [M_\nu M_\nu^\dagger]_{wv} [M_\nu]_{ru} + 16 \lwc{\nu\gamma}{*}[][wv] \lwc{\nu\gamma}{}[][ws] [M_\nu M_\nu^\dagger M_\nu]_{rv} \nn
		&\quad -96  \lwc{\nu \gamma}{}[][rv]  \lwc{\nu \gamma}{}[][ws]  [M_\nu^\dagger M_\nu M_\nu^\dagger]_{vw}
\end{align}
for the Majorana neutrino mass and 
\begin{align}
	\label{eq:RGEemass}
	[ \dot M_e ]_{rs}
		&= -  6\, \q_e^2 e^2 [ M_e ]_{rs}  + 12\, \q_e e \lwc{e \gamma}{*}[][vr] [M_e^\dagger M_e ]_{vs} + 12\, \q_e e \lwc{e\gamma}{*}[][sv]  [M_e M_e^\dagger ]_{rv}  \nn
		&\quad + 4 \lwc{\nu e}{LL}[S][v w r s]  [ M^\dagger_\nu M_\nu M^\dagger_\nu]_{wv} + 4 \lwc{\nu e}{LR*}[S][w v s r]  [ M_\nu M_\nu^\dagger M_\nu]_{wv} \nn
		&\quad + \Bigl( 8 \lwc{ee}{RR*}[S][srwv] -4 \lwc{ee}{RR*}[S][wrsv] \Bigr) [M_e^\dagger M_e M_e^\dagger ]_{wv}   - 8 \lwc{ee}{LR}[V][vsrw] [M_e M_e^\dagger M_e]_{wv}  \nn
		&\quad + 4 N_c \lwc{eu}{RR*}[S][srwv] [M_u^\dagger M_u M_u^\dagger ]_{wv}  +  4 N_c \lwc{eu}{RL*}[S][srwv] [M_u M_u^\dagger M_u]_{wv}  \nn
		&\quad + 4 N_c \lwc{ed}{RR*}[S][srwv] [M_d^\dagger M_d M_d^\dagger ]_{wv}  + 4 N_c \lwc{ed}{RL*}[S][srwv] [M_d M_d^\dagger M_d]_{wv}  \nn
		&\quad -12 \lwc{e\gamma}{*}[][wr] \lwc{e\gamma}{}[][vu] [M_e^\dagger M_e]_{wv} [M_e]_{us} + 4 \lwc{e\gamma}{*}[][wr] \lwc{e\gamma}{}[][wv] [M_e M_e^\dagger M_e]_{vs}   \nn
		&\quad -12 \lwc{e\gamma}{}[][uw] \lwc{e\gamma}{*}[][sv] [M_e M_e^\dagger]_{wv} [M_e]_{ru} + 4 \lwc{e\gamma}{}[][vw] \lwc{e\gamma}{*}[][sw] [M_e M_e^\dagger M_e]_{rv}  - 24  \lwc{e \gamma}{*}[][vr]  \lwc{e \gamma}{*}[][sw]  [M_e^\dagger M_e M_e^\dagger]_{vw}  \, , \\
\nnn
	\label{eq:RGEumass}
	[ \dot M_u ]_{rs}
		&= -6\left( C_F g^2 +  \q_u^2 e^2 \right) [ M_u ]_{rs} \nn
		&\quad +12 \Bigl(C_F g \lwc{u G}{*}[][vr] + \q_u e \lwc{u \gamma}{*}[][vr] \Bigr) [M_u^\dagger M_u ]_{vs} + 12 \Bigl(C_F g \lwc{u G}{*}[][sv] + \q_u e \lwc{u \gamma}{*}[][sv]  \Bigr) [M_u M_u^\dagger ]_{rv} \nn
		&\quad + 4 \lwc{\nu u}{LL}[S][v w r s]  [ M^\dagger_\nu M_\nu M^\dagger_\nu]_{wv} + 4 \lwc{\nu u}{LR*}[S][w v s r]  [ M_\nu M_\nu^\dagger M_\nu]_{wv}  \nn
		&\quad + 4 \lwc{eu}{RR*}[S][wvsr] [M_e^\dagger M_e M_e^\dagger ]_{wv}  + 4  \lwc{eu}{RL}[S][vwrs] [M_e M_e^\dagger M_e ]_{wv}   \nn
		&\quad + \Bigl( 8 N_c \lwc{uu}{RR*}[S1][srwv] - 4 \lwc{uu}{RR*}[S1][svwr] - 4 C_F \lwc{uu}{RR*}[S8][svwr]  \Bigr) [M_u^\dagger M_u M_u^\dagger ]_{wv}    \nn
		&\quad - 8 \Bigl( \lwc{uu}{LR}[V1][vsrw] + C_F \lwc{uu}{LR}[V8][vsrw]  \Bigr) [M_u M_u^\dagger M_u]_{wv}  \nn
		&\quad + \Bigl( 4 N_c \lwc{ud}{RR*}[S1][srwv]  -2 \lwc{uddu}{RR*}[S1][svwr]-2C_F \lwc{uddu}{RR*}[S8][svwr] \Bigr) [M_d^\dagger M_d M_d^\dagger ]_{wv}    \nn
		&\quad - 8 \Bigl( \lwc{uddu}{LR*}[V1][svwr] + C_F \lwc{uddu}{LR*}[V8][svwr]  \Bigr) [M_d M_d^\dagger M_d]_{wv}  \nn
		&\quad -12 \lwc{u\gamma}{*}[][wr] \lwc{u\gamma}{}[][vu] [M_u^\dagger M_u]_{wv} [M_u]_{us} + 4 \lwc{u\gamma}{*}[][wr] \lwc{u\gamma}{}[][wv] [M_u M_u^\dagger M_u]_{vs} \nn
		&\quad -12 \lwc{u\gamma}{}[][uw] \lwc{u\gamma}{*}[][sv] [M_u M_u^\dagger]_{wv} [M_u]_{ru} + 4 \lwc{u\gamma}{}[][vw] \lwc{u\gamma}{*}[][sw] [M_u M_u^\dagger M_u]_{rv} \nn
		&\quad -12 C_F \lwc{uG}{*}[][wr] \lwc{uG}{}[][vu] [M_u^\dagger M_u]_{wv} [M_u]_{us} + 4 C_F \lwc{uG}{*}[][wr] \lwc{uG}{}[][wv] [M_u M_u^\dagger M_u]_{vs} \nn
		&\quad -12 C_F \lwc{uG}{}[][uw] \lwc{uG}{*}[][sv] [M_u M_u^\dagger]_{wv} [M_u]_{ru} + 4 C_F \lwc{uG}{}[][vw] \lwc{uG}{*}[][sw] [M_u M_u^\dagger M_u]_{rv} \nn
		&\quad - 24   \lwc{u \gamma}{*}[][vr] \lwc{u \gamma}{*}[][sw] [M_u^\dagger M_u M_u^\dagger]_{vw} - 24 C_F  \lwc{u G}{*}[][vr] \lwc{u G}{*}[][sw] [M_u^\dagger M_u M_u^\dagger]_{vw} \, , \\
\nnn
	\label{eq:RGEdmass}
	[ \dot M_d ]_{rs}
		&=  -6 \left(C_F g^2 + \q_d^2 e^2 \right) [ M_d ]_{rs} \nn
		&\quad +12 \Bigl(C_F g \lwc{d G}{*}[][vr] + \q_d e \lwc{d \gamma}{*}[][vr] \Bigr) [M_d^\dagger M_d]_{vs} +12  \Bigl(C_F g \lwc{d G}{*}[][sv] + \q_d e \lwc{d \gamma}{*}[][sv]  \Bigr)  [M_d M_d^\dagger ]_{rv}  \nn
		&\quad + 4 \lwc{\nu d}{LL}[S][v w r s]  [ M^\dagger_\nu M_\nu M^\dagger_\nu]_{wv} + 4 \lwc{\nu d}{LR*}[S][w v s r]  [ M_\nu M_\nu^\dagger M_\nu]_{wv}  \nn
		&\quad +  4 \lwc{ed}{RR*}[S][wvsr] [M_e^\dagger M_e M_e^\dagger ]_{wv}  + 4  \lwc{ed}{RL}[S][vwrs] [M_e M_e^\dagger M_e ]_{wv}  \nn
		&\quad + \Bigl( 4 N_c \lwc{ud}{RR*}[S1][wvsr]  - 2 \lwc{uddu}{RR*}[S1][wrsv]-2C_F \lwc{uddu}{RR*}[S8][wrsv]  \Bigr) [M_u^\dagger M_u M_u^\dagger ]_{wv}    \nn
		&\quad - 8 \Bigl( \lwc{uddu}{LR}[V1][vsrw] + C_F \lwc{uddu}{LR}[V8][vsrw]  \Bigr) [M_u M_u^\dagger M_u]_{wv} \nn
		&\quad + \Bigl( 8 N_c \lwc{dd}{RR*}[S1][wvsr] - 4 \lwc{dd}{RR*}[S1][wrsv]  - 4 C_F \lwc{dd}{RR*}[S8][wrsv]  \Bigr) [M_d^\dagger M_d M_d^\dagger ]_{wv}    \nn
		&\quad - 8 \Bigl( \lwc{dd}{LR}[V1][vsrw] + C_F \lwc{dd}{LR}[V8][vsrw]  \Bigr) [M_d M_d^\dagger M_d]_{wv}  \nn 
		&\quad -12 \lwc{d\gamma}{*}[][wr] \lwc{d\gamma}{}[][vu] [M_d^\dagger M_d]_{wv} [M_d]_{us} + 4 \lwc{d\gamma}{*}[][wr] \lwc{d\gamma}{}[][wv] [M_d M_d^\dagger M_d]_{vs}  \nn
		&\quad -12 \lwc{d\gamma}{}[][uw] \lwc{d\gamma}{*}[][sv] [M_d M_d^\dagger]_{wv} [M_d]_{ru} + 4 \lwc{d\gamma}{}[][vw] \lwc{d\gamma}{*}[][sw] [M_d M_d^\dagger M_d]_{rv}  \nn
		&\quad -12 C_F \lwc{dG}{*}[][wr] \lwc{dG}{}[][vu] [M_d^\dagger M_d]_{wv} [M_d]_{us} + 4 C_F \lwc{dG}{*}[][wr] \lwc{dG}{}[][wv] [M_d M_d^\dagger M_d]_{vs}  \nn
		&\quad -12 C_F \lwc{dG}{}[][uw] \lwc{dG}{*}[][sv] [M_d M_d^\dagger]_{wv} [M_d]_{ru} + 4 C_F \lwc{dG}{}[][vw] \lwc{dG}{*}[][sw] [M_d M_d^\dagger M_d]_{rv}  \nn
		&\quad - 24  \lwc{d \gamma}{*}[][vr] \lwc{d \gamma}{*}[][sw]  [M_d^\dagger M_d M_d^\dagger]_{vw}  - 24  C_F  \lwc{d G}{*}[][vr] \lwc{d G}{*}[][sw]  [M_d^\dagger M_d M_d^\dagger]_{vw} 
\end{align}
for the other fermion masses. Each first term in Eqs.~\eqref{eq:RGEemass}--\eqref{eq:RGEdmass} is the SM anomalous dimension. The remaining terms are corrections from insertions of one
dimension-five or dimension-six operator, as well as from insertions of two dimension-five operators.

\subsection{Dimension 4: Gauge Couplings}
\label{sec:GaugeCouplingRGEs}

The gauge coupling $\beta$-functions are
\begin{align}
	\dot e &= -b_{0,e}\,  e^3    - 8 e^2 \q_e (\lwc{e \gamma}{}[][rs] [M_e]_{sr} +  [M_e^\dagger]_{rs} \lwc{e \gamma}{*}[][rs]) \nn
		&\quad - 8 e^2 N_c \q_u (\lwc{u \gamma}{}[][rs] [M_u]_{sr} +  [M_u^\dagger]_{rs} \lwc{u\gamma}{*}[][rs]) - 8 e^2 N_c \q_d (\lwc{d \gamma}{}[][rs] [M_d]_{sr} +  [M_d^\dagger]_{rs} \lwc{d \gamma}{*}[][rs]) \nn
		&\quad + 8e \Bigl( 2 [M_\nu^\dagger]_{ts}  \lwc{\nu \gamma}{}[][sr] [M_\nu^\dagger]_{rp} \lwc{\nu \gamma}{}[][pt] + 2 [M_\nu]_{pr}  \lwc{\nu \gamma}{*}[][sr]  [M_\nu]_{st} \lwc{\nu \gamma}{*}[][pt] \nn
				&\qquad\quad + [M_e]_{ts}  \lwc{e \gamma}{}[][sr] [M_e]_{rp} \lwc{e \gamma}{}[][pt] + [M_e^\dagger]_{pr}   \lwc{e \gamma}{*}[][sr] [M_e^\dagger]_{st} \lwc{e \gamma}{*}[][pt] \nn
				&\qquad\quad +  N_c [M_u]_{ts}  \lwc{u \gamma}{}[][sr] [M_u]_{rp} \lwc{u \gamma}{}[][pt] + N_c [M_u^\dagger]_{pr}  \lwc{u \gamma}{*}[][sr] [M_u^\dagger]_{st} \lwc{u \gamma}{*}[][pt] \nn
				&\qquad\quad +  N_c [M_d]_{ts}  \lwc{d \gamma}{}[][sr] [M_d]_{rp} \lwc{d \gamma}{}[][pt] + N_c [M_d^\dagger]_{pr}  \lwc{d \gamma}{*}[][sr] [M_d^\dagger]_{st} \lwc{d \gamma}{*}[][pt] \Bigr) \, ,  \\
\nnn
	\dot g &= -b_{0,g} \, g^3  -4 g^2 (\lwc{uG}{}[][rs] [M_u]_{sr} +  [M_u^\dagger]_{rs} \lwc{uG}{*}[][rs])-4 g^2 (\lwc{dG}{}[][rs] [M_d]_{sr}  +  [M_d^\dagger]_{rs} \lwc{dG}{*}[][rs]) \nn
		&\quad + 4 g \Bigl( [M_u]_{ts}  \lwc{u G}{}[][sr] [M_u]_{rp} \lwc{u G}{}[][pt] + [M_u^\dagger]_{pr}  \lwc{u G}{*}[][sr] [M_u^\dagger]_{st} \lwc{u G}{*}[][pt] \nn
			&\qquad\quad +  [M_d]_{ts}  \lwc{d G}{}[][sr] [M_d]_{rp} \lwc{d G}{}[][pt] + [M_d^\dagger]_{pr}  \lwc{d G}{*}[][sr] [M_d^\dagger]_{st} \lwc{d G}{*}[][pt] \Bigr) 
\end{align}
and the $\theta$ anomalous dimensions are
\begin{align}
	\dot \theta_{\text{QCD}} &= \frac{64 \pi^2}{g} (i \lwc{uG}{}[][rs] [M_u]_{sr} -i[M_u^\dagger]_{rs} \lwc{uG}{*}[][rs])+\frac{64 \pi^2}{g}  (i \lwc{dG}{}[][rs] [M_d]_{sr}  -i[M_d^\dagger]_{rs} \lwc{dG}{*}[][rs])  \nn
		&\quad + \frac{32 \pi^2}{g^2}2i \Bigl( - [M_u]_{ts}  \lwc{u G}{}[][sr] [M_u]_{rp} \lwc{u G}{}[][pt] + [M_u]_{ts}^*  \lwc{u G}{*}[][sr] [M_u]_{rp}^* \lwc{u G}{*}[][pt] \nns
		&\qquad\qquad\qquad - [M_d]_{ts}  \lwc{d G}{}[][sr] [M_d]_{rp} \lwc{d G}{}[][pt] + [M_d]_{ts}^*  \lwc{d G}{*}[][sr] [M_d]_{rp}^* \lwc{d G}{*}[][pt] \Bigr) \, , \\
\nnn
\nonumber\\[-1.5cm]
	\dot \theta_{\text{QED}} &=  \frac{128  \pi^2 \q_e}{e} (i \lwc{e\gamma}{}[][rs] [M_e]_{sr} -i[M_e^\dagger]_{rs} \lwc{e\gamma}{*}[][rs])+\frac{128 N_c \pi^2 \q_u}{e} (i \lwc{u\gamma}{}[][rs] [M_u]_{sr} -i[M_u^\dagger]_{rs} \lwc{u\gamma}{*}[][rs]) \nn
		&\quad + \frac{128 N_c \pi^2 \q_d}{e}  (i \lwc{d\gamma}{}[][rs] [M_d]_{sr}  -i[M_d^\dagger]_{rs} \lwc{d\gamma}{*}[][rs]) \nn
		&\quad + \frac{32 \pi^2}{e^2}  4i \Bigl( 2 [M_\nu^\dagger]_{ts}  \lwc{\nu \gamma}{}[][sr] [M_\nu^\dagger]_{rp} \lwc{\nu \gamma}{}[][pt] - 2 [M_\nu]_{pr}  \lwc{\nu \gamma}{*}[][sr] [M_\nu]_{st} \lwc{\nu \gamma}{*}[][pt] \nn
		&\qquad\qquad\qquad - [M_e]_{ts}  \lwc{e \gamma}{}[][sr] [M_e]_{rp} \lwc{e \gamma}{}[][pt] + [M_e^\dagger]_{pr} \lwc{e \gamma}{*}[][sr] [M_e^\dagger]_{st} \lwc{e \gamma}{*}[][pt] \nn
		&\qquad\qquad\qquad -  N_c [M_u]_{ts}  \lwc{u \gamma}{}[][sr] [M_u]_{rp} \lwc{u \gamma}{}[][pt] + N_c [M_u^\dagger]_{pr} \lwc{u \gamma}{*}[][sr] [M_u^\dagger]_{st} \lwc{u \gamma}{*}[][pt] \nn
		&\qquad\qquad\qquad -  N_c [M_d]_{ts}  \lwc{d \gamma}{}[][sr] [M_d]_{rp} \lwc{d \gamma}{}[][pt] + N_c [M_d^\dagger]_{pr}  \lwc{d \gamma}{*}[][sr] [M_d^\dagger]_{st} \lwc{d \gamma}{*}[][pt] \Bigr) \, .
\end{align}

\subsection{Dimension 5: Dipole Operators}
\label{sec:DipoleRGEs}

The anomalous dimension of the $\Delta L=2$ neutrino dipole operator is
\begin{align}
	\dlwc{\nu \gamma}{}[][rs] &= -b_{0,e} e^2 \lwc{\nu \gamma}{}[][rs] \nn
		&\quad { {} - 8 e N_c \q_d [M_d^\dagger]_{wv} \lwc{\nu d}{LL}[T][rsvw] - 8 e N_c \q_u [M_u^\dagger]_{wv} \lwc{\nu u}{LL}[T][rsvw] - 8 e \q_e [M_e^\dagger]_{wv} \lwc{\nu e}{LL}[T][rsvw]  } \nn
		&\quad - 8 e \q_e  \lwc{\nu\gamma}{}[][rs] \Bigl( \lwc{e\gamma}{}[][vw] [M_e]_{wv} +  [M_e^\dagger]_{vw} \lwc{e\gamma}{*}[][vw] \Bigr) 
		 - 8 e N_c \q_u  \lwc{\nu\gamma}{}[][rs] \Bigl( \lwc{u\gamma}{}[][vw] [M_u]_{wv} +  [M_u^\dagger]_{vw} \lwc{u\gamma}{*}[][vw] \Bigr)  \nn
		&\quad - 8 e N_c \q_d  \lwc{\nu\gamma}{}[][rs] \Bigl( \lwc{d\gamma}{}[][vw] [M_d]_{wv} +  [M_d^\dagger]_{vw} \lwc{d\gamma}{*}[][vw] \Bigr)  \, .
\end{align}
The anomalous dimensions of the other photonic dipole operators are
\begin{align}
	\dlwc{e\gamma}{}[][rs] &=  \Bigl[ \left(10 \q_e^2 -b_{0,e}\right) e^2 \Bigr] \lwc{e\gamma}{}[][rs]  \nn
		&\quad - 8 e N_c \q_d [M_d]_{wv} \lwc{e d}{RR}[T][rsvw] - 8 e N_c \q_u [M_u]_{wv} \lwc{e u}{RR}[T][r s v w] + 2 e \q_e [M_e]_{wv} \lwc{e e}{RR}[S][rwvs]   \nn
		&\quad -12 e \q_e  \lwc{e \gamma}{}[][rw] [M_e]_{wv} \lwc{e \gamma}{}[][vs]   -8 e \q_e \lwc{e \gamma}{*}[][vw] \Bigl( \lwc{e \gamma}{}[][rw]  [M_e^\dagger]_{vs} + \lwc{e \gamma}{}[][vs]  [M_e^\dagger]_{rw} \Bigr) \nn
		&\quad - 8 e \q_e  \lwc{e\gamma}{}[][rs] \Bigl( \lwc{e\gamma}{}[][vw] [M_e]_{wv} +  [M_e^\dagger]_{vw} \lwc{e\gamma}{*}[][vw] \Bigr)  - 8 e N_c \q_u  \lwc{e\gamma}{}[][rs] \Bigl( \lwc{u\gamma}{}[][vw] [M_u]_{wv} +  [M_u^\dagger]_{vw} \lwc{u\gamma}{*}[][vw] \Bigr) \nn
		&\quad - 8 e N_c \q_d  \lwc{e\gamma}{}[][rs] \Bigl( \lwc{d\gamma}{}[][vw] [M_d]_{wv} +  [M_d^\dagger]_{vw} \lwc{d\gamma}{*}[][vw] \Bigr)  \, , \\
\nnn
	\dlwc{u\gamma}{}[][rs] &=  \Bigl[ 2 C_{F}  g^2  + \left(10 \q_u^2 -b_{0,e}\right) e^2\Bigr] \lwc{u\gamma}{}[][rs] + 8 C_{F} e g\q_u  \lwc{uG}{}[][rs] \nn
		&\quad - 8 e \q_e [M_e]_{wv} \lwc{e u}{RR}[T][vwrs] + 2 e \q_u [M_u]_{wv} \lwc{uu}{RR}[S1][rwvs]    + 2 e C_F \q_u [M_u]_{wv} \lwc{uu}{RR}[S8][rwvs]  \nn
		&\quad  + e \q_d  [M_d]_{wv} \lwc{uddu}{RR}[S1][rwvs] + e C_F \q_d  [M_d]_{wv} \lwc{uddu}{RR}[S8][rwvs] \nn
		&\quad  - \lwc{u \gamma}{}[][rw] \Bigl( 4 g C_F \lwc{u G}{*}[][vw] + 8 e \q_u \lwc{u \gamma}{*}[][vw] \Bigr) [M_u^\dagger]_{vs} - [M_u^\dagger]_{rw}  \Bigl( 4 g C_F \lwc{u G}{*}[][vw] + 8 e \q_u \lwc{u \gamma}{*}[][vw] \Bigr) \lwc{u \gamma}{}[][vs]  \nn
		&\quad - 8 e \q_e  \lwc{u\gamma}{}[][rs] \Bigl( \lwc{e\gamma}{}[][vw] [M_e]_{wv} +  [M_e^\dagger]_{vw} \lwc{e\gamma}{*}[][vw] \Bigr)  - 8 e N_c \q_u  \lwc{u\gamma}{}[][rs] \Bigl( \lwc{u\gamma}{}[][vw] [M_u]_{wv} +  [M_u^\dagger]_{vw} \lwc{u\gamma}{*}[][vw] \Bigr)  \nn
		&\quad - 8 e N_c \q_d  \lwc{u\gamma}{}[][rs] \Bigl( \lwc{d\gamma}{}[][vw] [M_d]_{wv} +  [M_d^\dagger]_{vw} \lwc{d\gamma}{*}[][vw] \Bigr)   - 12 e \q_u  \lwc{u \gamma}{}[][rw] [M_u]_{wv} \lwc{u \gamma}{}[][vs]   \nn
		&\quad - 4 e \q_u [M_u]_{wv} C_F \lwc{u G}{}[][rw] \lwc{u G}{}[][vs] - 4 g [M_u]_{wv} C_F \Bigl( \lwc{u \gamma}{}[][rw] \lwc{u G}{}[][vs] + \lwc{u G}{}[][rw] \lwc{u \gamma}{}[][vs] \Bigr)  \nn
		&\quad - 4 C_F e \q_u \lwc{uG}{*}[][wv] \lwc{uG}{}[][ws] [M_u^\dagger]_{rv} - 4 C_F e \q_u \lwc{uG}{}[][rw] \lwc{uG}{*}[][vw] [M_u^\dagger]_{vs} \, , \\
\nnn
\nonumber\\[-1cm]
	\dlwc{d\gamma}{}[][rs] &=  \Bigl[ 2 C_{F}  g^2  + \left(10 \q_d^2 -b_{0,e}\right) e^2 \Bigr] \lwc{d\gamma}{}[][rs] + 8 C_{F} e g\q_d  \lwc{dG}{}[][rs]  \nn
		&\quad - 8 e \q_e [M_e]_{wv} \lwc{e d}{RR}[T][vwrs] + 2 e \q_d [M_d]_{wv} \lwc{dd}{RR}[S1][rwvs]   + 2 e C_F \q_d [M_d]_{wv} \lwc{dd}{RR}[S8][rwvs]  \nn
		&\quad + e \q_u  [M_u]_{wv} \lwc{uddu}{RR}[S1][vsrw] + e C_F \q_u  [M_u]_{wv} \lwc{uddu}{RR}[S8][vsrw]  \nn
		&\quad  - \lwc{d \gamma}{}[][rw] \Bigl( 4 g C_F \lwc{d G}{*}[][vw] + 8 e \q_d \lwc{d \gamma}{*}[][vw]\Bigr) [M_d^\dagger]_{vs} - [M_d^\dagger]_{rw}  \Bigl( 4 g C_F \lwc{d G}{*}[][vw] + 8 e \q_d \lwc{d \gamma}{*}[][vw] \Bigr) \lwc{d \gamma}{}[][vs]  \nn
		&\quad - 8 e \q_e  \lwc{d\gamma}{}[][rs] \Bigl( \lwc{e\gamma}{}[][vw] [M_e]_{wv} +  [M_e^\dagger]_{vw} \lwc{e\gamma}{*}[][vw] \Bigr)  - 8 e N_c \q_u  \lwc{d\gamma}{}[][rs] \Bigl( \lwc{u\gamma}{}[][vw] [M_u]_{wv} +  [M_u^\dagger]_{vw} \lwc{u\gamma}{*}[][vw] \Bigr) \nn
		&\quad - 8 e N_c \q_d  \lwc{d\gamma}{}[][rs] \Bigl( \lwc{d\gamma}{}[][vw] [M_d]_{wv} +  [M_d^\dagger]_{vw} \lwc{d\gamma}{*}[][vw] \Bigr)  -12 e \q_d  \lwc{d \gamma}{}[][rw] [M_d]_{wv} \lwc{d \gamma}{}[][vs]  \nn
		&\quad - 4 e \q_d [M_d]_{wv} C_F \lwc{d G}{}[][rw] \lwc{d G}{}[][vs] - 4 g [M_d]_{wv} C_F \Bigl( \lwc{d \gamma}{}[][rw] \lwc{d G}{}[][vs] + \lwc{d G}{}[][rw] \lwc{d \gamma}{}[][vs] \Bigr)  \nn
		&\quad - 4 C_F e \q_d \lwc{dG}{*}[][wv] \lwc{dG}{}[][ws] [M_d^\dagger]_{rv} - 4 C_F e \q_d \lwc{dG}{}[][rw] \lwc{dG}{*}[][vw] [M_d^\dagger]_{vs}  \, .
\end{align}
The anomalous dimensions for the gluonic dipole operators are
\begin{align}
	\dlwc{uG}{}[][rs] &=  \Bigl[ \left(10 C_{F} - 4C_{A} -b_{0,g}\right) g^2 + 2 e^2 \q_u^2  \Bigr] \lwc{uG}{}[][rs]  +  8 e g \q_u \lwc{u\gamma}{}[][rs]  + 3 g^2 C_A [M^\dagger_u]_{rs} (L_G + i L_{\widetilde G}) \nn
		&\quad + 2 g [M_u]_{wv} \lwc{uu}{RR}[S1][rwvs]   + 2 g \Bigl( C_F - \frac{1}{2} C_A \Bigr) [M_u]_{wv} \lwc{uu}{RR}[S8][rwvs]  \nn
		&\quad + g  [M_d]_{wv} \lwc{uddu}{RR}[S1][rwvs] + g \Bigl( C_F - \frac{1}{2} C_A \Bigr) [M_d]_{wv} \lwc{uddu}{RR}[S8][rwvs] \nn
		&\quad - \lwc{u G}{}[][rw] \Bigl( (8 C_F - 2 C_A) g \lwc{u G}{*}[][vw] + 4 e \q_u \lwc{u \gamma}{*}[][vw] \Bigr) [M_u^\dagger]_{vs}  \nn
		&\quad - [M_u^\dagger]_{rw}  \Bigl( (8 C_F - 2 C_A) g \lwc{u G}{*}[][vw] + 4 e \q_u \lwc{u \gamma}{*}[][vw] \Bigr) \lwc{u G}{}[][vs]  \nn
		&\quad - 4 g \lwc{uG}{}[][rs] \Bigl( \lwc{uG}{}[][vw] [M_u]_{wv} +  [M_u^\dagger]_{vw} \lwc{uG}{*}[][vw] \Bigr)  - 4 g \lwc{uG}{}[][rs] \Bigl( \lwc{dG}{}[][vw] [M_d]_{wv} +  [M_d^\dagger]_{vw} \lwc{dG}{*}[][vw] \Bigr)  \nn
		&\quad + (10 C_A-12 C_F) g  \lwc{u G}{}[][r w] [M_u]_{wv} \lwc{u G}{}[][v s] - 4 e \q_u [M_u]_{wv} \Bigl( \lwc{u \gamma}{}[][rw] \lwc{u G}{}[][vs] + \lwc{u G}{}[][rw] \lwc{u \gamma}{}[][vs] \Bigr)  \nn
		&\quad - 4 g [M_u]_{wv} \lwc{u\gamma}{}[][rw] \lwc{u\gamma}{}[][vs]  - 4 g \lwc{u\gamma}{*}[][wv] \lwc{u\gamma}{}[][ws] [M_u^\dagger]_{rv} - 4 g \lwc{u\gamma}{}[][rw] \lwc{u\gamma}{*}[][vw] [M_u^\dagger]_{vs}  \, , \\
\nnn
	\dlwc{dG}{}[][rs] &=  \Bigl[ \left(10 C_{F} - 4C_{A} -b_{0,g}\right) g^2 + 2 e^2 \q_d^2  \Bigr] \lwc{dG}{}[][rs]  +  8 e g \q_d \lwc{d\gamma}{}[][rs]  + 3 g^2 C_A [M^\dagger_d]_{rs} (L_G + i L_{\widetilde G}) \nn
		&\quad + 2 g [M_d]_{wv} \lwc{dd}{RR}[S1][rwvs]  + 2 g \Bigl( C_F - \frac{1}{2} C_A \Bigr) [M_d]_{wv} \lwc{dd}{RR}[S8][rwvs]  \nn
		&\quad + g  [M_u]_{wv} \lwc{uddu}{RR}[S1][vsrw] + g \Bigl( C_F - \frac{1}{2} C_A \Bigr) [M_u]_{wv} \lwc{uddu}{RR}[S8][vsrw]  \nn
		&\quad - \lwc{d G}{}[][rw] \Bigl( (8 C_F - 2 C_A) g \lwc{d G}{*}[][vw] + 4 e \q_d \lwc{d \gamma}{*}[][vw] \Bigr) [M_d^\dagger]_{vs}  \nn
		&\quad - [M_d^\dagger]_{rw}  \Bigl( (8 C_F - 2 C_A) g \lwc{d G}{*}[][vw] + 4 e \q_d \lwc{d \gamma}{*}[][vw] \Bigr) \lwc{d G}{}[][vs]  \nn
		&\quad - 4 g \lwc{dG}{}[][rs] \Bigl( \lwc{uG}{}[][vw] [M_u]_{wv} +  [M_u^\dagger]_{vw} \lwc{uG}{*}[][vw] \Bigr)  - 4 g \lwc{dG}{}[][rs] \Bigl( \lwc{dG}{}[][vw] [M_d]_{wv} +  [M_d^\dagger]_{vw} \lwc{dG}{*}[][vw] \Bigr)  \nn
		&\quad + (10 C_A-12 C_F) g   \lwc{d G}{}[][r w] [M_d]_{wv} \lwc{d G}{}[][v s] - 4 e \q_d [M_d]_{wv} \Bigl( \lwc{d \gamma}{}[][rw] \lwc{d G}{}[][vs] + \lwc{d G}{}[][rw] \lwc{d \gamma}{}[][vs] \Bigr)  \nns
		&\quad - 4 g [M_d]_{wv} \lwc{d\gamma}{}[][rw] \lwc{d\gamma}{}[][vs]  - 4 g \lwc{d\gamma}{*}[][wv] \lwc{d\gamma}{}[][ws] [M_d^\dagger]_{rv} - 4 g \lwc{d\gamma}{}[][rw] \lwc{d\gamma}{*}[][vw] [M_d^\dagger]_{vs}   \, .
\end{align}

\subsection{Dimension 6}

The anomalous dimensions of the dimension-six operators grouped by type are listed below.

\subsubsection{$X^3$}

\begin{align}
	\dlwc{G}{}[][] &= \left( 12 c_{A} - 3 b_{0,g}\right) g^2 \lwc{G}{}[][]  \, , \\
\nnn
	\dlwc{\widetilde G}{}[][]  &=\left( 12 c_{A} - 3 b_{0,g}\right) g^2 \lwc{\widetilde G}{}[][] \, .
\end{align}

\subsubsection{$\psi^4: (\overline L L)(\overline L L)$}

\begin{align}
	\dlwc{\nu\nu}{LL}[V][prst] &= 0 \, , \\
\nnn
	\dlwc{ee}{LL}[V][prst] &= 12 e^2 \q_e^2 \lwc{ee}{LL}[V][prst] \nn
		&\quad + \frac{1}{3} e^2 \q_e \delta_{pr} \begin{aligned}[t]
			&\bigg[ N_c \q_d \Bigl( \lwc{ed}{LL}[V][stww]+\lwc{ed}{LR}[V][stww] \Bigr) + N_c \q_u \Bigl( \lwc{eu}{LL}[V][stww]+\lwc{eu}{LR}[V][stww] \Bigr) \nn
			& + \q_e \Bigl( 4 \lwc{ee}{LL}[V][stww]+\lwc{ee}{LR}[V][stww] \Bigr)  \bigg] \end{aligned} \nn
		&\quad + \frac{1}{3} e^2 \q_e \delta_{st} \begin{aligned}[t]
			&\bigg[ N_c \q_d \Bigl( \lwc{ed}{LL}[V][prww]+\lwc{ed}{LR}[V][prww] \Bigr) + N_c \q_u \Bigl(\lwc{eu}{LL}[V][prww]+\lwc{eu}{LR}[V][prww] \Bigr) \nn
			& + \q_e \Bigl( 4 \lwc{ee}{LL}[V][prww]+\lwc{ee}{LR}[V][prww] \Bigr) \bigg] \end{aligned} \nn
		&\quad + \frac{1}{3} e^2 \q_e \delta_{pt} \begin{aligned}[t]
			&\bigg[ N_c \q_d \Bigl( \lwc{ed}{LL}[V][srww]+\lwc{ed}{LR}[V][srww] \Bigr) + N_c \q_u \Bigl( \lwc{eu}{LL}[V][srww]+\lwc{eu}{LR}[V][srww] \Bigr) \nn
			& + \q_e \Bigl( 4 \lwc{ee}{LL}[V][srww]+\lwc{ee}{LR}[V][srww] \Bigr)  \bigg] \end{aligned} \nn
		&\quad + \frac{1}{3} e^2 \q_e \delta_{sr} \begin{aligned}[t]
			&\bigg[ N_c \q_d \Bigl( \lwc{ed}{LL}[V][ptww]+\lwc{ed}{LR}[V][ptww] \Bigr) + N_c \q_u \Bigl(\lwc{eu}{LL}[V][ptww]+\lwc{eu}{LR}[V][ptww] \Bigr) \nn
			& + \q_e \Bigl( 4 \lwc{ee}{LL}[V][ptww]+\lwc{ee}{LR}[V][ptww] \Bigr) \bigg] \end{aligned} \nn
		&\quad + \frac{26}{3} e^2 \q_e^2 \Bigl(  \lwc{e \gamma}{}[][pw] \lwc{e \gamma}{*}[][rw] \delta_{st} +  \lwc{e \gamma}{}[][sw] \lwc{e \gamma}{*}[][tw] \delta_{pr} 
			+  \lwc{e \gamma}{}[][pw] \lwc{e \gamma}{*}[][tw] \delta_{sr} +  \lwc{e \gamma}{}[][sw] \lwc{e \gamma}{*}[][rw] \delta_{pt} \Bigr)  \nn
		&\quad  + { \frac{1}{2} e^2 \q_e^2 \zeta_e \left(  \delta_{pr} \delta_{st} + \delta_{pt} \delta_{sr} \right)} \, , \\
\nnn
	\dlwc{\nu e}{LL}[V][prst] &= \frac{4}{3} e^2 \q_e \delta_{st} \begin{aligned}[t]
			& \bigg[ N_c \q_u \Bigl( \lwc{\nu u}{LL}[V][prww] + \lwc{\nu u}{LR}[V][prww] \Bigr) + N_c \q_d \Bigl( \lwc{\nu d}{LL}[V][prww]+ \lwc{\nu d}{LR}[V][prww] \Bigr) \\
			& + \q_e \Bigl( \lwc{\nu e}{LL}[V][prww]+ \lwc{\nu e}{LR}[V][prww] \Bigr) \bigg] \end{aligned} \nn
		&\quad + 96  e^2 \q_e^2 \lwc{\nu \gamma}{}[][wr] \lwc{\nu \gamma}{*}[][wp] \delta_{st}  \, , \\
\nnn
	\dlwc{\nu u}{LL}[V][prst] &= \frac{4}{3} e^2 \q_u \delta_{st} \begin{aligned}[t]
			& \bigg[ N_c \q_u \Bigl( \lwc{\nu u}{LL}[V][prww] + \lwc{\nu u}{LR}[V][prww] \Bigr) + N_c \q_d \Bigl( \lwc{\nu d}{LL}[V][prww]+ \lwc{\nu d}{LR}[V][prww] \Bigr) \\
			& + \q_e \Bigl( \lwc{\nu e}{LL}[V][prww]+ \lwc{\nu e}{LR}[V][prww] \Bigr) \bigg] \end{aligned} \nns
		&\quad + 96  e^2 \q_u^2 \lwc{\nu \gamma}{}[][wr] \lwc{\nu \gamma}{*}[][wp] \delta_{st}  \, , \\
\nnn
	\dlwc{\nu d}{LL}[V][prst] &= \frac{4}{3} e^2 \q_d \delta_{st} \begin{aligned}[t]
			& \bigg[ N_c \q_u \Bigl( \lwc{\nu u}{LL}[V][prww] + \lwc{\nu u}{LR}[V][prww] \Bigr) + N_c \q_d \Bigl( \lwc{\nu d}{LL}[V][prww]+ \lwc{\nu d}{LR}[V][prww] \Bigr) \\
			& + \q_e \Bigl( \lwc{\nu e}{LL}[V][prww]+ \lwc{\nu e}{LR}[V][prww] \Bigr) \bigg] \end{aligned} \nn
		&\quad + 96 e^2 \q_d^2  \lwc{\nu \gamma}{}[][wr] \lwc{\nu \gamma}{*}[][wp] \delta_{st} \, , \\
\nnn
	\dlwc{eu}{LL}[V][prst] &= \frac{4}{3} e^2 \q_e \delta_{pr} \begin{aligned}[t]
			&\bigg[ N_c \q_d \Bigl({{\lwc{ud}{LL}[V1][stww]}}+{{\lwc{ud}{LR}[V1][stww]}} \Bigr) + N_c \q_u \Bigl( 2 \lwc{uu}{LL}[V][stww]+{{\lwc{uu}{LR}[V1][stww]}} \Bigr) \nn
			& + \q_e \Bigl( \lwc{eu}{LL}[V][wwst]+\lwc{ue}{LR}[V][stww] \Bigr) + 2 \q_u \lwc{uu}{LL}[V][swwt] \bigg] \end{aligned} \nn
		&\quad + \frac{4}{3} e^2 \q_u \delta_{st} \begin{aligned}[t]
			& \bigg[ N_c \q_d \Bigl( \lwc{ed}{LL}[V][prww]+\lwc{ed}{LR}[V][prww] \Bigr) + N_c \q_u \Bigl( \lwc{eu}{LL}[V][prww]+\lwc{eu}{LR}[V][prww] \Bigr) \nn
			& + \q_e \Bigl( 4 \lwc{ee}{LL}[V][prww]+\lwc{ee}{LR}[V][prww] \Bigr) \bigg] \end{aligned} \nn
		&\quad + 12 e^2 \q_e \q_u \lwc{eu}{LL}[V][prst]  - \frac{16}{3} C_F e^2 \q_e \q_u \lwc{u G}{}[][sw] \lwc{u G}{*}[][tw] \delta_{pr}  + 8 C_F e g \q_e \Bigl( \lwc{uG}{}[][sw] \lwc{u \gamma}{*}[][tw] + \lwc{u \gamma}{}[][sw] \lwc{uG}{*}[][tw] \Bigr) \delta_{pr}  \nn
		&\quad + e^2  \Bigl( 24 \q_u^2 + \frac{32}{3} \q_e \q_u \Bigr)  \lwc{e \gamma}{}[][pw] \lwc{e \gamma}{*}[][rw] \delta_{st} + e^2 \Bigl( 24 \q_e^2 + \frac{32}{3} \q_e \q_u \Bigr)   \lwc{u \gamma}{}[][sw] \lwc{u \gamma}{*}[][tw] \delta_{pr} \nn
		&\quad + 2 e^2 \q_e \q_u   \zeta_e  \delta_{pr} \delta_{st} \, , \\
\nnn
	\dlwc{ed}{LL}[V][prst] &= \frac{4}{3} e^2 \q_e \delta_{pr} \begin{aligned}[t]
			& \bigg[ N_c \q_u \Bigl( {{\lwc{ud}{LL}[V1][wwst]}}+{{\lwc{du}{LR}[V1][stww]}} \Bigr) + N_c \q_d \Bigl( 2 \lwc{dd}{LL}[V][stww]+{{\lwc{dd}{LR}[V1][stww]}} \Bigr) \nn
			& + \q_e \Bigl( \lwc{ed}{LL}[V][wwst]+\lwc{de}{LR}[V][stww] \Bigr) + 2 \q_d \lwc{dd}{LL}[V][swwt] \bigg] \end{aligned} \nn
		&\quad + \frac{4}{3} e^2 \q_d \delta_{st} \begin{aligned}[t]
			& \bigg[ N_c \q_d \Bigl( \lwc{ed}{LL}[V][prww]+\lwc{ed}{LR}[V][prww] \Bigr) + N_c \q_u \Bigl( \lwc{eu}{LL}[V][prww]+\lwc{eu}{LR}[V][prww] \Bigr) \nn
			& + \q_e \Bigl( 4 \lwc{ee}{LL}[V][prww]+\lwc{ee}{LR}[V][prww] \Bigr) \bigg] \end{aligned} \nn
		&\quad + 12 e^2 \q_d \q_e \lwc{ed}{LL}[V][prst]  - \frac{16}{3} C_F e^2 \q_e \q_d \lwc{d G}{}[][sw] \lwc{d G}{*}[][tw] \delta_{pr}  + 8 C_F e g \q_e \Bigl( \lwc{dG}{}[][sw] \lwc{d \gamma}{*}[][tw] + \lwc{d \gamma}{}[][sw] \lwc{dG}{*}[][tw] \Bigr) \delta_{pr}  \nn
		&\quad + e^2  \Bigl( 24 \q_d^2 + \frac{32}{3} \q_e \q_d \Bigr)  \lwc{e \gamma}{}[][pw] \lwc{e \gamma}{*}[][rw] \delta_{st}  +  e^2  \Bigl( 24 \q_e^2 + \frac{32}{3} \q_e \q_d \Bigr)   \lwc{d \gamma}{}[][sw] \lwc{d \gamma}{*}[][tw] \delta_{pr} \nn
		&\quad + 2 e^2 \q_e \q_d   \zeta_e  \delta_{pr} \delta_{st}  \, , \\
\nnn
	\dlwc{\nu e du}{LL}[V][prst] &= 6 e^2 \q_e \q_u \lwc{\nu e du}{LL}[V][prst]  \, , \\
\nnn
	\dlwc{uu}{LL}[V][prst] &= \frac{2}{3} e^2 \q_u \delta_{pr} \begin{aligned}[t]
			&\bigg[ N_c \q_d \Bigl( \lwc{ud}{LL}[V1][stww]+\lwc{ud}{LR}[V1][stww] \Bigr) + N_c \q_u \Bigl( 2 \lwc{uu}{LL}[V][stww]+\lwc{uu}{LR}[V1][stww] \Bigr) \nn
			& + \q_e \Bigl( \lwc{eu}{LL}[V][wwst]+\lwc{ue}{LR}[V][stww] \Bigr) + 2 \q_u \lwc{uu}{LL}[V][swwt] \bigg] \end{aligned} \nn
		&\quad + \frac{2}{3} e^2 \q_u \delta_{st} \begin{aligned}[t]
			&\bigg[ N_c \q_d \Bigl( \lwc{ud}{LL}[V1][prww]+\lwc{ud}{LR}[V1][prww] \Bigr) + N_c \q_u \Bigl( 2 \lwc{uu}{LL}[V][prww]+\lwc{uu}{LR}[V1][prww] \Bigr) \\
			& + \q_e \Bigl( \lwc{eu}{LL}[V][wwpr]+\lwc{ue}{LR}[V][prww] \Bigr) + 2 \q_u \lwc{uu}{LL}[V][pwwr] \bigg] \end{aligned} \nn
		&\quad + \frac{g^2}{6 N_c} \begin{aligned}[t]
			&\bigg[ N_c \delta_{pt} \Bigl( \lwc{ud}{LL}[V8][srww]+\lwc{ud}{LR}[V8][srww]+4 \lwc{uu}{LL}[V][swwr]+\lwc{uu}{LR}[V8][srww] \Bigr) \nn
			& + N_c \delta_{rs} \Bigl( \lwc{ud}{LL}[V8][ptww]+\lwc{ud}{LR}[V8][ptww]+4 \lwc{uu}{LL}[V][pwwt]+\lwc{uu}{LR}[V8][ptww] \Bigr) \nn
			& - \delta_{pr} \Bigl( \lwc{ud}{LL}[V8][stww] + \lwc{ud}{LR}[V8][stww] + 4 \lwc{uu}{LL}[V][swwt] + \lwc{uu}{LR}[V8][stww] \Bigr) \nn
			& - \delta_{st} \Bigl( \lwc{ud}{LL}[V8][prww] + \lwc{ud}{LR}[V8][prww] + 4 \lwc{uu}{LL}[V][pwwr] + \lwc{uu}{LR}[V8][prww] \Bigr) \bigg] \end{aligned} \nn
		&\quad + \left[ 12 e^2 \q_u^2  - \frac{6 g^2}{N_c} \right] \lwc{uu}{LL}[V][prst] + 6 g^2 \lwc{uu}{LL}[V][ptsr]  \nn
		&\quad - \frac{8}{3} C_F e^2 \q_u^2 \Bigl( \lwc{uG}{}[][pw] \lwc{uG}{*}[][rw] \delta_{st} + \lwc{uG}{}[][sw] \lwc{uG}{*}[][tw] \delta_{pr} \Bigr) \nn
		&\quad - \frac{1}{2N_c} \Bigl(\frac{16}{3} C_F + 3 C_d - 5 C_A - 24 N_c C_1\Bigr) g^2 \Bigl( \lwc{uG}{}[][pw] \lwc{uG}{*}[][rw] \delta_{st} + \lwc{uG}{}[][sw] \lwc{uG}{*}[][tw] \delta_{pr} \Bigr) \nn
		&\quad + \frac{1}{2} \Bigl(\frac{16}{3} C_F + 3 C_d - 5 C_A\Bigr) g^2 \Bigl( \lwc{uG}{}[][pw] \lwc{uG}{*}[][tw] \delta_{sr} +  \lwc{uG}{}[][sw] \lwc{uG}{*}[][rw] \delta_{pt} \Bigr) \nn
		&\quad + 4 \Bigl( C_F - \frac{2}{N_c} \Bigr) e g \q_u \Bigl( \lwc{u \gamma}{}[][pw] \lwc{uG}{*}[][rw]  + \lwc{uG}{}[][pw] \lwc{u\gamma}{*}[][rw] \Bigr) \delta_{st} \nn
		&\quad + 4 \Bigl( C_F - \frac{2}{N_c} \Bigr) e g \q_u \Bigl(  \lwc{u \gamma}{}[][sw] \lwc{uG}{*}[][tw] + \lwc{uG}{}[][sw] \lwc{u\gamma}{*}[][tw] \Bigr) \delta_{pr}  \nn
		&\quad + 8 e g \q_u \Bigl( \lwc{u\gamma}{}[][pw] \lwc{uG}{*}[][tw] + \lwc{uG}{}[][pw] \lwc{u\gamma}{*}[][tw] \Bigr) \delta_{sr} + 8 e g \q_u \Bigl( \lwc{u\gamma}{}[][sw] \lwc{uG}{*}[][rw] + \lwc{uG}{}[][sw] \lwc{u\gamma}{*}[][rw] \Bigr) \delta_{pt} \nn
		&\quad + \left[ \frac{52}{3} e^2 \q_u^2 + \frac{4}{3N_c} g^2 \right] \Bigl(  \lwc{u \gamma}{}[][pw] \lwc{u \gamma}{*}[][rw] \delta_{st} +   \lwc{u \gamma}{}[][sw] \lwc{u \gamma}{*}[][tw] \delta_{pr} \Bigr) 
			 - \frac{4}{3} g^2 \Bigl( \lwc{u\gamma}{}[][pw] \lwc{u\gamma}{*}[][tw] \delta_{sr} + \lwc{u\gamma}{}[][sw] \lwc{u\gamma}{*}[][rw] \delta_{pt} \Bigr) \nn
		&\quad + e^2 \q_u^2  \zeta_e  \delta_{pr} \delta_{st}  - \frac{1}{2N_c} g^2   \zeta_g  \delta_{pr} \delta_{st} + \frac{1}{2} g^2   \zeta_g  \delta_{pt} \delta_{sr}   \, , \\
\nnn
	\dlwc{dd}{LL}[V][prst] &= \frac{2}{3} e^2 \q_d \delta_{pr} \begin{aligned}[t]
			&\bigg[  N_c \q_d \Bigl( 2 \lwc{dd}{LL}[V][stww]+\lwc{dd}{LR}[V1][stww] \Bigr) + N_c \q_u \Bigl( \lwc{ud}{LL}[V1][wwst]+\lwc{du}{LR}[V1][stww] \Bigr) \nn
			& + \q_e \Bigl(\lwc{ed}{LL}[V][wwst]+\lwc{de}{LR}[V][stww] \Bigr) + 2 \q_d \lwc{dd}{LL}[V][swwt] \bigg] \end{aligned} \nn
		&\quad + \frac{2}{3} e^2 \q_d \delta_{st} \begin{aligned}[t]
			&\bigg[ N_c \q_d \Bigl( 2 \lwc{dd}{LL}[V][prww]+\lwc{dd}{LR}[V1][prww]\Bigr) + N_c \q_u \Bigl(\lwc{ud}{LL}[V1][wwpr]+\lwc{du}{LR}[V1][prww] \Bigr) \nn
			& + \q_e \Bigl( \lwc{ed}{LL}[V][wwpr]+\lwc{de}{LR}[V][prww] \Bigr) + 2 \q_d \lwc{dd}{LL}[V][pwwr] \bigg] \end{aligned} \nn
		&\quad + \frac{g^2}{6 N_c} \begin{aligned}[t]
			&\bigg[ N_c \delta_{pt} \Bigl( 4 \lwc{dd}{LL}[V][swwr]+\lwc{ud}{LL}[V8][wwsr]+\lwc{du}{LR}[V8][srww]+\lwc{dd}{LR}[V8][srww] \Bigr) \nn
			& + N_c \delta_{rs} \Bigl( 4 \lwc{dd}{LL}[V][pwwt]+\lwc{ud}{LL}[V8][wwpt]+\lwc{du}{LR}[V8][ptww]+\lwc{dd}{LR}[V8][ptww] \Bigr) \nn
			& - \delta_{pr} \Bigl( 4 \lwc{dd}{LL}[V][swwt] + \lwc{ud}{LL}[V8][wwst] + \lwc{du}{LR}[V8][stww] + \lwc{dd}{LR}[V8][stww] \Bigr) \nn
			& - \delta_{st} \Bigl( 4 \lwc{dd}{LL}[V][pwwr] + \lwc{ud}{LL}[V8][wwpr] + \lwc{du}{LR}[V8][prww] + \lwc{dd}{LR}[V8][prww] \Bigr) \bigg] \end{aligned} \nn
		&\quad + \left[ 12 e^2 \q_d^2 - \frac{6 g^2}{N_c} \right] \lwc{dd}{LL}[V][prst] + 6 g^2 \lwc{dd}{LL}[V][ptsr]  \nn
		&\quad - \frac{8}{3} C_F e^2 \q_d^2  \Bigl( \lwc{dG}{}[][pw] \lwc{dG}{*}[][rw] \delta_{st} + \lwc{dG}{}[][sw] \lwc{dG}{*}[][tw] \delta_{pr}  \Bigr) \nn
		&\quad - \frac{1}{2N_c} \Bigl(\frac{16}{3} C_F + 3 C_d - 5 C_A - 24 N_c C_1 \Bigr) g^2 \Bigl( \lwc{dG}{}[][pw] \lwc{dG}{*}[][rw] \delta_{st} + \lwc{dG}{}[][sw] \lwc{dG}{*}[][tw] \delta_{pr} \Bigr) \nn
		&\quad + \frac{1}{2} \Bigl(\frac{16}{3} C_F + 3 C_d - 5 C_A\Bigr) g^2 \Bigl( \lwc{dG}{}[][pw] \lwc{dG}{*}[][tw] \delta_{sr} + \lwc{dG}{}[][sw] \lwc{dG}{*}[][rw] \delta_{pt} \Bigr) \nn
		&\quad + 4 \Bigl( C_F - \frac{2}{N_c} \Bigr) e g \q_d \Bigl(\lwc{d \gamma}{}[][pw] \lwc{dG}{*}[][rw] + \lwc{dG}{}[][pw] \lwc{d\gamma}{*}[][rw] \Bigr) \delta_{st} \nn
		&\quad + 4 \Bigl( C_F - \frac{2}{N_c} \Bigr) e g \q_d \Bigl( \lwc{d \gamma}{}[][sw] \lwc{dG}{*}[][tw] + \lwc{dG}{}[][sw] \lwc{d\gamma}{*}[][tw] \Bigr) \delta_{pr}  \nn
		&\quad + 8 eg \q_d \Bigl(  \lwc{d \gamma}{}[][pw] \lwc{d G}{*}[][tw]+\lwc{d G}{}[][pw] \lwc{d \gamma}{*}[][tw] \Bigr) \delta_{sr} + 8 e g \q_d \Bigl(  \lwc{d \gamma}{}[][sw] \lwc{d G}{*}[][rw]+\lwc{d G}{}[][sw] \lwc{d \gamma}{*}[][rw] \Bigr) \delta_{pt}  \nn
		&\quad + \left[ \frac{52}{3} e^2 \q_d^2 + \frac{4}{3N_c} g^2 \right] \Bigl(  \lwc{d \gamma}{}[][pw] \lwc{d \gamma}{*}[][rw] \delta_{st}  +   \lwc{d \gamma}{}[][sw] \lwc{d \gamma}{*}[][tw] \delta_{pr} \Bigr) 
			 - \frac{4}{3} g^2 \Bigl( \lwc{d\gamma}{}[][pw] \lwc{d\gamma}{*}[][tw] \delta_{sr} + \lwc{d\gamma}{}[][sw] \lwc{d\gamma}{*}[][rw] \delta_{pt} \Bigr) \nn
		&\quad + e^2 \q_d^2    \zeta_e  \delta_{pr} \delta_{st}  - \frac{1}{2N_c} g^2   \zeta_g  \delta_{pr} \delta_{st} + \frac{1}{2} g^2   \zeta_g  \delta_{pt} \delta_{sr}  \, , \\
\nnn
	\dlwc{ud}{LL}[V1][prst] &= \frac{4}{3} e^2 \q_d \delta_{st} \begin{aligned}[t]
			&\bigg[ N_c \q_d \Bigl(\lwc{ud}{LL}[V1][prww]+\lwc{ud}{LR}[V1][prww]\Bigr) + N_c \q_u \Bigl( 2 \lwc{uu}{LL}[V][prww]+\lwc{uu}{LR}[V1][prww] \Bigr) \nn
			& + \q_e \Bigl(\lwc{eu}{LL}[V][wwpr]+\lwc{ue}{LR}[V][prww]\Bigr) + 2 \q_u \lwc{uu}{LL}[V][pwwr] \bigg] \end{aligned} \nn
		&\quad + \frac{4}{3} e^2 \q_u \delta_{pr} \begin{aligned}[t]
			&\bigg[  N_c \q_d \Bigl( 2 \lwc{dd}{LL}[V][stww]+\lwc{dd}{LR}[V1][stww] \Bigr) + N_c \q_u \Bigl( \lwc{ud}{LL}[V1][wwst]+\lwc{du}{LR}[V1][stww] \Bigr) \nn
			& + \q_e \Bigl( \lwc{ed}{LL}[V][wwst]+\lwc{de}{LR}[V][stww] \Bigr) + 2 \q_d \lwc{dd}{LL}[V][swwt] \bigg] \end{aligned} \nn
		&\quad + 12 e^2 \q_d \q_u \lwc{ud}{LL}[V1][prst] + 6 g^2 \frac{C_F}{N_c} \lwc{ud}{LL}[V8][prst]  \nn
		&\quad - \left[ \frac{16}{3} C_F e^2 \q_u \q_d - 24 g^2 C_1\right] \Bigl( \lwc{uG}{}[][pw] \lwc{uG}{*}[][rw] \delta_{st} + \lwc{dG}{}[][sw] \lwc{dG}{*}[][tw] \delta_{pr} \Bigr) \nn
		&\quad + 8 C_F e g \q_d \Bigl( \lwc{uG}{}[][pw] \lwc{u\gamma}{*}[][rw] + \lwc{u \gamma}{}[][pw] \lwc{uG}{*}[][rw] \Bigr) \delta_{st} + 8 C_F e g \q_u \Bigl( \lwc{dG}{}[][sw] \lwc{d\gamma}{*}[][tw] + \lwc{d \gamma}{}[][sw] \lwc{dG}{*}[][tw] \Bigr) \delta_{pr}  \nn
		&\quad + e^2 \Bigl( 24 \q_d^2 + \frac{32}{3} \q_u \q_d  \Bigr) \lwc{u \gamma}{}[][pw] \lwc{u \gamma}{*}[][rw] \delta_{st}   + e^2 \Bigl( 24 \q_u^2 + \frac{32}{3} \q_u \q_d  \Bigr)    \lwc{d \gamma}{}[][sw] \lwc{d \gamma}{*}[][tw] \delta_{pr}  \nn
		&\quad + 2 e^2 \q_u \q_d   \zeta_e  \delta_{pr} \delta_{st}  \, , \\
\nnn
	\dlwc{ud}{LL}[V8][prst] &=  \frac{2}{3} g^2 \delta_{pr} \begin{aligned}[t]
			&\bigg[ 4 \lwc{dd}{LL}[V][swwt]+\lwc{ud}{LL}[V8][wwst]+\lwc{du}{LR}[V8][stww]+\lwc{dd}{LR}[V8][stww] \bigg] \end{aligned} \nn
		&\quad + \frac{2}{3} g^2 \delta_{st} \begin{aligned}[t]
			&\bigg[ \lwc{ud}{LL}[V8][prww]+\lwc{ud}{LR}[V8][prww]+4 \lwc{uu}{LL}[V][pwwr]+\lwc{uu}{LR}[V8][prww] \bigg] \end{aligned} \nn
		&\quad + 12 g^2 \lwc{ud}{LL}[V1][prst] + \left[ 12 e^2 \q_d \q_u - \frac{12 g^2}{N_c} \right] \lwc{ud}{LL}[V8][prst]  \nn
		&\quad + \Bigl(\frac{32}{3}C_F + 6 C_d - 10 C_A\Bigr)  g^2 \Bigl( \lwc{u G}{}[][pw] \lwc{u G}{*}[][rw] \delta_{st} +  \lwc{d G}{}[][sw] \lwc{d G}{*}[][tw] \delta_{pr} \Bigr) \nn
		&\quad + 8 e g ( \q_u + 3 \q_d ) \Bigl( \lwc{u\gamma}{}[][pw] \lwc{uG}{*}[][rw] + \lwc{uG}{}[][pw] \lwc{u\gamma}{*}[][rw] \Bigr) \delta_{st} + 8 e g ( \q_d + 3 \q_u) \Bigl( \lwc{d\gamma}{}[][sw] \lwc{dG}{*}[][tw] + \lwc{dG}{}[][sw] \lwc{d\gamma}{*}[][tw] \Bigr) \delta_{pr}  \nn
		&\quad - \frac{16}{3} g^2 \Bigl( \lwc{u\gamma}{}[][pw] \lwc{u\gamma}{*}[][rw] \delta_{st} + \lwc{d\gamma}{}[][sw] \lwc{d\gamma}{*}[][tw] \delta_{pr} \Bigr)  + 2 g^2    \zeta_g  \delta_{pr} \delta_{st}  \, . 
\end{align}
%

\subsubsection{$\psi^4: (\overline R R)(\overline R R)$}

\begin{align}
	\dlwc{ee}{RR}[V][prst] &= 12 e^2 \q_e^2 \lwc{ee}{RR}[V][prst]  \nn
		&\quad + \frac{1}{3} e^2 \q_e \delta_{pr} \begin{aligned}[t]
			& \bigg[ N_c \q_d \Bigl(\lwc{de}{LR}[V][wwst]+\lwc{ed}{RR}[V][stww]\Bigr) + N_c \q_u \Bigl(\lwc{ue}{LR}[V][wwst]+\lwc{eu}{RR}[V][stww] \Bigr) \nn
			& + \q_e \Bigl( \lwc{ee}{LR}[V][wwst]+4\lwc{ee}{RR}[V][stww] \Bigr) \bigg] \end{aligned} \nn
		&\quad + \frac{1}{3} e^2 \q_e \delta_{st} \begin{aligned}[t]
			& \bigg[ N_c \q_d \Bigl( \lwc{de}{LR}[V][wwpr]+\lwc{ed}{RR}[V][prww] \Bigr) + N_c \q_u \Bigl( \lwc{ue}{LR}[V][wwpr]+\lwc{eu}{RR}[V][prww] \Bigr) \nn
			& + \q_e \Bigl( \lwc{ee}{LR}[V][wwpr]+4\lwc{ee}{RR}[V][prww] \Bigr) \bigg] \end{aligned} \nn
		&\quad + \frac{1}{3} e^2 \q_e \delta_{pt} \begin{aligned}[t]
			& \bigg[ N_c \q_d \Bigl(\lwc{de}{LR}[V][wwsr]+\lwc{ed}{RR}[V][srww]\Bigr) + N_c \q_u \Bigl(\lwc{ue}{LR}[V][wwsr]+\lwc{eu}{RR}[V][srww] \Bigr) \nn
			& + \q_e \Bigl( \lwc{ee}{LR}[V][wwsr]+4\lwc{ee}{RR}[V][srww] \Bigr) \bigg] \end{aligned} \nn
		&\quad + \frac{1}{3} e^2 \q_e \delta_{sr} \begin{aligned}[t]
			& \bigg[ N_c \q_d \Bigl( \lwc{de}{LR}[V][wwpt]+\lwc{ed}{RR}[V][ptww] \Bigr) + N_c \q_u \Bigl( \lwc{ue}{LR}[V][wwpt]+\lwc{eu}{RR}[V][ptww] \Bigr) \nn
			& + \q_e \Bigl( \lwc{ee}{LR}[V][wwpt]+4\lwc{ee}{RR}[V][ptww] \Bigr) \bigg] \end{aligned} \nn
		&\quad + \frac{26}{3} e^2 \q_e^2 \Bigl( \lwc{e \gamma}{*}[][wp] \lwc{e \gamma}{}[][wr] \delta_{st} + \lwc{e \gamma}{*}[][ws] \lwc{e \gamma}{}[][wt] \delta_{pr} 
			+ \lwc{e \gamma}{*}[][wp] \lwc{e \gamma}{}[][wt] \delta_{sr} + \lwc{e \gamma}{*}[][ws] \lwc{e \gamma}{}[][wr] \delta_{pt} \Bigr) \nns
		&\quad + { \frac{1}{2} e^2 \q_e^2 \zeta_e \left(  \delta_{pr} \delta_{st} + \delta_{pt} \delta_{sr} \right) } \, , \\
\nnn
	\dlwc{eu}{RR}[V][prst] &= \frac{4}{3} e^2 \q_e \delta_{pr} \begin{aligned}[t]
			& \bigg[ N_c \q_d \Bigl( {{\lwc{du}{LR}[V1][wwst]}}+{{\lwc{ud}{RR}[V1][stww]}} ) + N_c \q_u \Bigl( {{\lwc{uu}{LR}[V1][wwst]}}+2\lwc{uu}{RR}[V][stww] \Bigr) \nn
			& +\q_e \Bigl( \lwc{eu}{LR}[V][wwst]+\lwc{eu}{RR}[V][wwst] \Bigr) + 2 \q_u \lwc{uu}{RR}[V][swwt] \bigg] \end{aligned} \nn
		&\quad + \frac{4}{3} e^2 \q_u \delta_{st} \begin{aligned}[t]
			& \bigg[ N_c \q_d \Bigl( \lwc{de}{LR}[V][wwpr]+\lwc{ed}{RR}[V][prww] \Bigr) + N_c \q_u \Bigl( \lwc{ue}{LR}[V][wwpr]+\lwc{eu}{RR}[V][prww] \Bigr) \nn
			& +\q_e \Bigl( \lwc{ee}{LR}[V][wwpr]+4\lwc{ee}{RR}[V][prww] \Bigr) \bigg] \end{aligned} \nn
		&\quad + 12 e^2 \q_e \q_u \lwc{eu}{RR}[V][prst]  - \frac{16}{3} C_F e^2 \q_e \q_u \lwc{uG}{*}[][ws] \lwc{uG}{}[][wt] \delta_{pr} + 8 C_F e g \q_e \Bigl( \lwc{uG}{*}[][ws] \lwc{u \gamma}{}[][wt] + \lwc{u \gamma}{*}[][ws] \lwc{uG}{}[][wt] \Bigr) \delta_{pr} \nn
		&\quad + e^2 \Bigl( 24 \q_u^2 + \frac{32}{3} \q_e \q_u \Bigr) \lwc{e \gamma}{*}[][wp] \lwc{e \gamma}{}[][wr] \delta_{st}   + e^2 \Bigl( 24 \q_e^2 + \frac{32}{3} \q_e \q_u \Bigr)  \lwc{u \gamma}{*}[][ws] \lwc{u \gamma}{}[][wt] \delta_{pr}  \nn
		&\quad + 2 e^2 \q_e \q_u   \zeta_e  \delta_{pr} \delta_{st}  \, , \\
\nnn
	\dlwc{ed}{RR}[V][prst] &= \frac{4}{3} e^2 \q_e \delta_{pr} \begin{aligned}[t]
			& \bigg[ N_c \q_d \Bigl( {{\lwc{dd}{LR}[V1][wwst]}}+2\lwc{dd}{RR}[V][stww] \Bigr) + N_c \q_u \Bigl( {{\lwc{ud}{LR}[V1][wwst]}}+{{\lwc{ud}{RR}[V1][wwst]}} \Bigr) \nn
			&+ \q_e \Bigl(\lwc{ed}{LR}[V][wwst]+\lwc{ed}{RR}[V][wwst]\Bigr) + 2 \q_d \lwc{dd}{RR}[V][swwt] \bigg] \end{aligned} \nn
		&\quad + \frac{4}{3} e^2 \q_d \delta_{st} \begin{aligned}[t]
			& \bigg[ N_c \q_d \Bigl( \lwc{de}{LR}[V][wwpr]+\lwc{ed}{RR}[V][prww] \Bigr) + N_c \q_u \Bigl( \lwc{ue}{LR}[V][wwpr]+\lwc{eu}{RR}[V][prww] \Bigr) \nn
			& + \q_e \Bigl( \lwc{ee}{LR}[V][wwpr]+4\lwc{ee}{RR}[V][prww] \Bigr) \bigg] \end{aligned} \nn
		&\quad + 12 e^2 \q_d \q_e \lwc{ed}{RR}[V][prst]   - \frac{16}{3} C_F e^2 \q_e \q_d \lwc{dG}{*}[][ws] \lwc{dG}{}[][wt] \delta_{pr}   + 8 C_F e g \q_e \Bigl( \lwc{dG}{*}[][ws] \lwc{d \gamma}{}[][wt] + \lwc{d \gamma}{*}[][ws] \lwc{dG}{}[][wt] \Bigr) \delta_{pr}  \nn
		&\quad + e^2 \Bigl( 24 \q_d^2 + \frac{32}{3} \q_e \q_d \Bigr) \lwc{e \gamma}{*}[][wp] \lwc{e \gamma}{}[][wr] \delta_{st}  + e^2 \Bigl( 24 \q_e^2 + \frac{32}{3} \q_e \q_d \Bigr)  \lwc{d \gamma}{*}[][ws] \lwc{d \gamma}{}[][wt] \delta_{pr} \nn
		&\quad + 2 e^2 \q_e \q_d   \zeta_e  \delta_{pr} \delta_{st}   \, , \\
\nnn
	\dlwc{uu}{RR}[V][prst] &= \frac{2}{3} e^2 \q_u \delta_{pr} \begin{aligned}[t]
			& \bigg[ N_c \q_d \Bigl( {{\lwc{du}{LR}[V1][wwst]}}+{{\lwc{ud}{RR}[V1][stww]}} \Bigr) + N_c \q_u \Bigl( {{\lwc{uu}{LR}[V1][wwst]}}+2\lwc{uu}{RR}[V][stww] \Bigr) \nn
			& + \q_e \Bigl( \lwc{eu}{LR}[V][wwst]+\lwc{eu}{RR}[V][wwst] \Bigr) + 2 \q_u \lwc{uu}{RR}[V][swwt] \bigg] \end{aligned} \nn
		&\quad + \frac{2}{3} e^2 \q_u \delta_{st} \begin{aligned}[t]
			& \bigg[ N_c \q_d \Bigl( {{\lwc{du}{LR}[V1][wwpr]}}+{{\lwc{ud}{RR}[V1][prww]}} \Bigr) + N_c \q_u \Bigl( {{\lwc{uu}{LR}[V1][wwpr]}}+2\lwc{uu}{RR}[V][prww] \Bigr) \nn
			& + \q_e \Bigl( \lwc{eu}{LR}[V][wwpr]+\lwc{eu}{RR}[V][wwpr] \Bigr) + 2 \q_u \lwc{uu}{RR}[V][pwwr] \bigg] \end{aligned} \nn
		&\quad + \frac{g^2}{6 N_c} \begin{aligned}[t]
			& \bigg[ N_c \delta_{pt} \Bigl( \lwc{du}{LR}[V8][wwsr]+\lwc{ud}{RR}[V8][srww]+\lwc{uu}{LR}[V8][wwsr]+4 \lwc{uu}{RR}[V][swwr] \Bigr) \nn
			& + N_c \delta_{rs} \Bigl( \lwc{du}{LR}[V8][wwpt]+\lwc{ud}{RR}[V8][ptww]+\lwc{uu}{LR}[V8][wwpt]+4 \lwc{uu}{RR}[V][pwwt] \Bigr) \nn
			& - \delta_{pr} \Bigl( \lwc{du}{LR}[V8][wwst] + \lwc{ud}{RR}[V8][stww] + \lwc{uu}{LR}[V8][wwst] + 4 \lwc{uu}{RR}[V][swwt] \Bigr) \nn
			& - \delta_{st} \Bigl( \lwc{du}{LR}[V8][wwpr] + \lwc{ud}{RR}[V8][prww] + \lwc{uu}{LR}[V8][wwpr] + 4 \lwc{uu}{RR}[V][pwwr] \Bigr) \bigg] \end{aligned} \nn
		&\quad + \left[ 12 e^2 \q_u^2 - \frac{6 g^2}{N_c} \right] \lwc{uu}{RR}[V][prst] + 6 g^2 \lwc{uu}{RR}[V][ptsr]  \nn
		&\quad - \frac{8}{3} C_F e^2 \q_u^2 \Bigl( \lwc{uG}{*}[][wp] \lwc{uG}{}[][wr] \delta_{st} + \lwc{uG}{*}[][ws] \lwc{uG}{}[][wt] \delta_{pr} \Bigr) \nn
		&\quad - \frac{1}{2N_c} \Bigl( \frac{16}{3} C_F + 3 C_d - 5 C_A - 24 N_c C_1 \Bigr) g^2 \Bigl( \lwc{uG}{*}[][wp] \lwc{uG}{}[][wr] \delta_{st} + \lwc{uG}{*}[][ws] \lwc{uG}{}[][wt] \delta_{pr} \Bigr) \nn
		&\quad + \frac{1}{2} \Bigl( \frac{16}{3} C_F + 3 C_d - 5 C_A \Bigr) g^2 \Bigl( \lwc{uG}{*}[][wp] \lwc{uG}{}[][wt] \delta_{sr} +  \lwc{uG}{*}[][ws] \lwc{uG}{}[][wr] \delta_{pt} \Bigr) \nn
		&\quad + 4 \Bigl( C_F - \frac{2}{N_c} \Bigr) e g \q_u \Bigl( \lwc{u \gamma}{*}[][wp] \lwc{uG}{}[][wr] + \lwc{uG}{*}[][wp] \lwc{u\gamma}{}[][wr] \Bigr) \delta_{st} \nn
		&\quad + 4 \Bigl( C_F - \frac{2}{N_c} \Bigr) e g \q_u \Bigl(  \lwc{u \gamma}{*}[][ws] \lwc{uG}{}[][wt] + \lwc{uG}{*}[][ws] \lwc{u\gamma}{}[][wt] \Bigr) \delta_{pr}  \nn
		&\quad + 8 eg \q_u \Bigl( \lwc{u\gamma}{*}[][wp] \lwc{uG}{}[][wt]+\lwc{uG}{*}[][wp] \lwc{u\gamma}{}[][wt] \Bigr) \delta_{sr} + 8 eg \q_u \Bigl( \lwc{u\gamma}{*}[][ws] \lwc{uG}{}[][wr]+ \lwc{uG}{*}[][ws] \lwc{u\gamma}{}[][wr] \Bigr) \delta_{pt}  \nn
		&\quad + \left[ \frac{52}{3} e^2 \q_u^2 + \frac{4}{3N_c} g^2 \right] \Bigl( \lwc{u \gamma}{*}[][wp] \lwc{u \gamma}{}[][wr] \delta_{st} +  \lwc{u \gamma}{*}[][ws] \lwc{u \gamma}{}[][wt] \delta_{pr} \Bigr) 
			 - \frac{4}{3} g^2 \Bigl( \lwc{u\gamma}{*}[][wp] \lwc{u\gamma}{}[][wt] \delta_{sr} + \lwc{u\gamma}{*}[][ws] \lwc{u\gamma}{}[][wr] \delta_{pt} \Bigr) \nn
		&\quad + e^2 \q_u^2    \zeta_e  \delta_{pr} \delta_{st}  - \frac{1}{2N_c} g^2   \zeta_g  \delta_{pr} \delta_{st} + \frac{1}{2} g^2   \zeta_g  \delta_{pt} \delta_{sr}  \, , \\ 
\nnn
	\dlwc{dd}{RR}[V][prst] &= \frac{2}{3} e^2 \q_d \delta_{pr} \begin{aligned}[t]
			& \bigg[ N_c \q_d \Bigl( {{\lwc{dd}{LR}[V1][wwst]}}+2\lwc{dd}{RR}[V][stww] \Bigr) + N_c \q_u \Bigl( {{\lwc{ud}{LR}[V1][wwst]}}+{{\lwc{ud}{RR}[V1][wwst]}} \Bigr) \nn
			& + \q_e \Bigl( \lwc{ed}{LR}[V][wwst]+\lwc{ed}{RR}[V][wwst] \Bigr) + 2 \q_d \lwc{dd}{RR}[V][swwt] \bigg] \end{aligned} \nn
		&\quad + \frac{2}{3} e^2 \q_d \delta_{st} \begin{aligned}[t]
			& \bigg[ N_c \q_d \Bigl( {{\lwc{dd}{LR}[V1][wwpr]}}+2\lwc{dd}{RR}[V][prww] \Bigr) + N_c \q_u \Bigl( {{\lwc{ud}{LR}[V1][wwpr]}}+{{\lwc{ud}{RR}[V1][wwpr]}} \Bigr) \nn
			& + \q_e \Bigl( \lwc{ed}{LR}[V][wwpr]+\lwc{ed}{RR}[V][wwpr] \Bigr) + 2 \q_d \lwc{dd}{RR}[V][pwwr] \bigg] \end{aligned} \nn
		&\quad + \frac{g^2}{6 N_c} \begin{aligned}[t]
			& \bigg[ N_c \delta_{pt} \Bigl( \lwc{ud}{LR}[V8][wwsr]+\lwc{dd}{LR}[V8][wwsr]+4 \lwc{dd}{RR}[V][swwr]+\lwc{ud}{RR}[V8][wwsr] \Bigr) \nn
			& + N_c \delta_{rs} \Bigl( \lwc{ud}{LR}[V8][wwpt]+\lwc{dd}{LR}[V8][wwpt]+4 \lwc{dd}{RR}[V][pwwt]+\lwc{ud}{RR}[V8][wwpt] \Bigr) \nn
			& - \delta_{pr} \Bigl( \lwc{ud}{LR}[V8][wwst] + \lwc{dd}{LR}[V8][wwst] + 4 \lwc{dd}{RR}[V][swwt] + \lwc{ud}{RR}[V8][wwst] \Bigr) \nn
			& - \delta_{st} \Bigl( \lwc{ud}{LR}[V8][wwpr] + \lwc{dd}{LR}[V8][wwpr] + 4 \lwc{dd}{RR}[V][pwwr] + \lwc{ud}{RR}[V8][wwpr] \Bigr) \bigg] \end{aligned} \nn
		&\quad + \left[ 12 e^2 \q_d^2 - \frac{6 g^2}{N_c} \right]  \lwc{dd}{RR}[V][prst] + 6 g^2 \lwc{dd}{RR}[V][ptsr]  \nn
		&\quad - \frac{8}{3} C_F e^2 \q_d^2 \Bigl( \lwc{dG}{*}[][wp] \lwc{dG}{}[][wr] \delta_{st} + \lwc{dG}{*}[][ws] \lwc{dG}{}[][wt] \delta_{pr} \Bigr) \nn
		&\quad - \frac{1}{2N_c} \Bigl( \frac{16}{3} C_F + 3 C_d - 5 C_A - 24 N_c C_1 \Bigr) g^2 \Bigl( \lwc{dG}{*}[][wp] \lwc{dG}{}[][wr] \delta_{st} + \lwc{dG}{*}[][ws] \lwc{dG}{}[][wt] \delta_{pr} \Bigr) \nn
		&\quad + \frac{1}{2} \Bigl( \frac{16}{3} C_F + 3 C_d - 5 C_A \Bigr) g^2 \Bigl( \lwc{dG}{*}[][wp] \lwc{dG}{}[][wt] \delta_{sr} + \lwc{dG}{*}[][ws] \lwc{dG}{}[][wr] \delta_{pt} \Bigr) \nn
		&\quad + 4 \Bigl( C_F - \frac{2}{N_c} \Bigr) e g \q_d \Bigl(  \lwc{d \gamma}{*}[][wp] \lwc{dG}{}[][wr] + \lwc{dG}{*}[][wp] \lwc{d\gamma}{}[][wr] \Bigr) \delta_{st} \nn
		&\quad + 4 \Bigl( C_F - \frac{2}{N_c} \Bigr) e g \q_d \Bigl( \lwc{d \gamma}{*}[][ws] \lwc{dG}{}[][wt] + \lwc{dG}{*}[][ws] \lwc{d\gamma}{}[][wt] \Bigr) \delta_{pr}  \nn
		&\quad + 8 eg \q_d \Bigl( \lwc{d\gamma}{*}[][wp] \lwc{dG}{}[][wt]+\lwc{dG}{*}[][wp] \lwc{d\gamma}{}[][wt] \Bigr) \delta_{sr} + 8 eg \q_d \Bigl( \lwc{d\gamma}{*}[][ws] \lwc{dG}{}[][wr]+ \lwc{dG}{*}[][ws] \lwc{d\gamma}{}[][wr] \Bigr) \delta_{pt}  \nn
		&\quad + \left[ \frac{52}{3} e^2 \q_d^2 + \frac{4}{3N_c} g^2 \right] \Bigl( \lwc{d \gamma}{*}[][wp] \lwc{d \gamma}{}[][wr] \delta_{st}  +  \lwc{d \gamma}{*}[][ws] \lwc{d \gamma}{}[][wt] \delta_{pr} \Bigr) 
			 - \frac{4}{3} g^2 \Bigl( \lwc{d\gamma}{*}[][wp] \lwc{d\gamma}{}[][wt] \delta_{sr} + \lwc{d\gamma}{*}[][ws] \lwc{d\gamma}{}[][wr] \delta_{pt} \Bigr) \nn
		&\quad + e^2 \q_d^2    \zeta_e  \delta_{pr} \delta_{st}  - \frac{1}{2N_c} g^2   \zeta_g  \delta_{pr} \delta_{st} + \frac{1}{2} g^2   \zeta_g  \delta_{pt} \delta_{sr}  \, , \\
\nnn
	\dlwc{ud}{RR}[V1][prst] &= \frac{4}{3} e^2 \q_u \delta_{pr} \begin{aligned}[t]
			& \bigg[ N_c \q_d \Bigl( {{\lwc{dd}{LR}[V1][wwst]}}+2\lwc{dd}{RR}[V][stww] \Bigr) + N_c \q_u \Bigl( {{\lwc{ud}{LR}[V1][wwst]}}+{{\lwc{ud}{RR}[V1][wwst]}} \Bigr) \nn
			& + \q_e \Bigl( \lwc{ed}{LR}[V][wwst]+\lwc{ed}{RR}[V][wwst] \Bigr) + 2 \q_d \lwc{dd}{RR}[V][swwt] \bigg] \end{aligned} \nn
		&\quad + \frac{4}{3} e^2 \q_d \delta_{st} \begin{aligned}[t]
			& \bigg[ N_c \q_d \Bigl({{\lwc{du}{LR}[V1][wwpr]}}+{{\lwc{ud}{RR}[V1][prww]}} \Bigr) + N_c \q_u \Bigl( {{\lwc{uu}{LR}[V1][wwpr]}}+2\lwc{uu}{RR}[V][prww] \Bigr) \nn
			& + \q_e \Bigl( \lwc{eu}{LR}[V][wwpr]+\lwc{eu}{RR}[V][wwpr] \Bigr) + 2 \q_u \lwc{uu}{RR}[V][pwwr] \bigg] \end{aligned} \nn
		&\quad + 12 e^2 \q_d \q_u \lwc{ud}{RR}[V1][prst] + 6 g^2 \frac{C_F}{N_c} \lwc{ud}{RR}[V8][prst]  \nn
		&\quad - \left[ \frac{16}{3} C_F e^2 \q_u \q_d - 24 g^2 C_1 \right] \Bigl( \lwc{uG}{*}[][wp] \lwc{uG}{}[][wr] \delta_{st} + \lwc{dG}{*}[][ws] \lwc{dG}{}[][wt] \delta_{pr} \Bigr) \nn
		&\quad + 8 C_F e g \q_d \Bigl( \lwc{uG}{*}[][wp] \lwc{u\gamma}{}[][wr] + \lwc{u \gamma}{*}[][wp] \lwc{uG}{}[][wr] \Bigr) \delta_{st} + 8 C_F e g \q_u \Bigl( \lwc{dG}{*}[][ws] \lwc{d\gamma}{}[][wt] + \lwc{d \gamma}{*}[][ws] \lwc{dG}{}[][wt] \Bigr) \delta_{pr}  \nn
		&\quad + e^2 \Bigl( 24 \q_d^2 + \frac{32}{3} \q_u \q_d \Bigr) \lwc{u \gamma}{*}[][wp] \lwc{u \gamma}{}[][wr] \delta_{st}  + e^2 \Bigl( 24 \q_u^2 + \frac{32}{3} \q_u \q_d \Bigr) \lwc{d \gamma}{*}[][ws] \lwc{d \gamma}{}[][wt] \delta_{pr} \nn
		&\quad + 2 e^2 \q_u \q_d   \zeta_e  \delta_{pr} \delta_{st} \, , \\
\nnn
	\dlwc{ud}{RR}[V8][prst] &= \frac{2}{3} g^2 \delta_{pr} \bigg[ \lwc{ud}{LR}[V8][wwst]+\lwc{dd}{LR}[V8][wwst]+4 \lwc{dd}{RR}[V][swwt]+\lwc{ud}{RR}[V8][wwst] \bigg] \nn
		&\quad + \frac{2}{3} g^2 \delta_{st} \bigg[ \lwc{du}{LR}[V8][wwpr]+\lwc{ud}{RR}[V8][prww]+\lwc{uu}{LR}[V8][wwpr]+4 \lwc{uu}{RR}[V][pwwr] \bigg] \nn
		&\quad + 12 g^2 \lwc{ud}{RR}[V1][prst] + \left[ 12 e^2 \q_d \q_u - \frac{12 g^2}{N_c} \right] \lwc{ud}{RR}[V8][prst]  \nn
		&\quad + g^2 \Bigl(\frac{32}{3}C_F + 6C_d -10 C_A \Bigr) \Bigl( \lwc{u G}{*}[][wp] \lwc{u G}{}[][wr] \delta_{st} +  \lwc{d G}{*}[][ws] \lwc{d G}{}[][wt] \delta_{pr} \Bigr) \nn
		&\quad + 8 e g ( \q_u + 3 \q_d ) \Bigl( \lwc{u\gamma}{*}[][wp] \lwc{uG}{}[][wr] + \lwc{uG}{*}[][wp] \lwc{u\gamma}{}[][wr] \Bigr) \delta_{st} + 8 e g ( \q_d + 3 \q_u ) \Bigl( \lwc{d\gamma}{*}[][ws] \lwc{dG}{}[][wt] + \lwc{dG}{*}[][ws] \lwc{d\gamma}{}[][wt] \Bigr) \delta_{pr}  \nn
		&\quad - \frac{16}{3} g^2 \Bigl( \lwc{u\gamma}{*}[][wp] \lwc{u\gamma}{}[][wr] \delta_{st} + \lwc{d\gamma}{*}[][ws] \lwc{d\gamma}{}[][wt] \delta_{pr} \Bigr)  + 2 g^2    \zeta_g  \delta_{pr} \delta_{st}  \, .
\end{align}

\subsubsection{$\psi^4: (\overline L L)(\overline R R)$}

\begin{align}
	\dlwc{\nu e}{LR}[V][prst] &= \frac{4}{3} e^2 \q_e \delta_{st} \begin{aligned}[t]
				& \bigg[ N_c \q_d \Bigl( \lwc{\nu d}{LL}[V][prww]+\lwc{\nu d}{LR}[V][prww] \Bigr) + N_c \q_u \Bigl( \lwc{\nu u}{LL}[V][prww]+\lwc{\nu u}{LR}[V][prww] \Bigr) \nn
				& + \q_e \Bigl( \lwc{\nu e}{LL}[V][prww]+\lwc{\nu e}{LR}[V][prww] \Bigr) \bigg] \end{aligned} \nn
		&\quad - 96 e^2 \q_e^2 \lwc{\nu \gamma}{}[][wr] \lwc{\nu \gamma}{*}[][wp] \delta_{st}  \, , \\
\nnn
	\dlwc{ee}{LR}[V][prst] &= \frac{4}{3} e^2 \q_e \delta_{pr} \begin{aligned}[t]
			& \bigg[ N_c \q_d \Bigl(\lwc{de}{LR}[V][wwst]+\lwc{ed}{RR}[V][stww] \Bigr) + N_c \q_u \Bigl( \lwc{ue}{LR}[V][wwst]+\lwc{eu}{RR}[V][stww] \Bigr) \nn
			& +\q_e \Bigl( \lwc{ee}{LR}[V][wwst]+4\lwc{ee}{RR}[V][stww] \Bigr) \bigg] \end{aligned} \nn
		&\quad + \frac{4}{3} e^2 \q_e \delta_{st} \begin{aligned}[t]
			& \bigg[ N_c \q_d \Bigl( \lwc{ed}{LL}[V][prww]+\lwc{ed}{LR}[V][prww] \Bigr) + N_c \q_u \Bigl( \lwc{eu}{LL}[V][prww]+\lwc{eu}{LR}[V][prww] \Bigr) \nn
			& + \q_e \Bigl( 4 \lwc{ee}{LL}[V][prww]+\lwc{ee}{LR}[V][prww] \Bigr) \bigg] \end{aligned} \nn
		&\quad - 12 e^2 \q_e^2 \lwc{ee}{LR}[V][prst]  - \frac{40}{3} e^2 \q_e^2 \Bigl(  \lwc{e \gamma}{}[][pw] \lwc{e \gamma}{*}[][rw] \delta_{st} + \lwc{e \gamma}{*}[][ws] \lwc{e \gamma}{}[][wt]  \delta_{pr}  \Bigr) \nn
		&\quad +  96 e^2 \q_e^2   \lwc{e \gamma}{}[][pt] \lwc{e \gamma}{*}[][rs]  
			 + 2 e^2 \q_e^2 \zeta_e   \delta_{pr} \delta_{st}  \, , \\
\nnn
	\dlwc{\nu u}{LR}[V][prst] &= \frac{4}{3} e^2 \q_u \delta_{st}  \begin{aligned}[t]
			& \bigg[ N_c \q_d \Bigl( \lwc{\nu d}{LL}[V][prww]+\lwc{\nu d}{LR}[V][prww] \Bigr) + N_c \q_u \Bigl( \lwc{\nu u}{LL}[V][prww]+\lwc{\nu u}{LR}[V][prww] \Bigr) \nn
			& + \q_e \Bigl( \lwc{\nu e}{LL}[V][prww]+\lwc{\nu e}{LR}[V][prww] \Bigr) \bigg] \end{aligned}  \nn
		&\quad - 96 e^2 \q_u^2 \lwc{\nu \gamma}{}[][wr] \lwc{\nu \gamma}{*}[][wp] \delta_{st}  \, , \\
\nnn
	\dlwc{\nu d}{LR}[V][prst] &= \frac{4}{3} e^2 \q_d \delta_{st} \begin{aligned}[t]
			& \bigg[ N_c \q_d \Bigl( \lwc{\nu d}{LL}[V][prww]+\lwc{\nu d}{LR}[V][prww] \Bigr) + N_c \q_u \Bigl( \lwc{\nu u}{LL}[V][prww]+\lwc{\nu u}{LR}[V][prww] \Bigr) \nn
			& + \q_e \Bigl( \lwc{\nu e}{LL}[V][prww]+\lwc{\nu e}{LR}[V][prww] \Bigr) \bigg]  \end{aligned} \nn
		&\quad - 96 e^2 \q_d^2 \lwc{\nu \gamma}{}[][wr] \lwc{\nu \gamma}{*}[][wp] \delta_{st}  \, , \\
\nnn
	\dlwc{eu}{LR}[V][prst] &= \frac{4}{3} e^2 \q_e \delta_{pr} \begin{aligned}[t]
			& \bigg[ N_c \q_d \Bigl(\lwc{du}{LR}[V1][wwst]+\lwc{ud}{RR}[V1][stww]\Bigr) + N_c \q_u \Bigl(\lwc{uu}{LR}[V1][wwst]+2\lwc{uu}{RR}[V][stww] \Bigr) \nn
			& + \q_e \Bigl(\lwc{eu}{LR}[V][wwst]+\lwc{eu}{RR}[V][wwst]\Bigr) + 2 \q_u \lwc{uu}{RR}[V][swwt] \bigg] \end{aligned} \nn
		&\quad + \frac{4}{3} e^2 \q_u \delta_{st} \begin{aligned}[t]
			& \bigg[ N_c \q_d \Bigl(\lwc{ed}{LL}[V][prww]+\lwc{ed}{LR}[V][prww]\Bigr) + N_c \q_u \Bigl(\lwc{eu}{LL}[V][prww]+\lwc{eu}{LR}[V][prww]\Bigr) \nn
			& + \q_e \Bigl( 4 \lwc{ee}{LL}[V][prww]+\lwc{ee}{LR}[V][prww]\Bigr) \bigg] \end{aligned} \nn
		&\quad - 12 e^2 \q_e \q_u \lwc{eu}{LR}[V][prst]   - \frac{16}{3} C_F e^2 \q_e \q_u \lwc{uG}{*}[][ws] \lwc{uG}{}[][wt] \delta_{pr}  + 8 C_F e g \q_e \Bigl( \lwc{uG}{*}[][ws] \lwc{u\gamma}{}[][wt] + \lwc{u\gamma}{*}[][ws] \lwc{uG}{}[][wt] \Bigr) \delta_{pr}  \nn
		&\quad + e^2 \Bigl( \frac{32}{3} \q_e \q_u - 24 \q_u^2 \Bigr)  \lwc{e \gamma}{}[][pw] \lwc{e \gamma}{*}[][rw] \delta_{st}  + e^2 \Bigl( \frac{32}{3} \q_e \q_u - 24 \q_e^2 \Bigr)  \lwc{u \gamma}{*}[][ws] \lwc{u \gamma}{}[][wt] \delta_{pr}  \nn
		&\quad + 2 e^2 \q_u \q_e \zeta_e   \delta_{pr} \delta_{st}  \, , \\
\nnn
	\dlwc{ed}{LR}[V][prst] &= \frac{4}{3} e^2 \q_e \delta_{pr} \begin{aligned}[t]
			& \bigg[ N_c \q_d \Bigl(\lwc{dd}{LR}[V1][wwst]+2\lwc{dd}{RR}[V][stww]\Bigr) + N_c \q_u \Bigl(\lwc{ud}{LR}[V1][wwst]+\lwc{ud}{RR}[V1][wwst] \Bigr) \nn
			& + \q_e \Bigl(\lwc{ed}{LR}[V][wwst]+\lwc{ed}{RR}[V][wwst]\Bigr) + 2 \q_d \lwc{dd}{RR}[V][swwt] \bigg] \end{aligned} \nn
		&\quad + \frac{4}{3} e^2 \q_d \delta_{st} \begin{aligned}[t]
			& \bigg[ N_c \q_d \Bigl(\lwc{ed}{LL}[V][prww]+\lwc{ed}{LR}[V][prww]\Bigr) + N_c \q_u \Bigl(\lwc{eu}{LL}[V][prww]+\lwc{eu}{LR}[V][prww] \Bigr) \nn
			& + \q_e \Bigl( 4 \lwc{ee}{LL}[V][prww]+\lwc{ee}{LR}[V][prww]\Bigr) \bigg] \end{aligned} \nn
		&\quad - 12 e^2 \q_d \q_e \lwc{ed}{LR}[V][prst]  - \frac{16}{3} C_F e^2 \q_e \q_d \lwc{dG}{*}[][ws] \lwc{dG}{}[][wt] \delta_{pr}  + 8 C_F e g \q_e \Bigl( \lwc{dG}{*}[][ws] \lwc{d\gamma}{}[][wt] + \lwc{d\gamma}{*}[][ws] \lwc{dG}{}[][wt] \Bigr) \delta_{pr}  \nn
		&\quad + e^2 \Bigl( \frac{32}{3} \q_e \q_d - 24 \q_d^2 \Bigr) \lwc{e \gamma}{}[][pw] \lwc{e \gamma}{*}[][rw] \delta_{st}  + e^2 \Bigl( \frac{32}{3} \q_e \q_d - 24 \q_e^2 \Bigr)  \lwc{d \gamma}{*}[][ws] \lwc{d \gamma}{}[][wt] \delta_{pr}  \nn
		&\quad + 2 e^2 \q_d \q_e \zeta_e   \delta_{pr} \delta_{st}  \, , \\
\nnn
	\dlwc{ue}{LR}[V][prst] &= \frac{4}{3} e^2 \q_u \delta_{pr} \begin{aligned}[t]
			& \bigg[ N_c \q_d \Bigl(\lwc{de}{LR}[V][wwst]+\lwc{ed}{RR}[V][stww]\Bigr) + N_c \q_u \Bigl(\lwc{ue}{LR}[V][wwst]+\lwc{eu}{RR}[V][stww] \Bigr) \nn
			& + \q_e \Bigl(\lwc{ee}{LR}[V][wwst]+4\lwc{ee}{RR}[V][stww]\Bigr) \bigg] \end{aligned} \nn
		&\quad + \frac{4}{3} e^2 \q_e \delta_{st} \begin{aligned}[t]
			& \bigg[ N_c \q_d \Bigl(\lwc{ud}{LL}[V1][prww]+\lwc{ud}{LR}[V1][prww]\Bigr) + N_c \q_u \Bigl(2\lwc{uu}{LL}[V][prww]+\lwc{uu}{LR}[V1][prww] \Bigr) \nn
			& + \q_e \Bigl(\lwc{eu}{LL}[V][wwpr]+\lwc{ue}{LR}[V][prww]\Bigr) + 2 \q_u \lwc{uu}{LL}[V][pwwr] \bigg] \end{aligned} \nn
		&\quad - 12 e^2 \q_e \q_u \lwc{ue}{LR}[V][prst]  - \frac{16}{3} C_F e^2 \q_e \q_u \lwc{uG}{}[][pw] \lwc{uG}{*}[][rw] \delta_{st}  + 8 C_F e g \q_e \Bigl( \lwc{uG}{}[][pw] \lwc{u\gamma}{*}[][rw] + \lwc{u\gamma}{}[][pw] \lwc{uG}{*}[][rw] \Bigr) \delta_{st}  \nn
		&\quad + e^2 \Bigl( \frac{32}{3} \q_e \q_u - 24 \q_e^2 \Bigr)  \lwc{u \gamma}{}[][pw] \lwc{u \gamma}{*}[][rw] \delta_{st} + e^2 \Bigl( \frac{32}{3} \q_e \q_u - 24 \q_u^2 \Bigr)  \lwc{e \gamma}{*}[][ws] \lwc{e \gamma}{}[][wt]  \delta_{pr}  \nn
		&\quad + 2 e^2 \q_u \q_e \zeta_e   \delta_{pr} \delta_{st}  \, , \\
\nnn
	\dlwc{de}{LR}[V][prst] &= \frac{4}{3} e^2 \q_d \delta_{pr} \begin{aligned}[t]
			& \bigg[ N_c \q_d \Bigl(\lwc{de}{LR}[V][wwst]+\lwc{ed}{RR}[V][stww]\Bigr) + N_c \q_u \Bigl(\lwc{ue}{LR}[V][wwst]+\lwc{eu}{RR}[V][stww] \Bigr) \nn
			& + \q_e \Bigl(\lwc{ee}{LR}[V][wwst]+4\lwc{ee}{RR}[V][stww]\Bigr) \bigg] \end{aligned} \nn
		&\quad + \frac{4}{3} e^2 \q_e \delta_{st} \begin{aligned}[t]
			& \bigg[ N_c \q_d \Bigl(2\lwc{dd}{LL}[V][prww]+\lwc{dd}{LR}[V1][prww]\Bigr) + N_c \q_u \Bigl(\lwc{ud}{LL}[V1][wwpr]+\lwc{du}{LR}[V1][prww]\Bigr) \nn
			& + \q_e \Bigl(\lwc{ed}{LL}[V][wwpr]+\lwc{de}{LR}[V][prww]\Bigr) + 2 \q_d \lwc{dd}{LL}[V][pwwr] \bigg] \end{aligned} \nn
		&\quad -12 e^2 \q_d \q_e \lwc{de}{LR}[V][prst]   - \frac{16}{3} C_F e^2 \q_e \q_d \lwc{dG}{}[][pw] \lwc{dG}{*}[][rw] \delta_{st}  + 8 C_F e g \q_e \Bigl( \lwc{dG}{}[][pw] \lwc{d\gamma}{*}[][rw] + \lwc{d\gamma}{}[][pw] \lwc{dG}{*}[][rw] \Bigr) \delta_{st}  \nn
		&\quad + e^2 \Bigl( \frac{32}{3} \q_e \q_d - 24 \q_e^2 \Bigr) \lwc{d \gamma}{}[][pw] \lwc{d \gamma}{*}[][rw] \delta_{st}  + e^2 \Bigl( \frac{32}{3} \q_e \q_d - 24 \q_d^2 \Bigr) \lwc{e \gamma}{*}[][ws] \lwc{e \gamma}{}[][wt]  \delta_{pr} \nn
		&\quad + 2 e^2 \q_d \q_e \zeta_e   \delta_{pr} \delta_{st}  \, , \\
\nnn
	\dlwc{\nu edu}{LR}[V][prst] &= -6 e^2 \q_e \q_d \lwc{\nu e d u}{LR}[V][prst]  \, , \\
\nnn
	\dlwc{uu}{LR}[V1][prst] &= \frac{4}{3} e^2 \q_u \delta_{pr} \begin{aligned}[t]
			& \bigg[ N_c \q_d \Bigl(\lwc{du}{LR}[V1][wwst]+\lwc{ud}{RR}[V1][stww]\Bigr) + N_c \q_u \Bigl(\lwc{uu}{LR}[V1][wwst]+2\lwc{uu}{RR}[V][stww] \Bigr) \nn
			& + \q_e \Bigl(\lwc{eu}{LR}[V][wwst]+\lwc{eu}{RR}[V][wwst]\Bigr)+ 2 \q_u \lwc{uu}{RR}[V][swwt] \bigg] \end{aligned} \nn
		&\quad + \frac{4}{3} e^2 \q_u \delta_{st} \begin{aligned}[t]
			& \bigg[ N_c \q_d \Bigl(\lwc{ud}{LL}[V1][prww]+\lwc{ud}{LR}[V1][prww]\Bigr) + N_c \q_u \Bigl(2 \lwc{uu}{LL}[V][prww]+\lwc{uu}{LR}[V1][prww] \Bigr) \nn
			& + \q_e \Bigl(\lwc{eu}{LL}[V][wwpr]+\lwc{ue}{LR}[V][prww]\Bigr)+ 2 \q_u \lwc{uu}{LL}[V][pwwr] \bigg] \end{aligned} \nn
		&\quad - 12 e^2 \q_u^2 \lwc{uu}{LR}[V1][prst] - 6 g^2 \frac{C_F}{N_c} \lwc{uu}{LR}[V8][prst]  \nn
		&\quad + \left[ 96 g^2 C_1 \frac{1}{N_c}+24 g^2 C_d \frac{N_c^2-1}{2N_c^2}  + 48  e^2 \q_u^2  \frac{N_c^2-1}{2N_c^2}   \right]   \lwc{uG}{}[][pt] \lwc{u G}{*}[][rs]  \nn
		&\quad - \left[ \frac{16}{3} C_F e^2 \q_u^2 + 24 g^2 C_1 \right] \Bigl( \lwc{uG}{}[][pw] \lwc{uG}{*}[][rw] \delta_{st} + \lwc{uG}{*}[][ws] \lwc{uG}{}[][wt] \delta_{pr} \Bigr)  \nn
		&\quad +  48  eg \q_u \frac{N_c^2-1}{2N_c^2}  \Bigl( \lwc{u \gamma}{}[][pt] \lwc{u G}{*}[][rs] +  \lwc{u G}{}[][pt] \lwc{u \gamma}{*}[][rs] \Bigr) \nn
		&\quad + 8 C_F e g \q_u \Bigl( \lwc{uG}{}[][pw] \lwc{u\gamma}{*}[][rw] + \lwc{u\gamma}{}[][pw] \lwc{uG}{*}[][rw] \Bigr) \delta_{st} + 8 C_F e g \q_u \Bigl( \lwc{uG}{*}[][ws] \lwc{u\gamma}{}[][wt] + \lwc{u\gamma}{*}[][ws] \lwc{uG}{}[][wt] \Bigr) \delta_{pr}  \nn
		&\quad + \left[ 96 e^2 \q_u^2 \frac{1}{N_c} +  48  g^2  \frac{N_c^2-1}{2N_c^2}  \right]  \lwc{u \gamma}{}[][pt] \lwc{u \gamma}{*}[][rs] 
			 - \frac{40}{3} e^2 \q_u^2 \Bigl(  \lwc{u \gamma}{}[][pw] \lwc{u \gamma}{*}[][rw] \delta_{st} +  \lwc{u \gamma}{*}[][ws] \lwc{u \gamma}{}[][wt] \delta_{pr}  \Bigr) \nns
		&\quad + 2 e^2 \q_u^2 \zeta_e   \delta_{pr} \delta_{st}  \, , \\
\nnn
	\dlwc{uu}{LR}[V8][prst] &= \frac{2}{3} g^2 \delta_{pr} \bigg[ \lwc{du}{LR}[V8][wwst]+\lwc{ud}{RR}[V8][stww]+\lwc{uu}{LR}[V8][wwst]+4 \lwc{uu}{RR}[V][swwt] \bigg] \nn
		&\quad + \frac{2}{3} g^2 \delta_{st} \bigg[ \lwc{ud}{LL}[V8][prww]+\lwc{ud}{LR}[V8][prww]+4 \lwc{uu}{LL}[V][pwwr]+\lwc{uu}{LR}[V8][prww] \bigg] \nn
		&\quad - 12 g^2 \lwc{uu}{LR}[V1][prst] - \left[ 12 e^2 \q_u^2 + 6 g^2 N_c - \frac{12 g^2}{N_c} \right] \lwc{uu}{LR}[V8][prst]  \nn
		&\quad + \left[192 g^2 C_1 -24 g^2 C_d  \frac{1}{N_c}  - 48 e^2\q_u^2  \frac{1}{N_c} \right]  \lwc{uG}{}[][pt] \lwc{u G}{*}[][rs] \nn
		&\quad + \Bigl(\frac{32}{3}C_F-6C_d-10 C_A \Bigr)  g^2 \Bigl(   \lwc{u G}{}[][pw] \lwc{u G}{*}[][rw] \delta_{st} + \lwc{u G}{*}[][ws] \lwc{u G}{}[][wt] \delta_{pr} \Bigr) \nn
		&\quad - 48 e g \q_u  \frac{1}{N_c} \Bigl(  \lwc{u \gamma}{}[][pt] \lwc{u G}{*}[][rs] +  \lwc{u G}{}[][pt] \lwc{u \gamma}{*}[][rs] \Bigr) \nn
		&\quad - 16 e g \q_u \Bigl( \lwc{u\gamma}{}[][pw] \lwc{uG}{*}[][rw] + \lwc{uG}{}[][pw] \lwc{u\gamma}{*}[][rw] \Bigr) \delta_{st} - 16 e g \q_u \Bigl( \lwc{u\gamma}{*}[][ws] \lwc{uG}{}[][wt] + \lwc{uG}{*}[][ws] \lwc{u\gamma}{}[][wt] \Bigr) \delta_{pr}  \nn
		&\quad + \left[ 192 e^2 \q_u^2  - 48 g^2  \frac{1}{N_c} \right] \lwc{u \gamma}{}[][pt] \lwc{u \gamma}{*}[][rs]  - \frac{16}{3} g^2  \Bigl( \lwc{u\gamma}{}[][pw] \lwc{u\gamma}{*}[][rw] \delta_{st} + \lwc{u\gamma}{*}[][ws] \lwc{u\gamma}{}[][wt] \delta_{pr} \Bigr) \nn
		&\quad + 2 g^2  \zeta_g   \delta_{pr} \delta_{st}  \, , \\
\nnn
	\dlwc{ud}{LR}[V1][prst] &= \frac{4}{3} e^2 \q_u \delta_{pr} \begin{aligned}[t]
			& \bigg[ N_c \q_d \Bigl(\lwc{dd}{LR}[V1][wwst]+2\lwc{dd}{RR}[V][stww]\Bigr) + N_c \q_u \Bigl(\lwc{ud}{LR}[V1][wwst]+\lwc{ud}{RR}[V1][wwst] \Bigr) \nn
			& + \q_e \Bigl(\lwc{ed}{LR}[V][wwst]+\lwc{ed}{RR}[V][wwst]\Bigr) + 2 \q_d \lwc{dd}{RR}[V][swwt] \bigg] \end{aligned} \nn
		&\quad + \frac{4}{3} e^2 \q_d \delta_{st} \begin{aligned}[t]
			& \bigg[ N_c \q_d \Bigl(\lwc{ud}{LL}[V1][prww]+\lwc{ud}{LR}[V1][prww]\Bigr) + N_c \q_u \Bigl(2 \lwc{uu}{LL}[V][prww]+\lwc{uu}{LR}[V1][prww] \Bigr) \nn
			& + \q_e \Bigl(\lwc{eu}{LL}[V][wwpr]+\lwc{ue}{LR}[V][prww]\Bigr) + 2 \q_u \lwc{uu}{LL}[V][pwwr] \bigg] \end{aligned} \nn
		&\quad - 12 e^2 \q_d \q_u \lwc{ud}{LR}[V1][prst] - 6 g^2 \frac{C_F}{N_c} \lwc{ud}{LR}[V8][prst] \nn
		&\quad - \left[ \frac{16}{3} C_F e^2 \q_u \q_d + 24 g^2 C_1\right] \Bigl( \lwc{uG}{}[][pw] \lwc{uG}{*}[][rw] \delta_{st} + \lwc{dG}{*}[][ws] \lwc{dG}{}[][wt] \delta_{pr} \Bigr)  \nn
		&\quad + 8 C_F e g \q_d \Bigl( \lwc{uG}{}[][pw] \lwc{u\gamma}{*}[][rw] + \lwc{u\gamma}{}[][pw] \lwc{uG}{*}[][rw] \Bigr) \delta_{st} + 8 C_F e g \q_u \Bigl( \lwc{dG}{*}[][ws] \lwc{d\gamma}{}[][wt] + \lwc{d\gamma}{*}[][ws] \lwc{dG}{}[][wt] \Bigr) \delta_{pr}  \nn
		&\quad + e^2 \Bigl( \frac{32}{3} \q_u \q_d - 24 \q_d^2 \Bigr) \lwc{u \gamma}{}[][pw] \lwc{u \gamma}{*}[][rw] \delta_{st}  + e^2 \Bigl( \frac{32}{3} \q_u \q_d -24 \q_u^2 \Bigr) \lwc{d \gamma}{*}[][ws] \lwc{d \gamma}{}[][wt] \delta_{pr}  \nn
		&\quad + 2 e^2 \q_u \q_d \zeta_e   \delta_{pr} \delta_{st}  \, , \\
\nnn
	\dlwc{ud}{LR}[V8][prst] &= \frac{2}{3} g^2 \delta_{pr} \bigg[ \lwc{ud}{LR}[V8][wwst]+\lwc{dd}{LR}[V8][wwst]+4 \lwc{dd}{RR}[V][swwt]+\lwc{ud}{RR}[V8][wwst] \bigg] \nn
		&\quad + \frac{2}{3} g^2 \delta_{st} \bigg[ \lwc{ud}{LL}[V8][prww]+\lwc{ud}{LR}[V8][prww]+4 \lwc{uu}{LL}[V][pwwr]+\lwc{uu}{LR}[V8][prww] \bigg] \nn
		&\quad - 12 g^2 \lwc{ud}{LR}[V1][prst] - \left[ 12 e^2 \q_d \q_u + 6 g^2 N_c - \frac{12 g^2}{N_c} \right] \lwc{ud}{LR}[V8][prst]  \nn
		&\quad + \Bigl(\frac{32}{3}C_F-6C_d-10 C_A \Bigr) g^2 \Bigl(  \lwc{u G}{}[][pw] \lwc{u G}{*}[][rw] \delta_{st} +   \lwc{d G}{*}[][ws] \lwc{d G}{}[][wt] \delta_{pr} \Bigr) \nn
		&\quad - { 8 eg (3\q_d-\q_u)} \Bigl( \lwc{u\gamma}{}[][pw] \lwc{uG}{*}[][rw]+\lwc{uG}{}[][pw] \lwc{u\gamma}{*}[][rw] \Bigr) \delta_{st} -  { 8 eg (3\q_u-\q_d)} \Bigl( \lwc{d\gamma}{*}[][ws] \lwc{dG}{}[][wt]+\lwc{dG}{*}[][ws] \lwc{d\gamma}{}[][wt] \Bigr) \delta_{pr}  \nn
		&\quad - \frac{16}{3} g^2 \Bigl( \lwc{u\gamma}{}[][pw] \lwc{u\gamma}{*}[][rw] \delta_{st}  + \lwc{d\gamma}{*}[][ws] \lwc{d\gamma}{}[][wt] \delta_{pr} \Bigr)  + 2 g^2  \zeta_g   \delta_{pr} \delta_{st}  \, , \\
\nnn
	\dlwc{du}{LR}[V1][prst] &= \frac{4}{3} e^2 \q_d \delta_{pr} \begin{aligned}[t]
			& \bigg[ N_c \q_d \Bigl(\lwc{du}{LR}[V1][wwst]+\lwc{ud}{RR}[V1][stww]\Bigr) + N_c \q_u \Bigl(\lwc{uu}{LR}[V1][wwst]+2\lwc{uu}{RR}[V][stww] \Bigr) \nn
			& + \q_e \Bigl(\lwc{eu}{LR}[V][wwst]+\lwc{eu}{RR}[V][wwst]\Bigr) + 2 \q_u \lwc{uu}{RR}[V][swwt] \bigg] \end{aligned} \nn
		&\quad + \frac{4}{3} e^2 \q_u \delta_{st} \begin{aligned}[t]
			& \bigg[ N_c \q_d \Bigl( 2 \lwc{dd}{LL}[V][prww]+\lwc{dd}{LR}[V1][prww]\Bigr) + N_c \q_u \Bigl(\lwc{ud}{LL}[V1][wwpr]+\lwc{du}{LR}[V1][prww] \Bigr) \nn
			& + \q_e \Bigl(\lwc{ed}{LL}[V][wwpr]+\lwc{de}{LR}[V][prww]\Bigr) + 2 \q_d \lwc{dd}{LL}[V][pwwr] \bigg] \end{aligned} \nn
		&\quad - 12 e^2 \q_d \q_u \lwc{du}{LR}[V1][prst] - 6 g^2 \frac{C_F}{N_c} \lwc{du}{LR}[V8][prst]  \nn
		&\quad - \left[ \frac{16}{3} C_F e^2 \q_u \q_d + 24 g^2 C_1 \right] \Bigl( \lwc{dG}{}[][pw] \lwc{dG}{*}[][rw] \delta_{st} + \lwc{uG}{*}[][ws] \lwc{uG}{}[][wt] \delta_{pr} \Bigr)  \nn
		&\quad + 8 C_F e g \q_u \Bigl( \lwc{dG}{}[][pw] \lwc{d\gamma}{*}[][rw] + \lwc{d\gamma}{}[][pw] \lwc{dG}{*}[][rw] \Bigr) \delta_{st} + 8 C_F e g \q_d \Bigl( \lwc{uG}{*}[][ws] \lwc{u\gamma}{}[][wt] + \lwc{u\gamma}{*}[][ws] \lwc{uG}{}[][wt] \Bigr) \delta_{pr}  \nn
		&\quad + e^2 \Bigl( \frac{32}{3} \q_u \q_d - 24 \q_u^2 \Bigr) \lwc{d \gamma}{}[][pw] \lwc{d \gamma}{*}[][rw] \delta_{st}  + e^2 \Bigl( \frac{32}{3} \q_u \q_d - 24 \q_d^2 \Bigr) \lwc{u \gamma}{*}[][ws] \lwc{u \gamma}{}[][wt] \delta_{pr}  \nn
		&\quad + 2 e^2 \q_u \q_d \zeta_e   \delta_{pr} \delta_{st}  \, , \\
\nnn
	\dlwc{du}{LR}[V8][prst] &= \frac{2}{3} g^2 \delta_{pr} \bigg[ \lwc{du}{LR}[V8][wwst]+\lwc{ud}{RR}[V8][stww]+\lwc{uu}{LR}[V8][wwst]+4 \lwc{uu}{RR}[V][swwt] \bigg] \nn
		&\quad + \frac{2}{3} g^2 \delta_{st} \bigg[ 4 \lwc{dd}{LL}[V][pwwr]+\lwc{ud}{LL}[V8][wwpr]+\lwc{du}{LR}[V8][prww]+\lwc{dd}{LR}[V8][prww] \bigg] \nn
		&\quad - 12 g^2 \lwc{du}{LR}[V1][prst] - \left[ 12 e^2 \q_d \q_u + 6 g^2 N_c - \frac{12 g^2}{N_c} \right] \lwc{du}{LR}[V8][prst]  \nn
		&\quad + \Bigl(\frac{32}{3}C_F-6C_d-10 C_A \Bigr)g^2  \Bigl(  \lwc{d G}{}[][pw] \lwc{d G}{*}[][rw] \delta_{st} +  \lwc{u G}{*}[][ws] \lwc{u G}{}[][wt] \delta_{pr} \Bigr) \nn
		&\quad - { 8 eg (3\q_u-\q_d)} \Bigl( \lwc{d\gamma}{}[][pw] \lwc{dG}{*}[][rw]+\lwc{dG}{}[][pw] \lwc{d\gamma}{*}[][rw] \Bigr) \delta_{st} - { 8 eg (3\q_d-\q_u)} \Bigl( \lwc{u\gamma}{*}[][ws] \lwc{uG}{}[][wt]+\lwc{uG}{*}[][ws] \lwc{u\gamma}{}[][wt] \Bigr) \delta_{pr}  \nn
		&\quad - \frac{16}{3} g^2 \Bigl( \lwc{d\gamma}{}[][pw] \lwc{d\gamma}{*}[][rw] \delta_{st} + \lwc{u\gamma}{*}[][ws] \lwc{u\gamma}{}[][wt] \delta_{pr} \Bigr)  + 2 g^2  \zeta_g   \delta_{pr} \delta_{st}  \, , \\
\nnn
	\dlwc{dd}{LR}[V1][prst] &= \frac{4}{3} e^2 \q_d \delta_{pr} \begin{aligned}[t]
			& \bigg[ N_c \q_d \Bigl(\lwc{dd}{LR}[V1][wwst]+2\lwc{dd}{RR}[V][stww]\Bigr) + N_c \q_u \Bigl(\lwc{ud}{LR}[V1][wwst]+\lwc{ud}{RR}[V1][wwst] \Bigr) \nn
			& + \q_e \Bigl(\lwc{ed}{LR}[V][wwst]+\lwc{ed}{RR}[V][wwst]\Bigr) + 2 \q_d \lwc{dd}{RR}[V][swwt] \bigg] \end{aligned} \nn
		&\quad + \frac{4}{3} e^2 \q_d \delta_{st} \begin{aligned}[t]
			& \bigg[ N_c \q_d \Bigl(2\lwc{dd}{LL}[V][prww]+\lwc{dd}{LR}[V1][prww]\Bigr) + N_c \q_u \Bigl(\lwc{ud}{LL}[V1][wwpr]+\lwc{du}{LR}[V1][prww] \Bigr) \nn
			& + \q_e \Bigl(\lwc{ed}{LL}[V][wwpr]+\lwc{de}{LR}[V][prww]\Bigr) + 2 \q_d \lwc{dd}{LL}[V][pwwr] \bigg] \end{aligned} \nn
		&\quad - 12 e^2 \q_d^2 \lwc{dd}{LR}[V1][prst] - 6 g^2 \frac{C_F}{N_c} \lwc{dd}{LR}[V8][prst] \nn
		&\quad + \left[ 96 g^2 C_1 \frac{1}{N_c}+24 g^2 C_d \frac{N_c^2-1}{2N_c^2}  + 48  e^2 \q_d^2  \frac{N_c^2-1}{2N_c^2} \right]   \lwc{dG}{}[][pt] \lwc{d G}{*}[][rs] \nn
		&\quad - \left[ \frac{16}{3} C_F e^2 \q_d^2 + 24 g^2 C_1 \right] \Bigl( \lwc{dG}{}[][pw] \lwc{dG}{*}[][rw] \delta_{st} + \lwc{dG}{*}[][ws] \lwc{dG}{}[][wt] \delta_{pr} \Bigr)  \nn
		&\quad + 48  eg \q_d  \frac{N_c^2-1}{2N_c^2} \Bigl(  \lwc{d \gamma}{}[][pt] \lwc{d G}{*}[][rs] +   \lwc{d G}{}[][pt] \lwc{d \gamma}{*}[][rs] \Bigr) \nn
		&\quad + 8 C_F e g \q_d \Bigl( \lwc{dG}{}[][pw] \lwc{d\gamma}{*}[][rw] + \lwc{d\gamma}{}[][pw] \lwc{dG}{*}[][rw] \Bigr) \delta_{st} + 8 C_F e g \q_d \Bigl( \lwc{dG}{*}[][ws] \lwc{d\gamma}{}[][wt] + \lwc{d\gamma}{*}[][ws] \lwc{dG}{}[][wt] \Bigr) \delta_{pr}  \nn
		&\quad + \left[ 96 e^2 \q_d^2 \frac{1}{N_c} + 48  g^2  \frac{N_c^2-1}{2N_c^2}  \right]  \lwc{d \gamma}{}[][pt] \lwc{d \gamma}{*}[][rs] 
			 - \frac{40}{3} e^2 \q_d^2 \Bigl(  \lwc{d \gamma}{}[][pw] \lwc{d \gamma}{*}[][rw] \delta_{st} +   \lwc{d \gamma}{*}[][ws] \lwc{d \gamma}{}[][wt] \delta_{pr}  \Bigr)  \nns
		&\quad + 2 e^2 \q_d^2  \zeta_e   \delta_{pr} \delta_{st}  \, , \\
\nnn
	\dlwc{dd}{LR}[V8][prst] &= \frac{2}{3} g^2 \delta_{pr} \bigg[ \lwc{ud}{LR}[V8][wwst]+\lwc{dd}{LR}[V8][wwst]+4 \lwc{dd}{RR}[V][swwt]+\lwc{ud}{RR}[V8][wwst] \bigg] \nn
		&\quad + \frac{2}{3} g^2 \delta_{st} \bigg[ 4 \lwc{dd}{LL}[V][pwwr]+\lwc{ud}{LL}[V8][wwpr]+\lwc{du}{LR}[V8][prww]+\lwc{dd}{LR}[V8][prww] \bigg] \nn
		&\quad - 12 g^2 \lwc{dd}{LR}[V1][prst] - \left[ 12 e^2 \q_d^2 + 6 g^2 N_c - \frac{12 g^2}{N_c} \right] \lwc{dd}{LR}[V8][prst]  \nn
		&\quad + \left[192 g^2 C_1 -24 g^2 C_d  \frac{1}{N_c}  - 48 e^2\q_d^2  \frac{1}{N_c} \right]  \lwc{dG}{}[][pt] \lwc{d G}{*}[][rs]  \nn
		&\quad + \Bigl(\frac{32}{3}C_F-6C_d-10 C_A \Bigr) g^2  \Bigl( \lwc{d G}{}[][pw] \lwc{d G}{*}[][rw] \delta_{st} +   \lwc{d G}{*}[][ws] \lwc{d G}{}[][wt] \delta_{pr} \Bigr) \nn
		&\quad - 48 e g \q_d  \frac{1}{N_c} \Bigl( \lwc{d \gamma}{}[][pt] \lwc{d G}{*}[][rs] + \lwc{d G}{}[][pt] \lwc{d \gamma}{*}[][rs] \Bigr) \nn
		&\quad - 16 eg  \q_d (\lwc{d\gamma}{}[][pw] \lwc{dG}{*}[][rw]+\lwc{dG}{}[][pw] \lwc{d\gamma}{*}[][rw]) \delta_{st} - 16 eg \q_d ( \lwc{d\gamma}{*}[][ws] \lwc{dG}{}[][wt]+\lwc{dG}{*}[][ws] \lwc{d\gamma}{}[][wt] )\delta_{pr}  \nn
		&\quad + \left[ 192 e^2 \q_d^2  - 48 g^2  \frac{1}{N_c} \right] \lwc{d \gamma}{}[][pt] \lwc{d \gamma}{*}[][rs]  - \frac{16}{3} g^2 \Bigl( \lwc{d\gamma}{}[][pw] \lwc{d\gamma}{*}[][rw] \delta_{st} + \lwc{d\gamma}{*}[][ws] \lwc{d\gamma}{}[][wt] \delta_{pr} \Bigr)  \nn
		&\quad + 2 g^2   \zeta_g   \delta_{pr} \delta_{st}  \, , \\
\nnn
	\dlwc{uddu}{LR}[V1][prst] &= -6 e^2 ( \q_d^2 + \q_u^2 ) \lwc{uddu}{LR}[V1][prst] - 6 g^2 \frac{C_F}{N_c} \lwc{uddu}{LR}[V8][prst]  \nn
		&\quad + \left[ 96 g^2 C_1 \frac{1}{N_c}+24 g^2 C_d \frac{N_c^2-1}{2N_c^2} +  48  e^2 \q_u \q_d  \frac{N_c^2-1}{2N_c^2}  \right]   \lwc{uG}{}[][pt] \lwc{d G}{*}[][rs]   \nn
		&\quad + 48  eg \q_d  \frac{N_c^2-1}{2N_c^2}   \lwc{u \gamma}{}[][pt] \lwc{d G}{*}[][rs] +  48  eg \q_u  \frac{N_c^2-1}{2N_c^2}   \lwc{u G}{}[][pt] \lwc{d \gamma}{*}[][rs]  \nn
		&\quad + \left[ 96 e^2 \q_u \q_d \frac{1}{N_c} + 48  g^2  \frac{N_c^2-1}{2N_c^2} \right]  \lwc{u \gamma}{}[][pt] \lwc{d \gamma}{*}[][rs]  \, , \\
\nnn
	\dlwc{uddu}{LR}[V8][prst] &= -12 g^2 \lwc{uddu}{LR}[V1][prst] - \left[ 6 e^2 (\q_u^2 + \q_d^2) + 6 g^2 N_c - \frac{12 g^2}{N_c} \right] \lwc{uddu}{LR}[V8][prst]  \nn
		&\quad + \left[192 g^2 C_1 -24 g^2 C_d  \frac{1}{N_c}  - 48 e^2\q_u \q_d  \frac{1}{N_c}  \right]  \lwc{uG}{}[][pt] \lwc{d G}{*}[][rs]  - 48 e g \q_d  \frac{1}{N_c}   \lwc{u \gamma}{}[][pt] \lwc{d G}{*}[][rs]  \nn
		&\quad - 48 e g \q_u  \frac{1}{N_c}   \lwc{u G}{}[][pt] \lwc{d \gamma}{*}[][rs]   + \left[ 192 e^2 \q_u \q_d    - 48 g^2  \frac{1}{N_c}  \right] \lwc{u \gamma}{}[][pt] \lwc{d \gamma}{*}[][rs]   \, .
\end{align}
%

\subsubsection{$\psi^4: (\overline L R)(\overline L R)$}

\begin{align}
	\dlwc{ee}{RR}[S][prst] &= 16 e^2 \q_e^2 \lwc{ee}{RR}[S][ptsr]-4 e^2 \q_e^2 \lwc{ee}{RR}[S][prst]  -96 e^2 \q_e^2 \lwc{e \gamma}{}[][pr] \lwc{e \gamma}{}[][st]  \, , \\
\nnn
	\dlwc{eu}{RR}[S][prst] &= - \left[ 6 e^2 (\q_e^2 + \q_u^2) + 6 g^2 C_F \right] \lwc{eu}{RR}[S][prst] -96 e^2 \q_e \q_u \lwc{eu}{RR}[T][prst]  -192 e^2 \q_e \q_u \lwc{e \gamma}{}[][pr] \lwc{u \gamma}{}[][st]   \, , \\
\nnn
	\dlwc{eu}{RR}[T][prst] &= -2 e^2 \q_e \q_u \lwc{eu}{RR}[S][prst] + \left[ 2 e^2 (\q_e^2+ \q_u^2)+ 2 g^2 C_F \right] \lwc{eu}{RR}[T][prst]   \, , \\
\nnn
	\dlwc{ed}{RR}[S][prst] &= - \left[ 6 e^2 (\q_e^2 + \q_d^2) + 6 g^2 C_F \right] \lwc{ed}{RR}[S][prst] -96 e^2 \q_e \q_d \lwc{ed}{RR}[T][prst]  -192 e^2 \q_e \q_d \lwc{e \gamma}{}[][pr] \lwc{d \gamma}{}[][st]   \, , \\
\nnn
	\dlwc{ed}{RR}[T][prst] &= -2 e^2 \q_d \q_e \lwc{ed}{RR}[S][prst] + \left[ 2 e^2 (\q_e^2 + \q_d^2) + 2 g^2 C_F \right] \lwc{ed}{RR}[T][prst]   \, , \\
\nnn
	\dlwc{\nu edu}{RR}[S][prst]  &=  - \left( 6 e^2 \q_u \q_d + 6 g^2 C_F \right) \lwc{\nu edu}{RR}[S][prst] + 24 e^2 ( \q_u^2 - \q_d^2) \lwc{\nu edu}{RR}[T][prst]   \, , \\
\nnn
	\dlwc{\nu edu}{RR}[T][prst]  &=  \frac{1}{2} e^2 ( \q_u^2 - \q_d^2 ) \lwc{\nu edu}{RR}[S][prst] - \left[ 2 e^2 ( 2\q_e^2  - \q_e \q_u - \q_u^2) - 2 g^2 C_F \right] \lwc{\nu edu}{RR}[T][prst]   \, , \\
\nnn
	\dlwc{uu}{RR}[S1][prst] &= - \left( 4 e^2 \q_u^2 + 12 g^2 C_F \right) \lwc{uu}{RR}[S1][prst] + \frac{1}{N_c} \left( 16 e^2 \q_u^2 + 16 g^2 C_F \right) \lwc{uu}{RR}[S1][ptsr] \nn
		&\quad + 4 g^2 \frac{C_F}{N_c} \lwc{uu}{RR}[S8][prst] + \frac{1}{N_c} \left[ 16 e^2 \q_u^2 C_F + 2 g^2 \left(\frac{2}{N_c^2} + N_c^2 - 3\right) \right] \lwc{uu}{RR}[S8][ptsr] \nn
		&\quad - 96 g^2 C_1 \lwc{u G}{}[][pr] \lwc{u G}{}[][st]  + 4 g^2 C_A \frac{N_c^2-1}{N_c^2}  \lwc{u G}{}[][pt] \lwc{u G}{}[][sr]  - 96 e^2 \q_u^2 \lwc{u \gamma}{}[][pr] \lwc{u \gamma}{}[][st]   \, , \\
\nnn
	\dlwc{uu}{RR}[S8][prst] &= 8 g^2 \lwc{uu}{RR}[S1][prst] + \left[ 32 e^2 \q_u^2 - \frac{16 g^2}{N_c} \right] \lwc{uu}{RR}[S1][ptsr]  - \left(  4 e^2 \q_u^2 - 4 g^2 C_F \right)\lwc{uu}{RR}[S8][prst] \nn
		&\quad - \frac{1}{N_c} \left[ 16 e^2 \q_u^2 - 4 g^2 \left( \frac{2}{N_c} + N_c \right) \right] \lwc{uu}{RR}[S8][ptsr]  \nn
		&\quad + \Bigl[ g^2(-24 C_d + 4 C_A)  - 48 e^2 \q_u^2 \Bigr] \lwc{u G}{}[][pr] \lwc{u G}{}[][st]  - 8 g^2  \frac{C_A}{N_c} \lwc{u G}{}[][p t] \lwc{u G}{}[][sr] \nn
		&\quad - 48 e g \q_u \Bigl( \lwc{u G}{}[][pr] \lwc{u \gamma}{}[][st]  +  \lwc{u \gamma}{}[][pr] \lwc{u G}{}[][st] \Bigr)  - 48 g^2  \lwc{u \gamma}{}[][pr] \lwc{u \gamma}{}[][st]  \, , \\
\nnn
	\dlwc{ud}{RR}[S1][prst] &= - \Bigl[ 2e^2 ( 3 \q_u^2 - 4 \q_u \q_d + 3 \q_d^2) + 12 g^2 C_F \Bigr] \lwc{ud}{RR}[S1][prst] \nn
		&\quad + \frac{1}{N_c} \Bigl[ 4 e^2 (\q_u + \q_d)^2 + 16 g^2 C_F \Bigr] \lwc{uddu}{RR}[S1][ptsr] \nn
		&\quad + 4 g^2 \frac{C_F}{N_c} \lwc{ud}{RR}[S8][prst] + \frac{1}{N_c} \left[ 4 e^2 (\q_u + \q_d)^2 C_F + 2 g^2 \left( \frac{2}{N_c^2} + N_c^2 - 3 \right) \right] \lwc{uddu}{RR}[S8][ptsr]  \nn
		&\quad -192 g^2 C_1 \lwc{u G}{}[][pr] \lwc{d G}{}[][st]  -192 e^2 \q_u \q_d \lwc{u \gamma}{}[][pr] \lwc{d \gamma}{}[][st]  \, , \\
\nnn
	\dlwc{ud}{RR}[S8][prst] &= 8 g^2 \lwc{ud}{RR}[S1][prst] + \left[ 8 e^2 (\q_u + \q_d)^2 - \frac{16 g^2}{N_c} \right] \lwc{uddu}{RR}[S1][ptsr] \nn
		&\quad - \Bigl[ 2 e^2 ( 3 \q_u^2 - 4 \q_u \q_d + 3 \q_d^2) - 4 g^2 C_F \Bigr] \lwc{ud}{RR}[S8][prst] \nn
		&\quad - \frac{1}{N_c} \left[ 4 e^2 (\q_u + \q_d)^2 - 4 g^2 \left( \frac{2}{N_c} + N_c \right) \right] \lwc{uddu}{RR}[S8][ptsr] \nn
		&\quad + \Bigl[ g^2(-48 C_d + 8 C_A) - 96 e^2 \q_u \q_d \Bigr] \lwc{u G}{}[][pr] \lwc{d G}{}[][st] \nns
		&\quad - 96 e g \q_u  \lwc{u G}{}[][pr] \lwc{d \gamma}{}[][st]  -96 e g \q_d  \lwc{u \gamma}{}[][pr] \lwc{d G}{}[][st]  - 96 g^2  \lwc{u \gamma}{}[][pr] \lwc{d \gamma}{}[][st]  \, , \\
\nnn
	\dlwc{dd}{RR}[S1][prst] &= - \left( 4 e^2 \q_d^2 + 12 g^2 C_F \right) \lwc{dd}{RR}[S1][prst] + \frac{1}{N_c} \left( 16 e^2 \q_d^2 + 16 g^2 C_F \right) \lwc{dd}{RR}[S1][ptsr] \nn
		&\quad + 4 g^2 \frac{C_F}{N_c} \lwc{dd}{RR}[S8][prst] + \frac{1}{N_c} \left[ 16 e^2 \q_d^2 C_F + 2 g^2 \left( \frac{2}{N_c^2} + N_c^2 - 3 \right) \right] \lwc{dd}{RR}[S8][ptsr] \nn
		&\quad -96 g^2 C_1 \lwc{d G}{}[][pr] \lwc{d G}{}[][st] + 4 g^2 C_A \frac{N_c^2-1}{N_c^2}  \lwc{d G}{}[][pt] \lwc{d G}{}[][sr]  - 96 e^2 \q_d^2 \lwc{d \gamma}{}[][pr] \lwc{d \gamma}{}[][st]  \, , \\
\nnn
	\dlwc{dd}{RR}[S8][prst] &= 8 g^2 \lwc{dd}{RR}[S1][prst] + \left[ 32 e^2 \q_d^2- \frac{16 g^2}{N_c} \right] \lwc{dd}{RR}[S1][ptsr]   - \left( 4 e^2 \q_d^2 - 4 g^2 C_F \right) \lwc{dd}{RR}[S8][prst]  \nn
		&\quad - \frac{1}{N_c} \left[ 16 e^2 \q_d^2 - 4 g^2 \left( \frac{2}{N_c} + N_c \right) \right] \lwc{dd}{RR}[S8][ptsr]  \nn
		&\quad + \Bigl[ g^2(-24 C_d + 4 C_A)  - 48 e^2 \q_d^2 \Bigr] \lwc{d G}{}[][pr] \lwc{d G}{}[][st]  - 8 g^2 \frac{C_A}{N_c} \lwc{d G}{}[][pt] \lwc{d G}{}[][sr]  \nn
		&\quad - 48 e g \q_d \Bigl( \lwc{d G}{}[][pr] \lwc{d \gamma}{}[][st]  +  \lwc{d \gamma}{}[][pr] \lwc{d G}{}[][st] \Bigr)  -48 g^2  \lwc{d \gamma}{}[][pr] \lwc{d \gamma}{}[][st]  \, , \\
\nnn
	\dlwc{uddu}{RR}[S1][prst] &= \Bigl[ 2 e^2 ( \q_u^2 - 4 \q_u \q_d + \q_d^2 )  - 12 g^2 C_F \Bigr] \lwc{uddu}{RR}[S1][prst] + \frac{1}{N_c} \left( 16 e^2 \q_u \q_d + 16 g^2 C_F \right)\lwc{ud}{RR}[S1][ptsr] \nn
		&\quad + 4 g^2 \frac{C_F}{N_c} \lwc{uddu}{RR}[S8][prst] + \frac{1}{N_c} \left[ 16 e^2 \q_u \q_d C_F + 2 g^2 \left( \frac{2}{N_c^2} + N_c^2 - 3 \right) \right] \lwc{ud}{RR}[S8][ptsr]  \nn
		&\quad + 8 g^2 C_A \frac{N_c^2-1}{N_c^2}  \lwc{u G}{}[][pt] \lwc{d G}{}[][sr]  \, , \\
\nnn
	\dlwc{uddu}{RR}[S8][prst] &= 8 g^2 \lwc{uddu}{RR}[S1][prst]  + \Bigl[ 2 e^2 (\q_u^2 - 4 \q_u \q_d + \q_d^2) + 4 g^2 C_F \Bigr] \lwc{uddu}{RR}[S8][prst] \nn
		&\quad + \left[ 32 e^2 \q_d \q_u - \frac{16 g^2}{N_c} \right] \lwc{ud}{RR}[S1][ptsr]  - \frac{1}{N_c} \left[ 16 e^2 \q_d \q_u - 4 g^2 \left( \frac{2}{N_c} + N_c \right) \right] \lwc{ud}{RR}[S8][ptsr]  \nn
		&\quad - 16 g^2 \frac{C_A}{N_c} \lwc{u G}{}[][pt] \lwc{d G}{}[][sr] \, .
\end{align}
%

\subsubsection[$\psi^4: (\overline L R)(\overline R L)$]{\boldmath $\psi^4: (\overline L R)(\overline R L)$}

\begin{align}
	\dlwc{eu}{RL}[S][prst] &= - \left[ 6 e^2 (\q_e^2 + \q_u^2) + 6 g^2 C_F \right] \lwc{eu}{RL}[S][prst]   - 192 e^2 \q_e \q_u  \lwc{e \gamma}{}[][pr] \lwc{u \gamma}{*}[][ts]  \, , \\
\nnn
	\dlwc{ed}{RL}[S][prst] &= - \left[ 6 e^2 (\q_e^2 + \q_d^2) + 6 g^2 C_F \right] \lwc{ed}{RL}[S][prst]  - 192 e^2 \q_e \q_d   \lwc{e \gamma}{}[][pr] \lwc{d \gamma}{*}[][ts]  \, , \\
\nnn
	\dlwc{\nu edu}{RL}[S][prst] &= - \left( 6 e^2 \q_u \q_d + 6 g^2 C_F \right) \lwc{\nu edu}{RL}[S][prst]  \, .
\end{align}

\subsubsection[$\psi^4: \Delta L=4$]{\boldmath $\psi^4: \Delta L=4$}

\begin{align}
	\label{eq:DeltaL4RGE}
	\dlwc{\nu\nu}{LL}[S][prst] &= 0 \, .
\end{align}

\subsubsection[$\psi^4: \Delta L=2$]{\boldmath $\psi^4: \Delta L=2$}

\begin{align}
	\dlwc{\nu e}{LL}[S][prst] &= -  6 e^2 \q_e^2\, \lwc{\nu e}{LL}[S][prst]  \, , \\
\nnn
	\dlwc{\nu e}{LL}[T][prst] &=  2 e^2 \q_e^2\, \lwc{\nu e}{LL}[T][prst]  \, , \\
\nnn
	\dlwc{\nu e}{LR}[S][prst] &= -  6 e^2 \q_e^2\, \lwc{\nu e}{LR}[S][prst]  \, , \\
\nnn
	\dlwc{\nu u}{LL}[S][prst] &= -6\left( e^2 \q_u^2 + g^2 C_F \right)  \lwc{\nu u}{LL}[S][prst]  \, , \\
\nnn
	\dlwc{\nu u}{LL}[T][prst] &= 2\left( e^2 \q_u^2  + g^2 C_F\right)  \lwc{\nu u}{LL}[T][prst]  \, , \\
\nnn
	\dlwc{\nu u}{LR}[S][prst] &= -6\left( e^2 \q_u^2 + g^2 C_F \right)  \lwc{\nu u}{LR}[S][prst]  \, , \\
\nnn
	\dlwc{\nu d}{LL}[S][prst] &=  -6\left( e^2 \q_d^2  + g^2 C_F \right)  \lwc{\nu d}{LL}[S][prst]  \, , \\
\nnn
	\dlwc{\nu d}{LL}[T][prst] &=  2\left( e^2 \q_d^2  + g^2 C_F \right)  \lwc{\nu d}{LL}[T][prst]  \, , \\
\nnn
	\dlwc{\nu d}{LR}[S][prst] &= -6\left( e^2 \q_d^2  + g^2 C_F\right)  \lwc{\nu d}{LR}[S][prst]  \, , \\
\nnn
	\dlwc{\nu e d u}{LL}[S][prst] &= -6 \left( e^2 \q_u \q_d + g^2 C_F \right) \lwc{\nu e d u}{LL}[S][prst] + 24 e^2\left( \q_u^2 - \q_d^2 \right) \lwc{\nu e d u}{LL}[T][prst]  \, , \\
\nnn
	\dlwc{\nu e d u}{LL}[T][prst] &= \left[ 2 e^2(\q_u^2 + \q_u \q_e - 2 \q_e^2 ) + 2 g^2 C_F \right] \lwc{\nu e d u}{LL}[T][prst] + \frac{1}{2} e^2\left( \q_u^2 - \q_d^2 \right) \lwc{\nu e d u}{LL}[S][prst]  \, , \\
\nnn
	\dlwc{\nu e d u}{LR}[S][prst] &= -6 \left( e^2 \q_u \q_d + g^2 C_F \right) \lwc{\nu e d u}{LR}[S][prst]  \, , \\
\nnn
	\dlwc{\nu e d u}{RL}[V][prst] &=  -6 e^2 \q_e \q_d \lwc{\nu e d u}{RL}[V][prst]  \, , \\
\nnn
	\dlwc{\nu e d u}{RR}[V][prst] &= 6 e^2 \q_e \q_u \lwc{\nu e d u}{RR}[V][prst] \, .
\end{align}

\clearpage

\subsubsection[$\psi^4: \Delta B=\Delta L=1$]{\boldmath $\psi^4: \Delta B=\Delta L=1$}

\begin{align}
	\dlwc{udd}{LL}[S][prst] &= \left[-4 g^2 +e^2[6 \q_u \q_d - 2(\q_u-\q_d) \q_d] \right] \lwc{udd}{LL}[S][prst] + 4 e^2 (\q_u-\q_d) \q_d  \lwc{udd}{LL}[S][psrt] \, , \\
\nnn
	\dlwc{duu}{LL}[S][prst] &= \left[- 4 g^2 + e^2[6(\q_d \q_u + \q_u \q_e )-2 (\q_d - \q_u)(\q_u-\q_e)]\right] \lwc{duu}{LL}[S][prst]  \nns
		&\quad + 4 e^2 (\q_d - \q_u)(\q_u-\q_e) \lwc{duu}{LL}[S][psrt]  \, , \\
\nnn
	\dlwc{uud}{LR}[S][prst] &= \left[- 4 g^2  + 6 e^2 (\q_u^2+\q_d \q_e)\right]  \lwc{uud}{LR}[S][prst]  \, , \\
\nnn
	\dlwc{duu}{LR}[S][prst] &= \left[-4 g^2 + 6 e^2 (\q_d \q_u + \q_u \q_e )  \right] \lwc{duu}{LR}[S][prst]  \, , \\
\nnn
	\dlwc{uud}{RL}[S][prst] &= \left[- 4 g^2 + 6 e^2 (\q_u^2+\q_d \q_e)\right] \lwc{uud}{RL}[S][prst]  \, , \\
\nnn
	\dlwc{duu}{RL}[S][prst] &= \left[- 4 g^2 + 6 e^2 (\q_d \q_u + \q_u \q_e ) \right] \lwc{duu}{RL}[S][prst]   \, , \\
\nnn
	\dlwc{dud}{RL}[S][prst] &= \left( - 4 g^2 + 6e^2 \q_d \q_u  \right) \lwc{dud}{RL}[S][prst]  \, , \\
\nnn
	\dlwc{ddu}{RL}[S][prst] &= \left( - 4 g^2 + 6 e^2 \q_d^2  \right) {\lwc{ddu}{RL}[S][prst]}   \, , \\
\nnn
	\dlwc{duu}{RR}[S][prst] &= \left[- 4 g^2 + e^2[6 (\q_d \q_u +\q_u \q_e)-2(\q_d - \q_u)(\q_u-\q_e)] \right] \lwc{duu}{RR}[S][prst]  \nn
		&\quad + 4 e^2 (\q_d - \q_u)(\q_u-\q_e)  \lwc{duu}{RR}[S][psrt]   \, .
\end{align}

\subsubsection[$\psi^4: \Delta B = -\Delta L=1$]{\boldmath $\psi^4: \Delta B = -\Delta L=1$}

\begin{align}
	\dlwc{ddd}{LL}[S][prst] &= \left[ -4g^2+ 6 e^2  (\q_d^2-\q_e \q_d) \right]  \lwc{ddd}{LL}[S][prst]  \, , \\
\nnn
	\dlwc{udd}{LR}[S][prst] &= \left( -4g^2 + 6 e^2 \q_u \q_d \right)  \lwc{udd}{LR}[S][prst]  \, , \\
\nnn
	\dlwc{ddu}{LR}[S][prst] &= \left( -4g^2 + 6 e^2 \q_d^2 \right)  \lwc{ddu}{LR}[S][prst]  \, , \\
\nnn
	\dlwc{ddd}{LR}[S][prst] &=  \left[ -4g^2+ 6 e^2 (\q_d^2-\q_e \q_d)  \right]  \lwc{ddd}{LR}[S][prst]  \, , \\
\nnn
	\dlwc{ddd}{RL}[S][prst] &= \left[ -4g^2 + 6 e^2 (\q_d^2-\q_e \q_d) \right]  \lwc{ddd}{RL}[S][prst]  \, , \\
\nnn
	\dlwc{udd}{RR}[S][prst] &= \left[ -4g^2 + e^2 [6 \q_u \q_d - 2 (\q_u-\q_d)\q_d ] \right] \lwc{udd}{RR}[S][prst] + 4 e^2 (\q_u-\q_d)\q_d  \lwc{udd}{RR}[S][ptsr]  \, , \\
\nnn
	\dlwc{ddd}{RR}[S][prst] &=  \left[ -4g^2 + 6 e^2  (\q_d^2-\q_e \q_d) \right]  \lwc{ddd}{RR}[S][prst]  \, .
\end{align}


\clearpage
\bibliographystyle{JHEP}
\bibliography{RGL}


\end{document}